\documentclass[twocolumn]{aastex63}
\usepackage{CJK}
\usepackage{amsmath}
\usepackage{mathtools}
\usepackage[caption=false]{subfig}

\shorttitle{The Magnetic Field in G28.34}
\shortauthors{Liu et al.}

\begin{document}
\title{Magnetic fields in the early stages of massive star formation as revealed by ALMA}

\correspondingauthor{Keping Qiu}
\email{kpqiu@nju.edu.cn}

\author[0000-0002-4774-2998]{Junhao Liu}
\affil{School of Astronomy and Space Science, Nanjing University, 163 Xianlin Avenue, Nanjing, Jiangsu 210023, People's Republic of China}
\affil{Key Laboratory of Modern Astronomy and Astrophysics (Nanjing University), Ministry of Education, Nanjing, Jiangsu 210023, People's Republic of China}
\affil{Center for Astrophysics $\vert$ Harvard \& Smithsonian, 60 Garden Street, Cambridge, MA 02138, USA}

\author[0000-0003-2384-6589]{Qizhou Zhang}
\affil{Center for Astrophysics $\vert$ Harvard \& Smithsonian, 60 Garden Street, Cambridge, MA 02138, USA}

\author[0000-0002-5093-5088]{Keping Qiu}
\affil{School of Astronomy and Space Science, Nanjing University, 163 Xianlin Avenue, Nanjing, Jiangsu 210023, People's Republic of China}
\affil{Key Laboratory of Modern Astronomy and Astrophysics (Nanjing University), Ministry of Education, Nanjing, Jiangsu 210023, People's Republic of China}

\author[0000-0003-2300-2626]{Hauyu Baobab Liu}
\affil{Academia Sinica Institute of Astronomy and Astrophysics, P.O. Box 23-141, Taipei 10617, Taiwan, Republic of China}

\author{Thushara Pillai}
\affil{Max-Planck-Institut f{\"u}r Radioastronomie, Auf dem H{\"u}gel 69, 53121 Bonn, Germany}
\affil{Institute for Astrophysical Research, 725 Commonwealth Ave, Boston University Boston, MA 02215, USA}

\author[0000-0002-3829-5591]{Josep Miquel Girart}
\affil{Institut de Ci$\grave{e}$ncies de l'Espai (ICE, CSIC), Can Magrans s/n, E-08193 Cerdanyola del Vall$\grave{e}$s, Catalonia}
\affil{Institut d'Estudis Espacials de de Catalunya (IEEC), E-08034 Barcelona, Catalonia}

\author{Zhi-Yun Li}
\affil{Astronomy Department, University of Virginia, Charlottesville, VA 22904-4325, USA}

\author{Ke Wang}
\affil{Kavli Institute for Astronomy and Astrophysics, Peking University, 5 Yiheyuan Road, Haidian District, Beijing 100871,
People's Republic of China}

\begin{abstract}
We present 1.3 mm ALMA dust polarization observations at a resolution of $\sim$0.02 pc of three massive molecular clumps, MM1, MM4, and MM9, in the infrared dark cloud G28.34+0.06. With the sensitive and high-resolution continuum data, MM1 is resolved into a cluster of condensations. The magnetic field structure in each clump is revealed by the polarized emission. We found a trend of decreasing polarized emission fraction with increasing Stokes $I$ intensities in MM1 and MM4. Using the angular dispersion function method (a modified Davis-Chandrasekhar-Fermi method), the plane-of-sky magnetic field strength in two massive dense cores, MM1-Core1 and MM4-Core4, are estimated to be $\sim$1.6 mG and $\sim$0.32 mG, respectively. The total virial parameters in MM1-Core1 and MM4-Core4 are calculated to be $\sim$0.76 and $\sim$0.37, respectively, suggesting that massive star formation does not start in equilibrium. Using the polarization-intensity gradient-local gravity method, we found that the local gravity is closely aligned with intensity gradient in the three clumps, and the magnetic field tends to be aligned with the local gravity in MM1 and MM4 except for regions near the emission peak, which suggests that the gravity plays a dominant role in regulating the gas collapse. Half of the outflows in MM4 and MM9 are found to be aligned within 10$\degr$ of the condensation-scale ($<$0.05 pc) magnetic field, indicating that the magnetic field could play an important role from condensation to disk scale in the early stage of massive star formation. We also find that the fragmentation in MM1-Core1 cannot be solely explained by thermal Jeans fragmentation or turbulent Jeans fragmentation. 
\end{abstract}


\section{Introduction}
It is clear that stars are formed from the collapse of molecular dense cores\footnote{Following the nomenclature in \citet{2009ApJ...696..268Z}, we refer to a molecular clump as an entity of $\sim$ 1 pc, a dense core as an entity of $\sim$ 0.1 pc, and a condensation as an entity of $\sim$ 0.01 pc that forms one star or a group of stars.} that are created by the contraction of over dense regions in molecular clouds when gravity overcomes the internal pressure such as magnetic fields or turbulence \citep{1987ARA&A..25...23S}. In the formation of massive stars (mass $>$ 8 $M_{\odot}$), turbulence is believed to play an important role because massive star-forming regions are usually found to be supersonic and have large line widths. However, the role of magnetic fields in massive star formation is still not well understood \citep{2014PhR...539...49K, 2014prpl.conf..149T, 2019FrASS...6....3H, 2019FrASS...6...15P}. 

The Radiative Alignment Torque (RAT) theory predicts that asymmetrical dust grains with sizes $>$0.05 $\mu$m are expected to spin-up due to the radiative torque in the presence of a radiation field and the short axis of these dust grains would tend to be aligned with magnetic field lines \citep{2007JQSRT.106..225L, 2007MNRAS.378..910L}. Theoretical and observational evidence has supported the RAT mechanism as the dominant grain alignment mechanism to produce polarized thermal emission at millimeter/sub-millimeter (mm/submm) wavelengths in diffuse intersteller medium and molecular clouds with $\mu$m or sub-$\mu$m size grains \citep{2015ARA&A..53..501A}. The exception is that the grain size grows to millimeter/sub-millimeter in protoplanetary disks and the observed polarization would be dominated by those produced from dust self-scattering \citep{2015ApJ...809...78K, 2016MNRAS.456.2794Y} or grains aligned by strong radiation fields \citep{2017ApJ...839...56T}. Thus, observing the polarized dust emission at mm/sub-mm offers a powerful way to trace the plane-of-sky magnetic field in star-forming molecular clouds at scales greater than the disk scales. 

In the past two decades, there has been an increasing number of high-resolution and high-sensitivity observational studies of the dust polarization in high-mass star-forming regions \citep[for a detailed review see ][]{2019FrASS...6....3H}. A variety of field structures from hourglass-like shapes \citep[e.g., G31.41, G240.31, OMC 1, ][]{2009Sci...324.1408G, 2014ApJ...794L..18Q, 2017ApJ...842...66W} to more complex and chaotic morphologies \citep[e.g., Orion KL, G5.89, ][]{ 1998ApJ...502L..75R, 2009ApJ...695.1399T} have been reported by these studies. The magnetic field in massive star-forming regions is found to be dynamically important from the clump scale to the core scale and the core scale magnetic field does not show strong correlation with the outflow axis \citep{2014ApJ...792..116Z}. Quantitatively, the magnetic field plays an important role in supporting the massive dense cores against gravitational collapse \citep{2013ApJ...772...69G, 2014A&A...567A.116F}. 

Despite the significant progress made by the recent polarization observations of massive star formation regions, most of these studies of magnetic fields targeted evolved star-forming regions. Due to the relatively weak polarized dust emission and the limitation of instrumental sensitivity, there are only a handful of single-dish case studies about the magnetic field in early massive star formation regions \citep{2015ApJ...799...74P, 2018A&A...620A..26J,  2018ApJ...859..151L, 2019ApJ...878...10T, 2019ApJ...883...95S} and there is only one inteferometric study of the magnetic field at the onset of massive star formation in the infrared dark cloud 18310-4 where the magnetic field structure is not well resolved \citep{2018A&A...614A..64B}. The role of magnetic fields in the initial stage of massive star formation remains an open question.

There are two distinct models of massive star formation. The turbulent core accretion model \citep{2002Natur.416...59M} envisioned that massive stars are formed via the monolithic collapse of massive dense cores in virial equilibrium, where the pressure support comes from turbulence and the magnetic field \citep{2013ApJ...779...96T}. Alternatively, the competitive accretion model \citep{1997MNRAS.285..201B} proposed that a cluster of low-mass protostars compete with one another to accrete from the natal gas reservoir and the protostars near the center of gravitational potential accreting at higher rates can thus form massive stars. \citet{2005Natur.438..332K} showed that a sub-virial state is required for the competitive accretion. Thus, the dynamical state of massive dense cores needs to be measured to distinguish between the two models.

Infrared dark clouds (IRDCs), which were first identified as dark regions against the diffuse mid-infrared emission in the Galactic plane \citep{1996A&A...315L.165P}, are believed to harbor the early phase of massive star formation. Recent observations and stability analyses of IRDCs show that the turbulence and the thermal pressure are too weak to provide enough support against the gravity in dense clumps and dense cores \citep{2011A&A...530A.118P, 2015ApJ...804..141Z, 2016ApJ...833..209O, 2017ApJ...841...97S, 2018ApJ...855....9L}, suggesting that magnetic fields must provide significant support to bring the core to equilibrium, or that massive star formation are not in equilibrium. Thus, observational studies of the magnetic field in IRDCs are critical to address the question: does massive star formation start in equilibrium?

At a kinematic distance of $\sim$4.8 kpc, G28.34+0.06 (hereafter G28.34) is a massive ($>10^4 M_{\odot}$) filamentary IRDC that harbors more than ten massive ($10^2 - 10^3 M_{\odot}$) molecular clumps \citep{1998ApJ...508..721C, 2006A&A...450..569P, 2006ApJ...641..389R,  2006ApJ...653.1325S, 2009ApJ...696..484B, 2017ApJ...840...22L, 2018RNAAS...2...52W}. Three prominent clumps, MM1 (also named as P2, C9, or Dragon-Head), MM4 (also named as P1, C2, or Dragon-Belly), and MM9 (also named as S, C1, or Dragon-Tail) are revealed from millimetre (mm) dust continuum and mid-infrared extinction maps \citep{2006ApJ...641..389R, 2009ApJ...696..484B}. All the three clumps are associated with water masers \citep{2006ApJ...651L.125W, 2008ApJ...672L..33W} and outflows \citep[][ and this work]{2011ApJ...735...64W, 2015ApJ...804..141Z, 2016ApJ...828..100F}, indicating that star formation has already began. The large gas reservoir makes them potential sites to study clustered massive star formation. 

The three massive clumps are found to be in different evolutionary stages. With low temperatures \citep[$\sim$15 K, ][]{2018RNAAS...2...52W}, high CO depletion \citep{2016A&A...592A..21F} and high level of deuteration \citep{2010ApJ...713L..50C}, the 70 $\micron$ dark clump MM9 appears to be in an early stage of massive star formation. MM9 is resolved into two dense cores \citep[C1-N and C1-S, also named as S-B and S-A, ][]{2013ApJ...779...96T, 2016A&A...592A..21F} where the south core, C1-S, further fragments into two protostellar condensations \citep[C1-Sa and C1-Sb, ][]{2016ApJ...821L...3T}. MM4 has a temperature of $\sim$16 K \citep{2008ApJ...672L..33W} and is associated with an IR-bright protostellar source with a luminosity of 152 $L_{\odot}$ \citep{2012A&A...547A..49R}. With high-resolution dust continuum observations, the MM4 clump is resolved into six dense cores (MM4-Core1 through MM4-Core6) along the main filament and each core further fragments into several condensations \citep{2009ApJ...696..268Z, 2011ApJ...735...64W, 2015ApJ...804..141Z, 2019ApJ...873...31K}. The more energetic outflow activities, CH$_3$OH maser detections, less deuterium fraction, and higher luminosity in MM4 indicate that MM4 is more evolved than MM9 \citep{2019ApJ...883..202F}. The MM1 clump is warmer \citep[$\sim$30 K, ][]{2008ApJ...672L..33W} than MM4 and MM9 and is associated with an IR-bright protostellar source with a luminosity of 2950 $L_{\odot}$ \citep{2012A&A...547A..49R}. Previous dust continuum observations have resolved MM1 into two fragments \citep[P2-SMA1 and P2-SMA2,][]{2009ApJ...696..268Z}. The rich organic molecular line emissions in P2-SMA1 indicate that it might be a hot molecular core \citep{2009ApJ...696..268Z}. These observations suggest that MM1 is at a later stage of massive star formation than MM4. 

We present 1.3 mm Atacama Large Millimeter/submillimeter Array (ALMA) polarization observations toward clumps MM1, MM4, and MM9 in this paper. In Section \ref{section:observation}, we summarize the observation and data reduction. In Section \ref{section:results}, we present maps of dust continuum, polarized dust emission, and molecular line emission, and derive the magnetic field strength. In Section \ref{section:discussion}, we discuss the dynamical state, outflow-magnetic field alignment, and fragmentation, and compare the orientations of magnetic fields, local gravity, and intensity gradient. In Section \ref{section:summary}, we provide a summary of this paper.

\section{Observation} \label{section:observation}

\begin{deluxetable}{cccc}[t!]
\tablecaption{Source coordinates \label{tab:sources}}
\tablecolumns{4}
\tablewidth{0pt}
\tablehead{
\colhead{Source} &
\colhead{Field} &
\colhead{$\alpha_{\mathrm{J2000}}$} & 
\colhead{$\delta_{\mathrm{J2000}}$}
}
\startdata
MM1 & MM1 & $18^{\mathrm{h}}42^{\mathrm{m}}52^{\mathrm{s}}.06$ & $-03\degr59\arcmin54\arcsec.3$ \\\hline
MM4 & MM4\_1 & $18^{\mathrm{h}}42^{\mathrm{m}}51^{\mathrm{s}}.05$ & $-04\degr03\arcmin08\arcsec.6$ \\
& MM4\_2 & $18^{\mathrm{h}}42^{\mathrm{m}}50^{\mathrm{s}}.69$ & $-04\degr03\arcmin13\arcsec.4$ \\
& MM4\_3 & $18^{\mathrm{h}}42^{\mathrm{m}}50^{\mathrm{s}}.39$ & $-04\degr03\arcmin19\arcsec.0$ \\
& MM4\_4 & $18^{\mathrm{h}}42^{\mathrm{m}}49^{\mathrm{s}}.88$ & $-04\degr03\arcmin24\arcsec.5$ \\\hline
MM9 & MM9\_1 & $18^{\mathrm{h}}42^{\mathrm{m}}46^{\mathrm{s}}.48$ & $-04\degr04\arcmin14\arcsec.3$ \\
& MM9\_2 & $18^{\mathrm{h}}42^{\mathrm{m}}46^{\mathrm{s}}.91$ & $-04\degr04\arcmin09\arcsec.3$ \\
\enddata

\end{deluxetable}


\begin{deluxetable*}{cccccccc}[t!]
\tablecaption{Observational parameters \label{tab:observation}}
\tablecolumns{8}
\tablewidth{0pt}
\tablehead{
\colhead{Date\tablenotemark{a}} &
\colhead{Configuration} &
\colhead{$N_{ant}$\tablenotemark{b}} & 
\colhead{Bandpass} & 
\colhead{Gain} &  
\colhead{Flux} &  
\colhead{Pol} & 
\colhead{Sources} \\
\colhead{} & \colhead{} &
\colhead{} & 
\colhead{calibrator} & \colhead{calibrator} & \colhead{calibrator} & \colhead{calibrator} & \colhead{}
}
\startdata
2017 Apr 18 & C40-3 & 48   & J1751+0939 & J1851+0035 & Titan & J1751+0939 & MM1, MM4, MM9\\
2018 Apr 29 & C43-3 & 43   & J1751+0939 & J1851+0035 & Titan & J1751+0939  & MM1, MM4, MM9\\
2018 Jun 23 & C43-1 & 47   &J1751+0939 & J1851+0035 & J1751+0939 & J1924-2914 & MM4, MM9\\
2018 Jun 29 & C43-1 & 45  &J1751+0939 & J1851+0035 & J1751+0939 & J1924-2914 & MM4, MM9\\
\enddata
\tablenotetext{a}{Observations were under precipitable water vapor (PWV) ranging from 1.3 to 2.1 mm.}
\tablenotetext{b}{Number of antennas.}
\end{deluxetable*}

\begin{figure}[htbp]
\centering
\includegraphics[scale=.6]{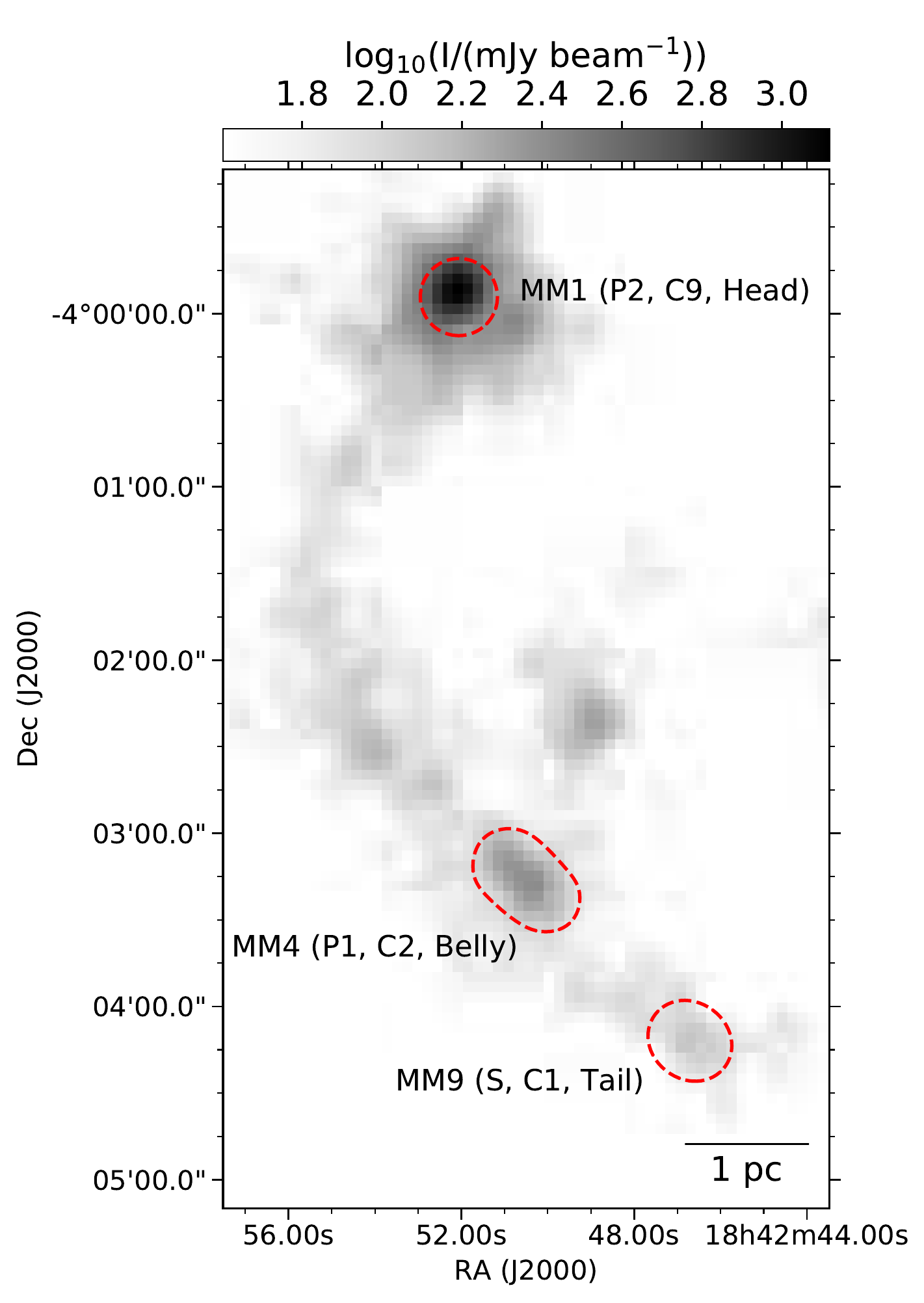}
\caption{Overview of the IRDC G28.34 (the Dragon Nebula). The gray scale shows the IRAM-30m 1.2 mm dust continuum of G28.34 with a resolution of 11$\arcsec$ \citep{2006ApJ...641..389R}. The red dashed contours indicate the Full Width at Half-Maximum (FWHM) field of view of our ALMA observations.  \label{fig:figG28}}
\end{figure}

Figure \ref{fig:figG28} presents an overview of G28.34 in IRAM-30m 1.3mm dust continuum \citep{2006ApJ...641..389R}. The observations of three massive clumps (MM1, MM4, MM9) in G28.34 were carried out with the ALMA between 2017 Apr 18 to 2018 Jun 29 under projects 2016.1.00248.S (Cycle 4; PI: Zhang) and 2017.1.00793.S (Cycle 5; PI: Zhang). An ALMA execution on 2017 April 21 under project 2016.1.00248.S was failed due to some correlator issues and was not used for our analysis. Tables \ref{tab:sources} and \ref{tab:observation} list the detailed information of the observations. The total on-source time is 17 minutes for MM1 and 17 minutes (in C40-3 or C43-3 configuration) + 8 minutes (in C43-1 configuration) for each fields of MM4 and MM9. The receiver was tuned to cover frequencies $\sim$215.5-219.5 GHz and $\sim$232.5-234.5 GHz (band 6), with a total bandwidth of 5.6 GHz (three basebands, with 1.875 GHz effective bandwidth each) for the dust continuum emission in the full polarization mode. Four spectral windows in another baseband were set to cover the CO (2-1), OCS (19-18), $^{13}$CS (5-4), and N$_2$D$^+$ (3-2) lines with a channel width of 122 kHz (0.16 km s$^{-1}$) over a bandwidth of 58.6 MHz ($\sim$76 km s$^{-1}$). 

The data taken in 2017 Apr 18 were manually calibrated by the authors using the Common Astronomy Software Applications \citep[CASA, ][]{2007ASPC..376..127M} and the rest of the data were calibrated by the ALMA supporting staff. The systematic flux uncertainty of ALMA at Band 6 due to calibration is $\sim$10 \%. We performed two iterations of phase-only self-calibrations on the continuum data using the CASA. The molecular line cubes and Stokes $I$, $Q$, and $U$ maps of dust continuum were produced from the visibility data using the CASA task \textit{TCLEAN} with a briggs weighting parameter of robust = 0.5. The maps for MM4 and MM9 are constructed from four-pointing mosaic of MM4\_1, MM4\_2, MM4\_3, and MM4\_4, and two-pointing mosaic of MM9\_1 and MM9\_2, respectively. The synthesized beams of the maps are 0$\arcsec$.8-0$\arcsec$.9 $\times$ 0$\arcsec$.6-0$\arcsec$.7 ($\sim$0.02-0.015 pc at a source distance of 4.8 kpc). The maximum recoverable scale\footnote{https://almascience.eso.org/observing/prior-cycle-observing-and-configuration-schedule} is $\sim$7$\arcsec$ ($\sim$0.14 pc at 4.8 kpc) for MM1 and $\sim$13$\arcsec$ ($\sim$0.3 pc at 4.8 kpc) for MM4 and MM9. After primary beam correction, the 1$\sigma$ rms noises of the Stokes $I$ maps of dust continuum reach $\sim$300, 80, and 60 $\mu$Jy/beam for MM1, MM4, and MM9, respectively, while the the Stokes $Q$/$U$ maps have rms noises of $\sim$35 $\mu$Jy/beam for MM1 and $\sim$15 $\mu$Jy/beam for MM4 and MM9. The sensitivity of the primary beam response corrected spectral line cubes for MM1, MM4, and MM9 is $\sim$2.2, 1.2, and 1.4 mJy/beam, respectively, with a velocity channel width of 0.5 km s$^{-1}$. 

Because the polarized intensity and polarized percentage are defined as positive values, the measured quantities of the two parameters are biased toward larger values \citep{2006PASP..118.1340V}. The debiased polarized intensity $PI$ and its corresponding uncertainty $\sigma_{PI}$ are calculated as:
\begin{equation}
PI = \sqrt{Q^2 + U^2 - \sigma_{Q/U}^2},
\end{equation} 
and
\begin{equation}
\sigma_{PI} = \sqrt{2}\sigma_{Q/U},
\end{equation} 
where $\sigma_{Q/U}$ is the 1$\sigma$ rms noise of the $Q$/$U$ maps. The debiased polarization percentage $P$ and its uncertainty $\delta P$ are therefore derived by:
\begin{equation}
P = \frac{PI}{I},
\end{equation}
and
\begin{equation}
\delta P = \sqrt{(\frac{\sigma_{PI}^2}{I^2} + \frac{\sigma_{I}^2 (Q^2 + U^2)}{I^4})},
\end{equation}
where $\sigma_{I}$ is the 1$\sigma$ rms noise of the $I$ map.

Finally, the polarization position angle $\theta$ and its uncertainty $\delta \theta$ \citep{1993A&A...274..968N} are estimated to be:
\begin{equation}
\theta = \frac{1}{2} \tan ^{-1} (\frac{U}{Q}),
\end{equation}
and 
\begin{equation}
\delta \theta = \frac{1}{2} \sqrt{\frac{\sigma_{Q/U}^2}{(Q^2 + U^2)}},
\end{equation}


\section{Results}\label{section:results}

\subsection{Dust continuum} \label{section:cont}
The 1.3 mm continuum emissions of three clumps are shown in Figure \ref{fig:figG28B}. In this subsection, we briefly overview the dust continuum emissions in MM4 and MM9 and focus on interpreting the dust continuum of MM1. 

\begin{figure*}[!tbp]
\includegraphics[scale=1.4]{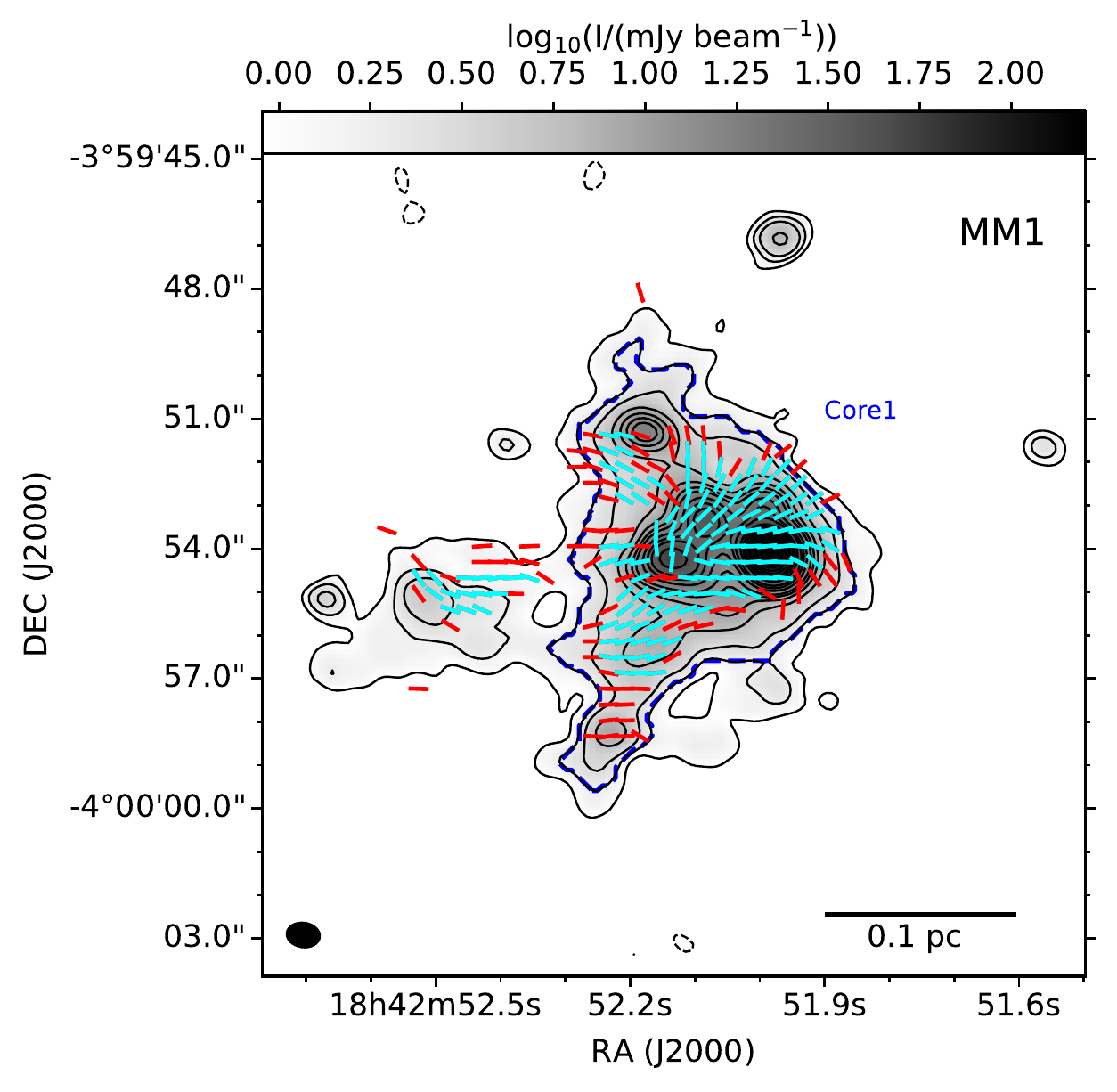}
\caption{Dust continuum and magnetic orientation maps. The Stokes $I$ of the ALMA 1.3 mm continuum is shown in gray scales and in contours. The ALMA 1.3 mm continuum of three clumps are also shown in contours. The contour levels are ($\pm$3, 6, 10, 20, 30, 40, 50, 70, 90, 110, 150, 180, 210, 250, 290, 340, 390, 450) $\times \sigma_{I}$, where $\sigma_{I}$ is the rms noise of the Stokes $I$ maps (see Section \ref{section:observation}). Line segments represent the orientation of the plane-of-sky magnetic field and have arbitrary length. Red and cyan line segments correspond to data with $PI/\sigma_{PI} > 2$ and $PI/\sigma_{PI} > 3$, respectively. Blue dashed regions indicate the areas of MM1-Core1 and MM4-Core4 in which the polarization data are used for the angular dispersion function analysis. The size of the synthesized beam are indicated in the lower left corner of each panel.  \label{fig:figG28B}}
\end{figure*}

\begin{figure*}[!tbp]
\ContinuedFloat
\includegraphics[scale=1.4]{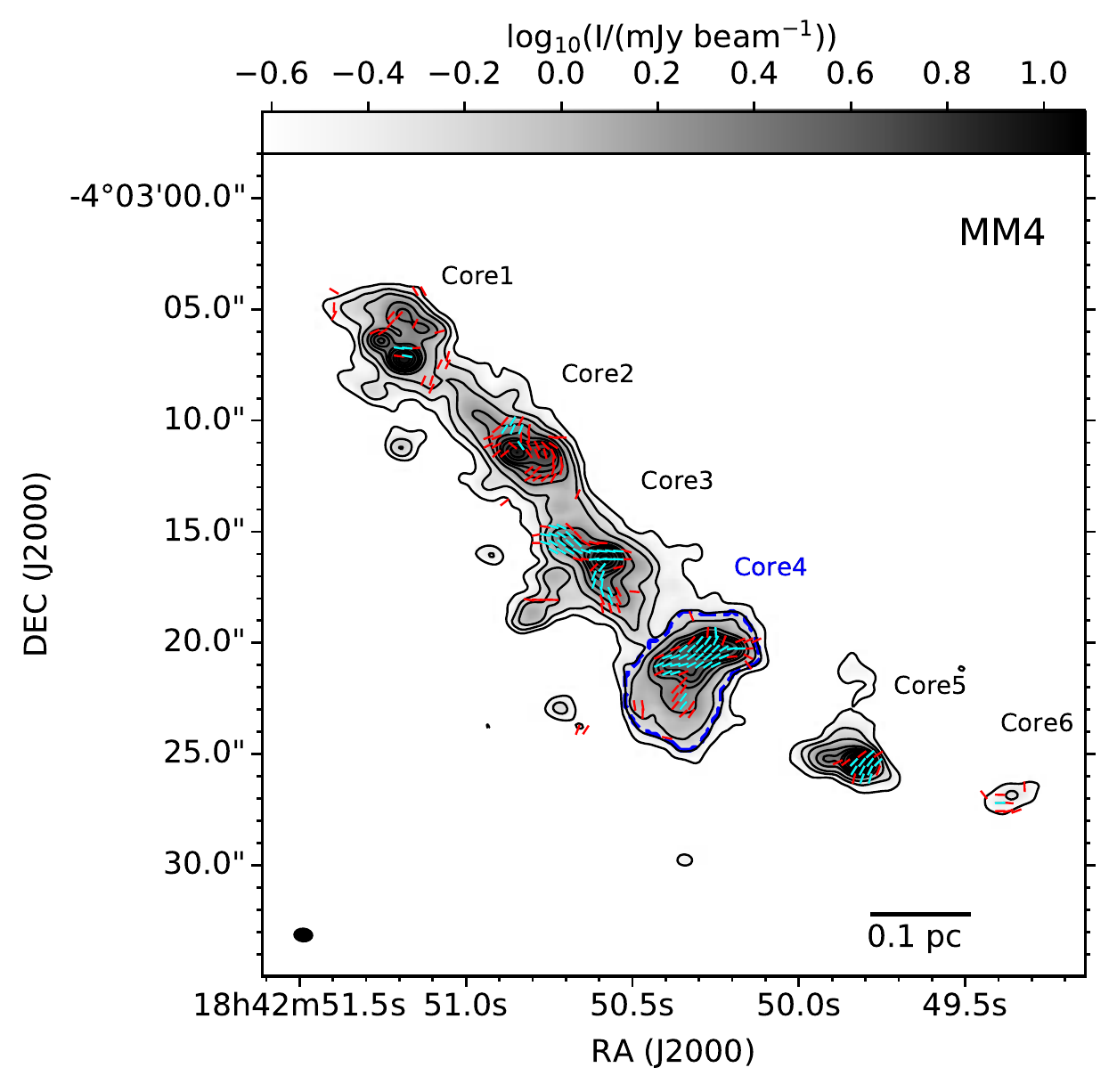}
\caption{(Continued)}
\end{figure*}

\begin{figure*}[!tbp]
\ContinuedFloat
\includegraphics[scale=1.4]{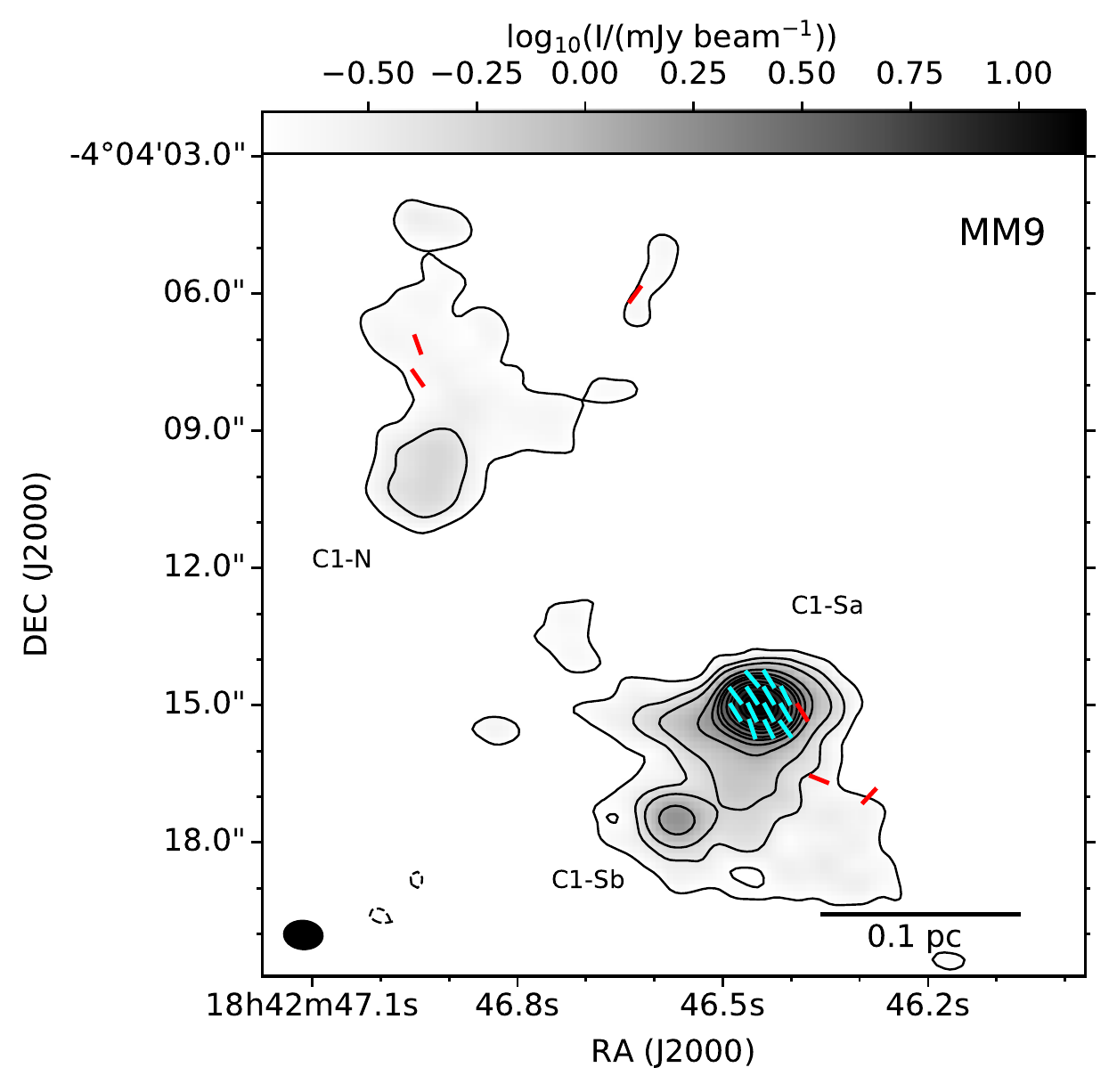}
\caption{(Continued)}
\end{figure*}

The MM4 region was studied by \citet{2015ApJ...804..141Z}, \citet{2018ApJ...855L..25K}, and \citet{2019ApJ...873...31K}, which revealed six cores (MM4-Core1 to MM4-Core6 from northeast to southwest) along the natal filament and identified the condensation-scale fragment structures in the continuum emission. We adopt the condensations identified by \citet{2015ApJ...804..141Z} and \citet{2019ApJ...873...31K} in our study. In MM9, two massive dense cores (C1-N and C1-S) were identified by \citet{2013ApJ...779...96T}. C1-S was further resolved into two protostellar condensations \citep[C1-Sa and C1-Sb: ][]{2016ApJ...821L...3T, 2018ApJ...867...94K}. Due to sensitivity limitation, the two more candidate protostellar sources \citep[C1a and C1b: ][]{2016ApJ...821L...3T} away from C1-N and C1-S are not detected in our ALMA 1.3 mm continuum.

The continuum emission of MM1 is resolved down to condensation scales (see Figure \ref{fig:figG28B}). The emission is dominated by a major core (hereafter MM1-Core1) in the west, which further fragments into several condensation structures. Several more condensations are detected in the east and in the southwest, which are connected to the main core. Another strong condensation is seen to the northwest of the main core. There appears to be two additional 6$\sigma$ continuum peaks to the northeast and to the east of MM1-Core1. 

To characterize the dense structures in MM1, we applied the dendrogram \citep{2008ApJ...679.1338R} technique on the dust continuum using the Python package \textit{astrodendro}. The dendrogram abstracts the changing topology of the isosurfaces as a function of contour levels and tracks the hierarchical structure over a range of scales \citep{2008ApJ...679.1338R}. This algorithm mainly has three parameters: the minimum value for the structure to be considered (min\_value), the minimum height required for a structure to be retained as independent (min\_delta), and the minimum number of pixels for a structure to be considered (min\_npix). Following similar works in \citet{2018ApJ...853..160C} and \citet{2018ApJ...862..105L}, we adopt min\_value = 4$\sigma$, min\_delta = 1$\sigma$, and set min\_npix to be the number of pixels in half beam area. We note that the combination of min\_value and min\_delta ensures that the identified structures have peak fluxes $>5 \sigma$. 

\begin{figure}[htbp]
\centering
\includegraphics[scale=.65]{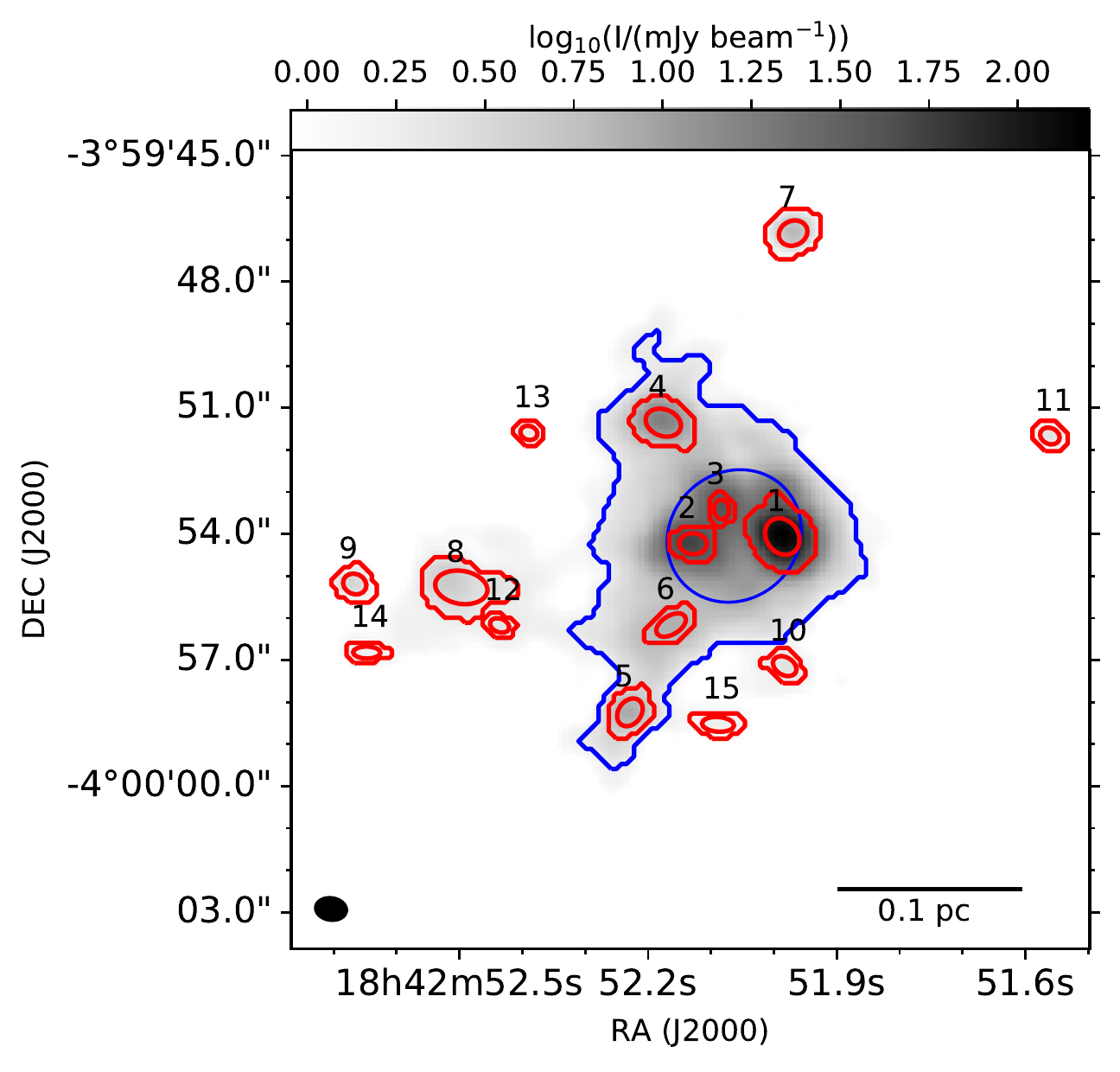}
\caption{Structures identified by dendrogram in MM1. The gray scale shows the 1.3 mm continuum emission. Red and blue contours indicate the boundries of the identified leaves (i.e., condensations) and MM1-Core1, respectively. Red and blue ellipses indicate the FWHM and position angle of the condensations and MM1-Core1, respectively, computed by dendrogram. \label{fig:figMM1dendro}}
\end{figure}

The structures identified by dendrogram in MM1 are shown in Figure \ref{fig:figMM1dendro}. The dendrogram reports the coordinates, the integrated flux ($S_{\mathrm{int}}$), the FWHM sizes along the major and minor axes, and the position angles ($\theta_{\mathrm{dendro}}$) of the structures. The parameters derived by dendrogram are reported in Table \ref{tab:dendro}. The brightest two sources, Condensation 1 and Condensation 2, corresponds to P2-SMA1 and P2-SMA2 in \citet{2009ApJ...696..268Z}. We note that nearly all condensations in MM1 are smaller than the beam area except for Condensation 8, which is marginally resolved. 

\begin{deluxetable*}{cccccccccc}[t!]
\tablecaption{Physical parameters of dense structures\label{tab:dendro}}
\tablecolumns{10}
\tablewidth{0pt}
\tablehead{
\colhead{Source} &
\colhead{RA} &
\colhead{DEC} & 
\colhead{$S_{\mathrm{int}}$\tablenotemark{a}} & 
\colhead{FWHM\tablenotemark{b}} &  
\colhead{$\theta_{\mathrm{dendro}}$} &
\colhead{$T$} &  
\colhead{$M$} &  
\colhead{$N_{\mathrm{H_2}}$} &  
\colhead{$n_{\mathrm{H_2}}$}  \\
\colhead{} & \colhead{(h:m:s)} &
\colhead{(d:m:s)} & \colhead{(mJy)} & \colhead{$\arcsec \times \arcsec$} & \colhead{($\degr$)} & \colhead{(K)}& \colhead{($M_{\odot}$)}& \colhead{($10^{23}$ cm$^{-2}$)}& \colhead{($10^6$ cm$^{-3}$)}
}
\startdata
MM1-Core1&  18:42:52.06&  -3:59:54.04&  745.4& 2.8 $\times$ 2.6& -138.8 & 30.0 & 212.4 & 5.0 & 3.2 \\ \hline
Condensation  1&  18:42:51.98&  -3:59:54.06&  258.0& 0.8 $\times$ 0.6& 126.2& 57.0& 35.5& $>$13.6& $>$29.2\\ 
Condensation  2&  18:42:52.12&  -3:59:54.23&   57.7& 0.6 $\times$ 0.4& 177.8& 23.6& 21.9& $>$23.5& $>$56.5\\ 
Condensation  3&  18:42:52.08&  -3:59:53.41&   25.0& 0.4 $\times$ 0.3&  97.0& 23.6& 9.5& $>$20.4& $>$65.6\\ 
Condensation  4&  18:42:52.17&  -3:59:51.35&   26.8& 0.7 $\times$ 0.5& 159.5& 24.1& 9.9& $>$5.7& $>$11.8\\ 
Condensation  5&   18:42:52.2&  -3:59:58.24&    9.7& 0.6 $\times$ 0.5&  53.9& 17.3& 5.5& $>$4.7& $>$11.0\\ 
Condensation  6&  18:42:52.16&  -3:59:56.16&    9.8& 0.7 $\times$ 0.4& -148.4& 23.6& 3.7& $>$4.0& $>$8.8\\ 
Condensation  7&  18:42:51.96&   -3:59:46.8&    7.0& 0.6 $\times$ 0.5& -155.7& 15.2& 4.7& $>$3.2& $>$8.8\\ 
Condensation  8&  18:42:52.49&  -3:59:55.26&   12.2& 1.1 $\times$ 0.7& 170.7& 21.9& 5.1& $>$1.9& $>$2.5\\ 
Condensation  9&  18:42:52.66&  -3:59:55.19&    2.8& 0.5 $\times$ 0.4& 156.2& 17.7& 1.6& $>$1.8& $>$5.2\\ 
Condensation 10&  18:42:51.98&  -3:59:57.14&    2.2& 0.5 $\times$ 0.4& 149.3& 26.4& 0.7& $>$1.1& $>$2.8\\ 
Condensation 11&  18:42:51.56&  -3:59:51.66&    1.5& 0.4 $\times$ 0.3& 154.8& 20.2& 0.7& $>$1.2& $>$4.2\\ 
Condensation 12&  18:42:52.43&  -3:59:56.17&    1.4& 0.4 $\times$ 0.3& 158.5& 19.4& 0.7& $>$1.6& $>$5.4\\ 
Condensation 13&  18:42:52.38&   -3:59:51.5&    0.9& 0.4 $\times$ 0.3& 161.6& 19.8& 0.4& $>$1.1& $>$4.2\\ 
Condensation 14&  18:42:52.64&  -3:59:56.81&    1.2& 0.6 $\times$ 0.2& 179.3& 17.7& 0.6& $>$1.3& $>$3.9\\ 
Condensation 15&  18:42:52.08&  -3:59:58.52&    1.7& 0.7 $\times$ 0.3& 176.0& 21.1& 0.7& $>$1.0& $>$2.6\\  \hline
MM4-Core4&  18:42:50.31&  -4:03:21.07&    68.3& 2.9 $\times$ 1.8& 45.9& 16.0& 43.0& 1.5& 1.1\\ 
\enddata
\tablenotetext{a}{Integrated flux.}
\tablenotetext{b}{Without deconvolution.}
\end{deluxetable*}

With the assumptions of the optically thin dust emission, isothermal, and a \textbf{gas-to-dust ratio $\Lambda$ of 100:1} \citep{1991ApJ...381..250B}, the gas mass $M$ of the dense structures can be calculated with 
\begin{equation}
M = \frac{\Lambda S_{\mathrm{int}} d^2}{ \kappa_{\nu} B_{\nu} (T)}, \label{eq:M}
\end{equation}
where $d = 4.8$ kpc is the distance to G28.34, $\kappa_{\nu} = (\nu / 1 \mathrm{THz})^{\beta}$ is the dust opacity \citep{1983QJRAS..24..267H} in m$^2$ kg$^{-1}$, and $B_{\nu} (T)$ is the Planck function at temperature $T$. Multi-wavelength observations toward massive star-forming regions usually found dust emissivity indexes $\beta$ of $\sim$1.5 \citep{2007A&A...466.1065B, 2007ApJ...654L..87C}. Adopting this value, the $\kappa_{\nu}$ is estimated to be 0.11 m$^2$ kg$^{-1}$. The gas temperature map of the entire IRDC G28.34 was derived by \citet{2018RNAAS...2...52W} with combined VLA and Effelsberg 100m data \citep{2008ApJ...672L..33W} of the NH$_3$ (1,1) and (2,2) lines at a resolution of 6$\arcsec$.5 $\times$ 3$\arcsec$.6. The temperatures of the condensations in MM1 ranges from 15 K to 27 K. Considering that Condensation 1 is a hot molecular core \citep{2009ApJ...696..268Z}, the NH$_3$ (1,1) and (2,2) lines, which are cold gas tracers \citep{1983ARA&A..21..239H}, might have underestimated the gas temperature in this condensation. As a good approximation of dust heated by a central source with luminosity $L_{*}$, the dependence of dust temperature $T$ on radius $r$ behaves like a simple power-law \citep{1993ApJ...414..759T}: 
\begin{equation}
T = T_{0} (\frac{r}{r_{0}})^{-q} (\frac{L_{*}}{L_{0}})^{q/2}, \label{eq:T}
\end{equation}
where $T_0$ is the fiducial dust temperature at a fiducial radius $r_0$ from a central source with luminosity $L_0$ and $q$ is the power-law index. The luminosity of the embedded protostellar object in Condensation 1 is $L_{*} = 2950 L_{\odot}$ \citep{2012A&A...547A..49R}. Adopting $T_0 = 38$ K, $r_0 = $ 100 AU, $L_0 = $ 1 $L_{\odot}$, $q=0.4$ \citep{2001A&A...365..440M}, and a background temperature of 20 K \citep{2018RNAAS...2...52W}, the average temperature of Condensation 1 within a radius of mean FWHM (FWHM$_{mean}$) is measured to be 57.0 K. A temperature of 30 K \citep{2008ApJ...672L..33W} is adopted for MM1-Core1. With a similar approach and adopting a temperature of $\sim$16 K \citep{2008ApJ...672L..33W}, we also derived the mass for MM4-Core4. The estimated masses and temperatures are listed in Table \ref{tab:dendro}.

The average column density $N_{\mathrm{H_2}}$ and volume density $n_{\mathrm{H_2}}$ within each core and condensation are measured using a radius equal to the FWHM$_{\mathrm{mean}}$:
\begin{equation}
N_{\mathrm{H_2}} = \frac{M}{\mu_{\mathrm{H_2}} m_{\mathrm{H}} \pi \mathrm{FWHM^2_{mean}}},
\end{equation}
\begin{equation}
n_{\mathrm{H_2}} = \frac{3M}{4\mu_{\mathrm{H_2}} m_{\mathrm{H}} \pi \mathrm{FWHM^3_{mean}}},
\end{equation}
where $\mu_{\mathrm{H_2}} = 2.86$ is the mean molecular weight per hydrogen molecule \citep{2013MNRAS.432.1424K, 2015MNRAS.450.1094P} and $m_{\mathrm{H}}$ is the atomic mass of hydrogen. The derived column and volumn densities are also listed in Table \ref{tab:dendro}. The estimated $n_{\mathrm{H_2}}$ and $N_{\mathrm{H_2}}$ of the condensations in MM1 and the two cores are generally comparable to those in other IRDCs and in more evolved massive star-forming regions \citep[$n_{ \mathrm{H_2} } \sim 10^5$-$10^7$ cm$^{-3}$ and $N_{\mathrm{H_2}} \sim 10^{22}$-$10^{24}$ cm$^{-2}$. e.g., ][]{2018ApJ...862..105L, 2018ApJ...855....9L, 2019ApJS..241....1C, 2019ApJ...886..102S}.  The main uncertainty of the calculations comes from the uncertainty of $\kappa_{\nu}$ \citep{1995P&SS...43.1333H}. We conservatively adopt a fractional uncertainty of 50\% \citep{2014A&A...562A.138R} for $M$, $N_{\mathrm{H_2}}$, and $n_{\mathrm{H_2}}$. We note that since most condensations are unresolved, the derived $N_{\mathrm{H_2}}$ and $n_{\mathrm{H_2}}$ for these condensations should be regarded as lower limits. 

\subsection{Polarized emission of dust}
\begin{figure}[!tbp]
\gridline{\fig{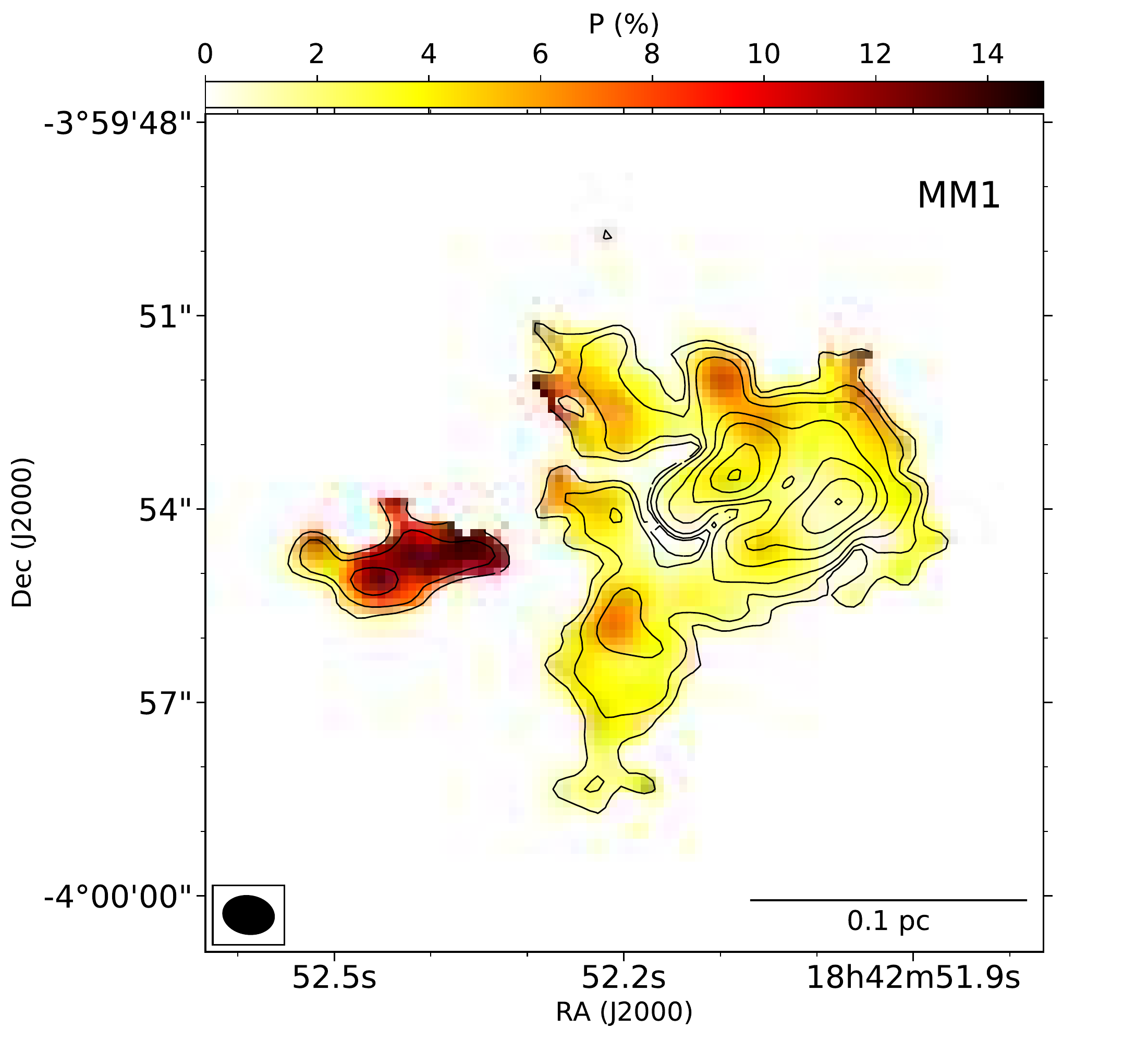}{0.5\textwidth}{}}
\gridline{\fig{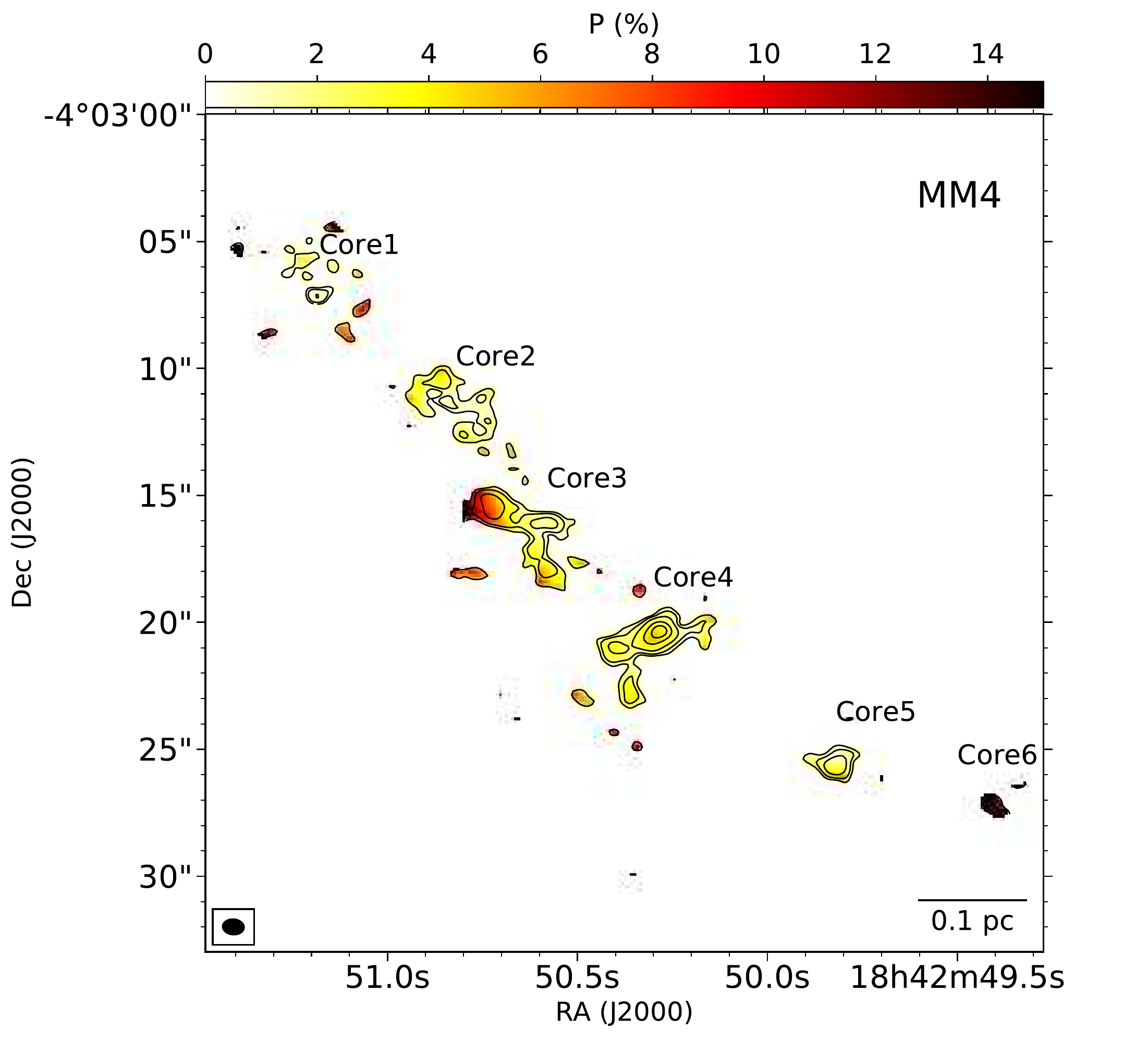}{0.5\textwidth}{}}
\caption{Dust polarization maps. The polarization percentage is shown in color scales. Black solid contours show the polarized intensity. Contour levels are (2, 3, 5, 10, 15, 25, 35, 50) $\times \sigma_{PI}$, where $\sigma_{PI}$ is the rms noise of the polarized intensity ($\sim$50 $\mu$Jy/beam for MM1 and $\sim$20 $\mu$Jy/beam for MM4 and MM9). \label{fig:figpolmap}}
\end{figure}

\begin{figure}[!tbp]
\ContinuedFloat
\includegraphics[width=.47\textwidth]{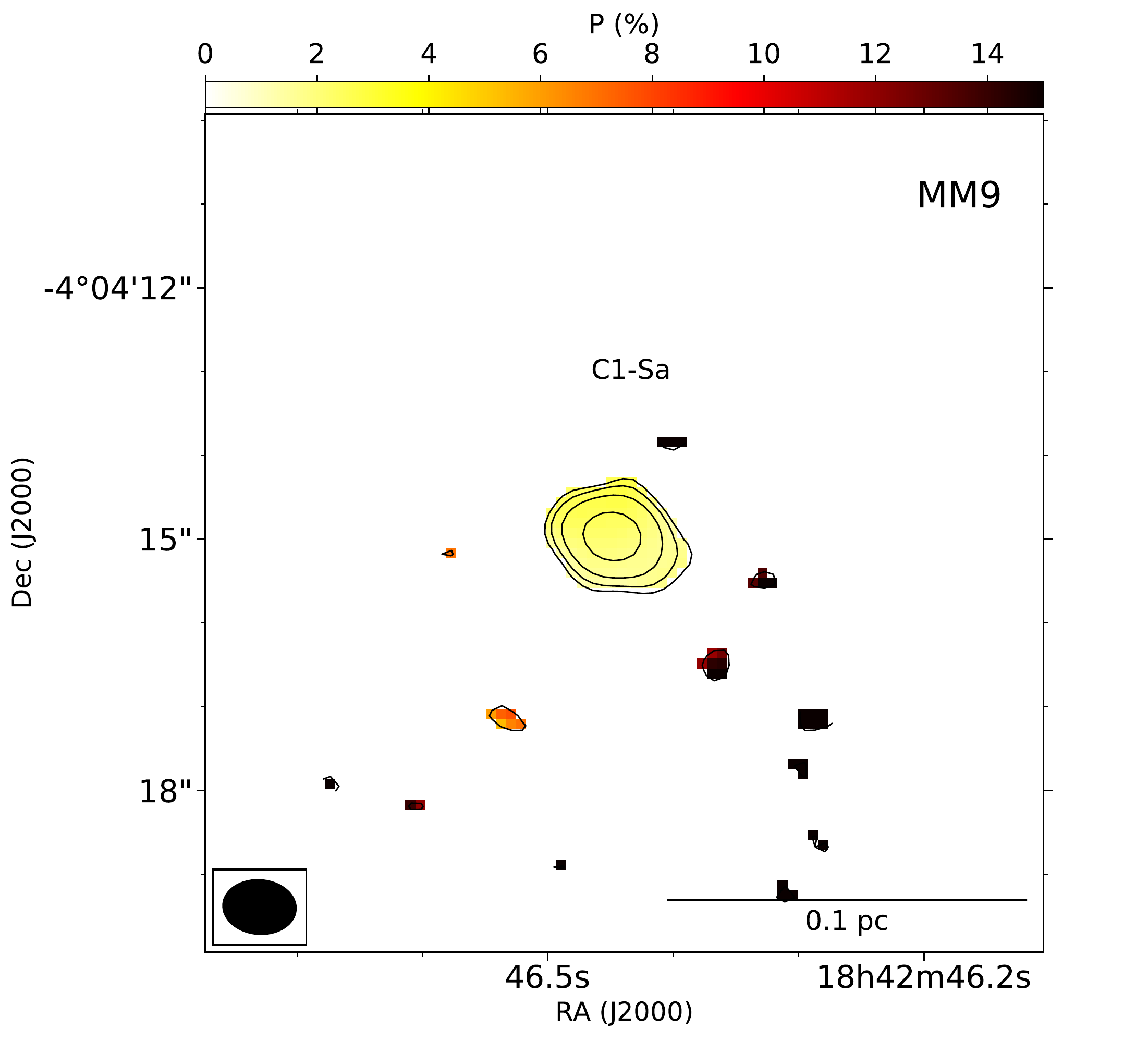}
\caption{(Continued)}
\end{figure}

\begin{figure}[!tbp]
\gridline{\fig{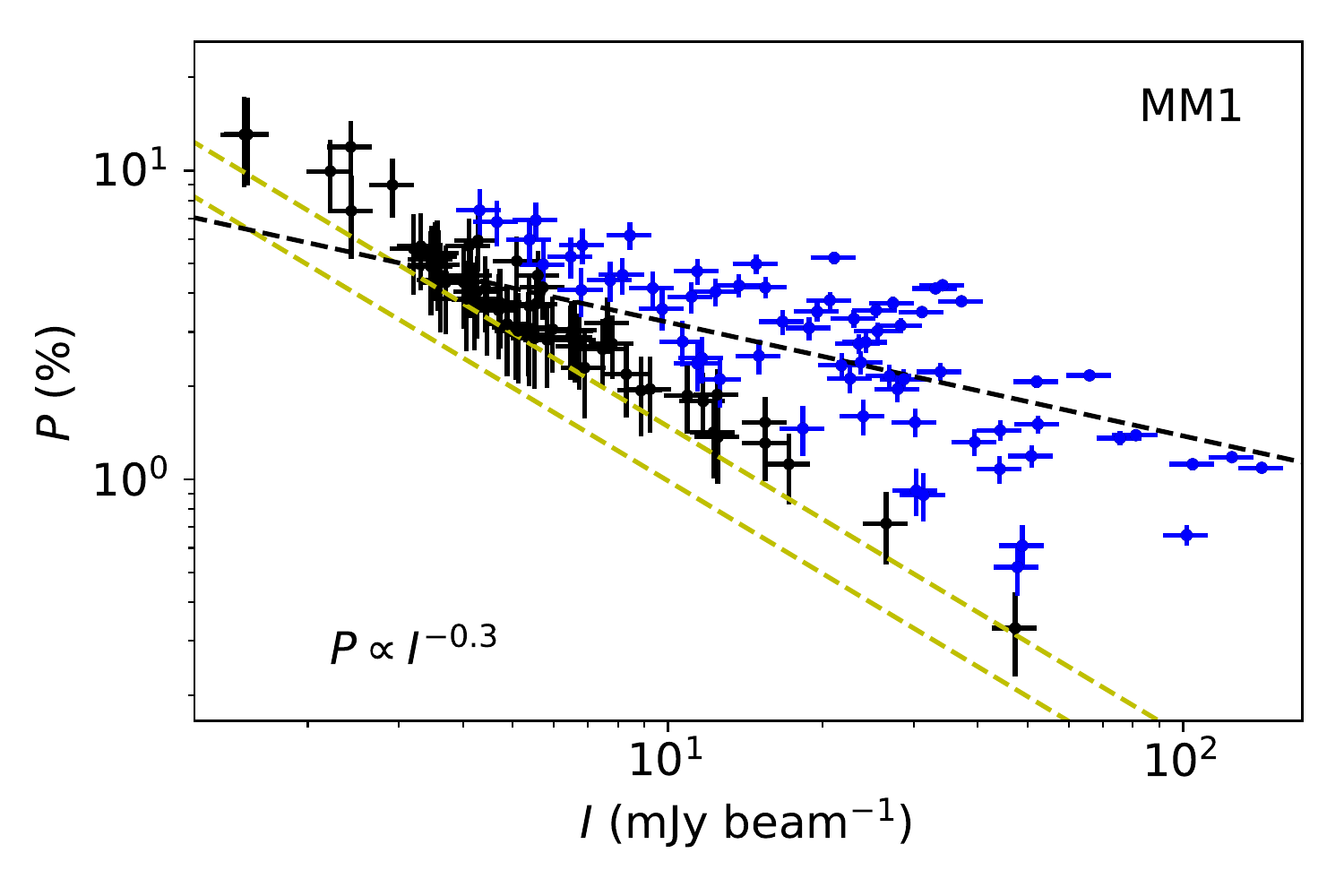}{0.47\textwidth}{}}
\vspace{-.5em}
\gridline{\fig{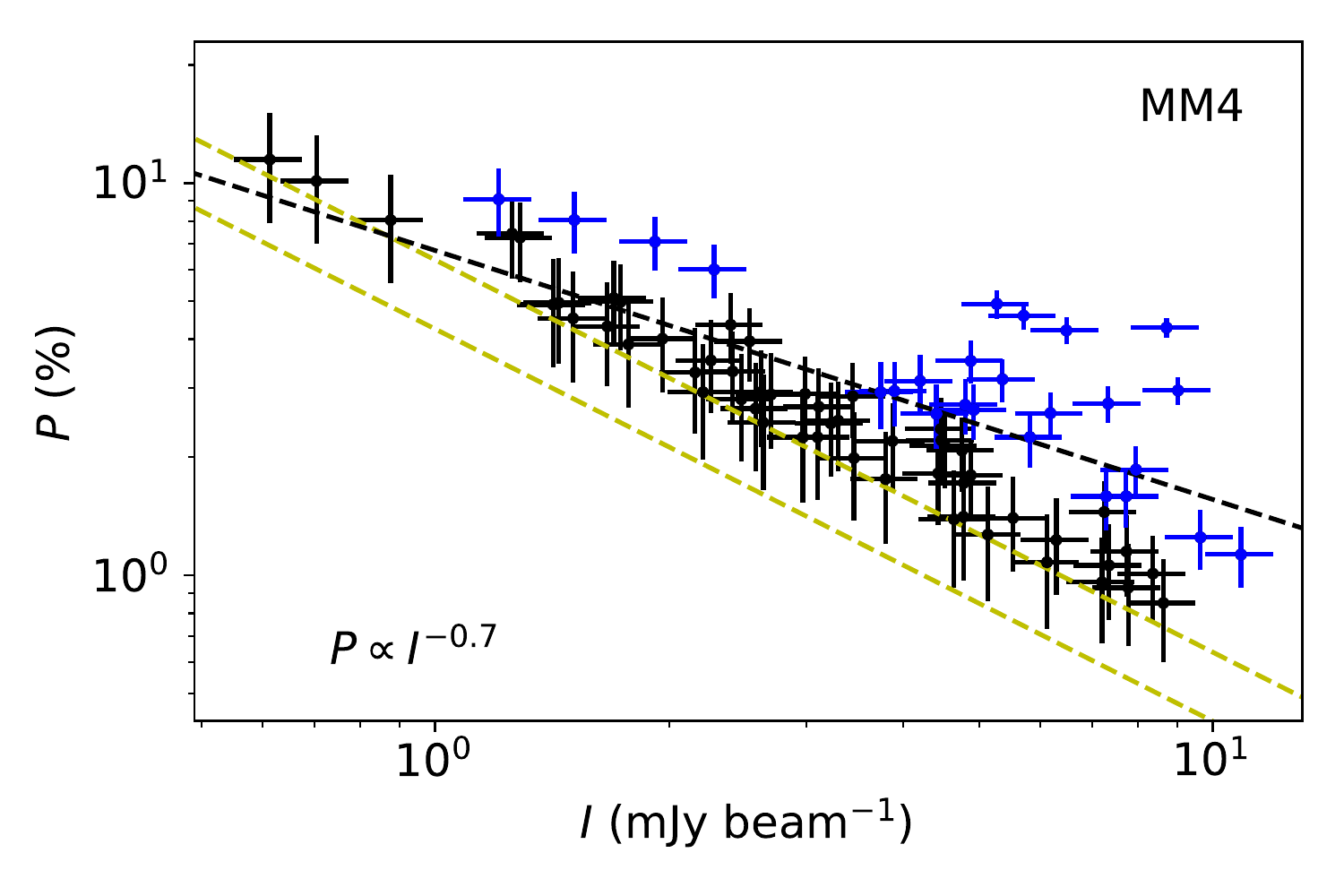}{0.47\textwidth}{}}
\vspace{-.5em}
\gridline{\fig{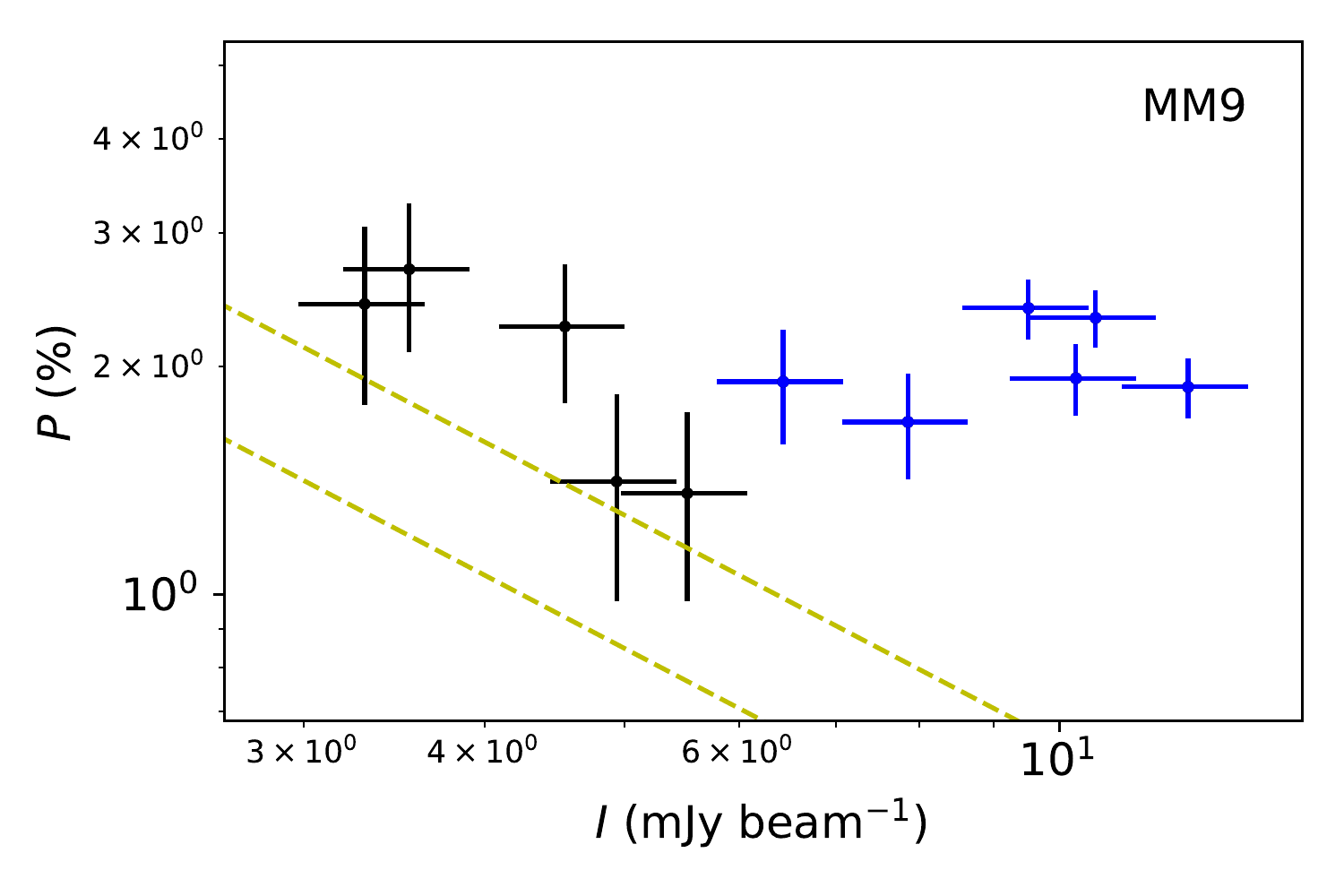}{0.47\textwidth}{}}
\vspace{-.5em}
\caption{Polarization percentage vs. Stokes $I$. Data points with $P/\delta P > 3$ and $P/\delta P > 5$ are shown with error bars in black and blue colors, respectively. Yellow dashed lines show the 2$\sigma_{PI}$ and 3$\sigma_{PI}$ noise levels. Results of the power-law fit for data with $P/\delta P > 3$ are shown in black dashed lines. \label{fig:figpi}}
\end{figure}

The plane-of-sky magnetic field orientation can by derived by rotating the orientation of the observed linear dust polarization of the electric field by 90$\degr$ with the assumption that the shortest axis of dust grains is perfectly aligned with the magnetic field. Figure \ref{fig:figG28B} shows the inferred magnetic field orientations overlaid on the Stokes $I$ maps toward the massive clumps MM1, MM4, and MM9. The ALMA maps reveal compact structures in the clumps down to scales of $\sim$0.02 pc (condensation scale). 

The magnetic field structure in MM1 is well resolved and shows a radial pattern, indicating that the field might be dragged towards the center by gravity. We further compare the magnetic field with the local gravity in Section \ref{section:pil}. 

In MM4, strong polarized emissions are detected in MM4-Core3, MM4-Core4, and MM4-Core5, while the polarized emissions in MM4-Core1, MM4-Core2, and MM4-Core6 are only marginally detected. The polarized emission in Core5 is compact and centered at the continuum peak in the west of the core, showing a northwest-southeast orientated magnetic field that is orthogonal to the parent filament. The polarized emission in MM4-Core4 is the most extended in MM4 and the magnetic field in the northern part of this core is well resolved, while the magnetic field in the southern part is only partially resolved. Similar to Core5, the magnetic field in Core4 shows a northwest-southeast orientation. The magnetic field in Core3 can be divided into three groups: the magnetic field in the center shows a east-west orientation; the magnetic field in the northeastern part shows a northeast-southwest orientation; and the magnetic field in the southern part rotates smoothly, showing a C-shaped morphology. 

In MM9, the polarized emission is confined at the continuum peak (C1-Sa) and the magnetic field shows a northeast-southwest orientation, following the parent filament. 

In Figure \ref{fig:figpolmap}, we show contours of the polarization emission intensity overlaid on the polarization percentage maps. In Figure \ref{fig:figpi}, the $P$-$I$ relation in the three clumps is shown. The polarization percentage is detected down to 0.3\%, 0.8\%, and 1.3\% in MM1, MM4, and MM9, respectively. Clear trends of decreasing $P$ with increasing $I$ are seen in MM1 and MM4. For the $P$-$I$ relations in MM1 and MM4, the polarization percentage at a constant $I$ shows broad scatters up to a order of magnitude, which indicates that the depolarization seen in the two clumps are the combined results of different factors. Using $\chi^2$ minimization, we fit the $P$-$I$ anti-correlations in MM1 and MM4 where $P/\delta P >3 $ with a simple power-law $P \propto I^{\alpha}$ and the power-law slope index $\alpha$ is estimated to be approximately -0.3 and -0.7 for MM1 and MM4, respectively. Due to the large scattering in the $P$-$I$ relations, we refrain from interpreting the absolute value of the slope indexes. However, we notice that the $P$-$I$ relation in MM1 is shallower than that in MM4, which may suggest the grain alignment efficiency is improved by additional internal radiation in more evolved star formation regions. Similar trend of shallower $P$-$I$ relation in more evolved cores is also seen in the low-mass star formation dense cores in the Ophiuchus cloud \citep{2018ApJ...859....4K, 2018ApJ...861...65S, 2019ApJ...877...43L, 2019ApJ...880...27P}. In MM9, the polarized percentages is $\sim$2\%. 

\subsection{Molecular lines}
\begin{figure*}[!tbp]
\gridline{\fig{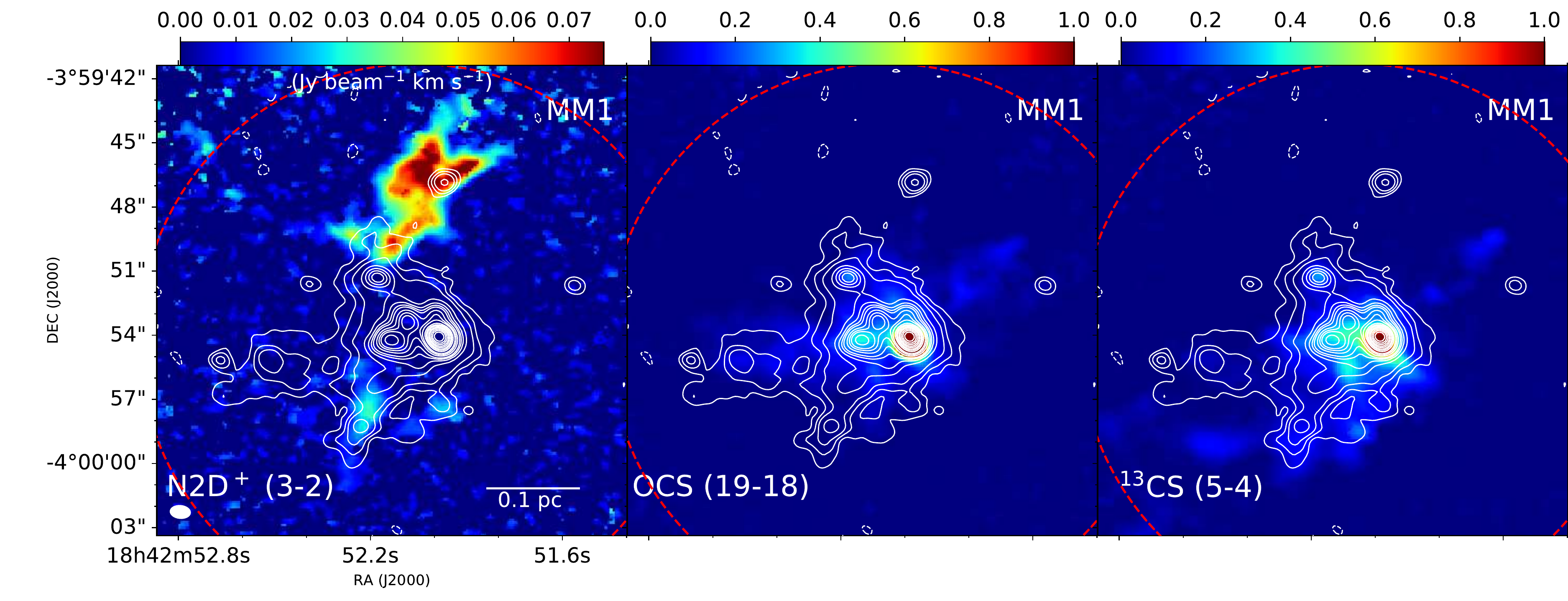}{0.95\textwidth}{}}
\gridline{\fig{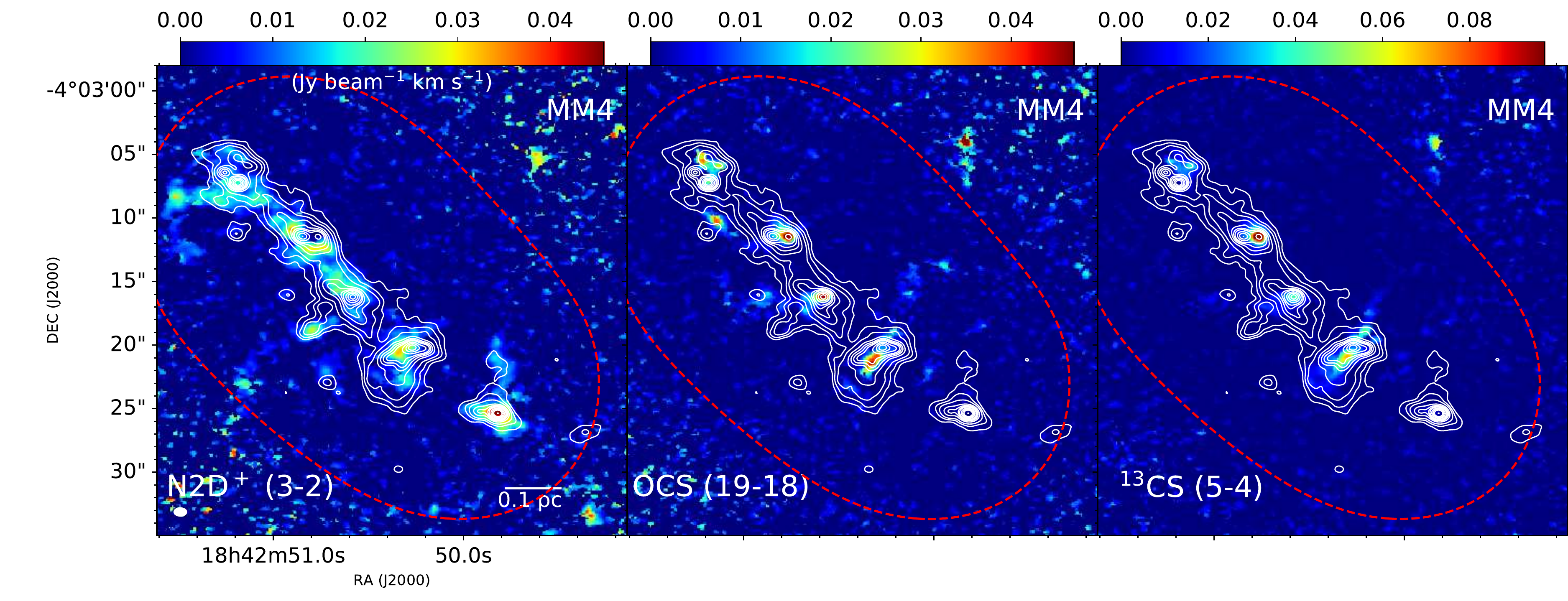}{0.95\textwidth}{}}
\gridline{\fig{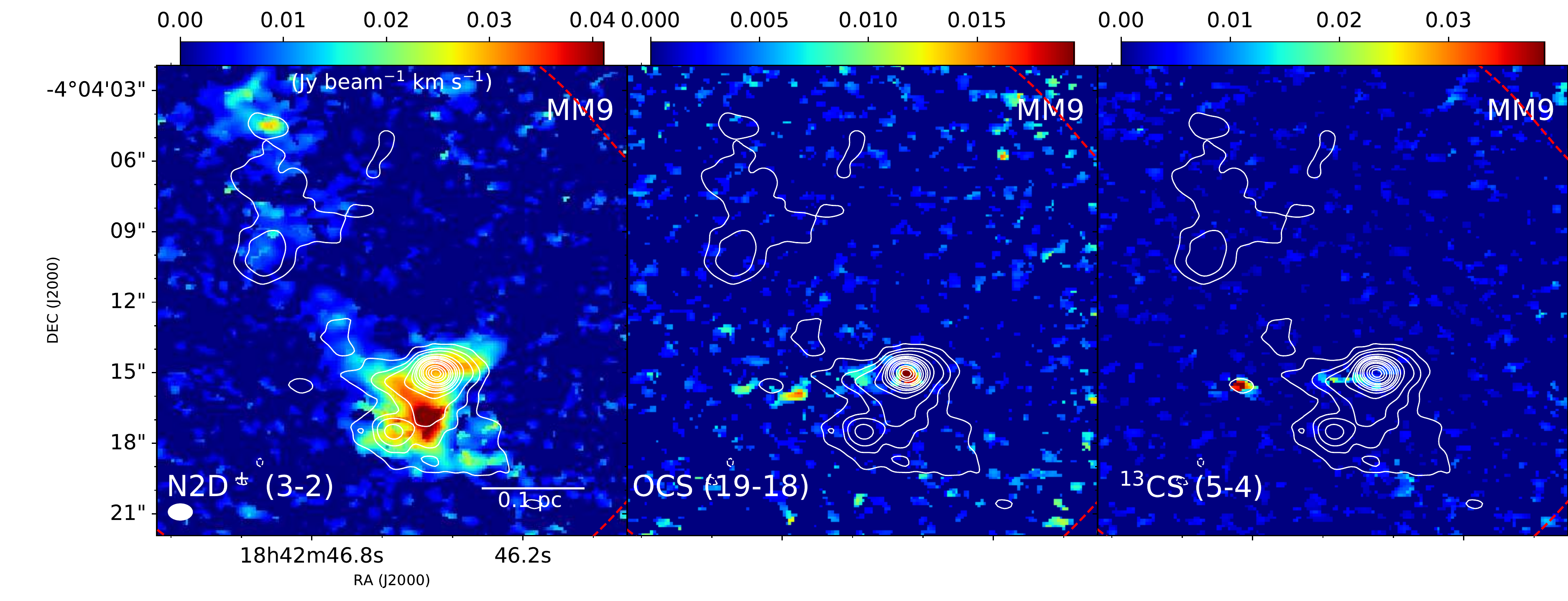}{0.95\textwidth}{}}
\caption{Momemt 0 maps of molecular lines from N$_2$D$^+$ (3-2), $^{13}$CS (5-4), and OCS (19-18) are shown in color scales. Only velocities with detection greater than 2$\sigma$ rms noise are integrated. White contour levels are the same as the contour levels in Figure \ref{fig:figG28B}. Red contours correspond to the FWHM field of view of the ALMA observations. \label{fig:figline}}
\end{figure*}

Figure \ref{fig:figline} show the zeroth-moment maps of N$_2$D$^+$ (3-2), $^{13}$CS (5-4), and OCS (19-18) in MM1, MM4, and MM9. The line emissions in three clumps share some similar patterns. The $^{13}$CS and OCS lines show similar emission morphology and present the strongest emission mostly near the dust continuum peaks. The distribution of the N$_2$D$^+$ emisison is almost anti-correlated with that of the $^{13}$CS and OCS emission, indicating that the $^{13}$CS and OCS molecules are likely depleted in the cold and dense gas highlighted by the N$_2$D$^+$ emisison.

In MM1, the emissions from $^{13}$CS and OCS peak at Condensation 1. The N$_2$D$^+$ line shows strong emissions to the north of Core1, which appears to be extended and probably links Condensation 7 with MM1-Core1. There are also faint N$_2$D$^+$ emissions in Condensation 10 and between Condensations 5 and 6. The lack of N$_2$D$^+$ detection in the dust emission region indicates that MM1 is in a later evolution stage of star formation.

In MM4, the N$_2$D$^+$ emission is detected in MM4-Core1 through Core5, while the $^{13}$CS and OCS emissions are detected in MM4-Core1 through Core4. The line emission of three molecules are mostly overlapping with dust continuum emissions, except that there appears to be an N$_2$D$^+$ emission peak to the east of MM4-Core1 and an OCS emission peak to the south of MM4-Core1, which are not associated with the dust emission above 3$\sigma$. The $^{13}$CS and N$_2$D$^+$ maps generally agree with previous observations reported in \citet{2015ApJ...804..141Z}, but show improvements in detecting fainter and more extended emissions. 

In MM9, the $^{13}$CS and OCS emissions are only marginally detected in C1-S. The N$_2$D$^+$ emission in core C1-S are consistent with previous observations \citep{2013ApJ...779...96T, 2016ApJ...821L...3T, 2018ApJ...867...94K}. Due to the limitation of sensitivity, the N$_2$D$^+$ emission in core C1-N is only marginally detected. 

\subsection{Molecular outflows in MM1}
\begin{figure}[!tbp]
\gridline{\fig{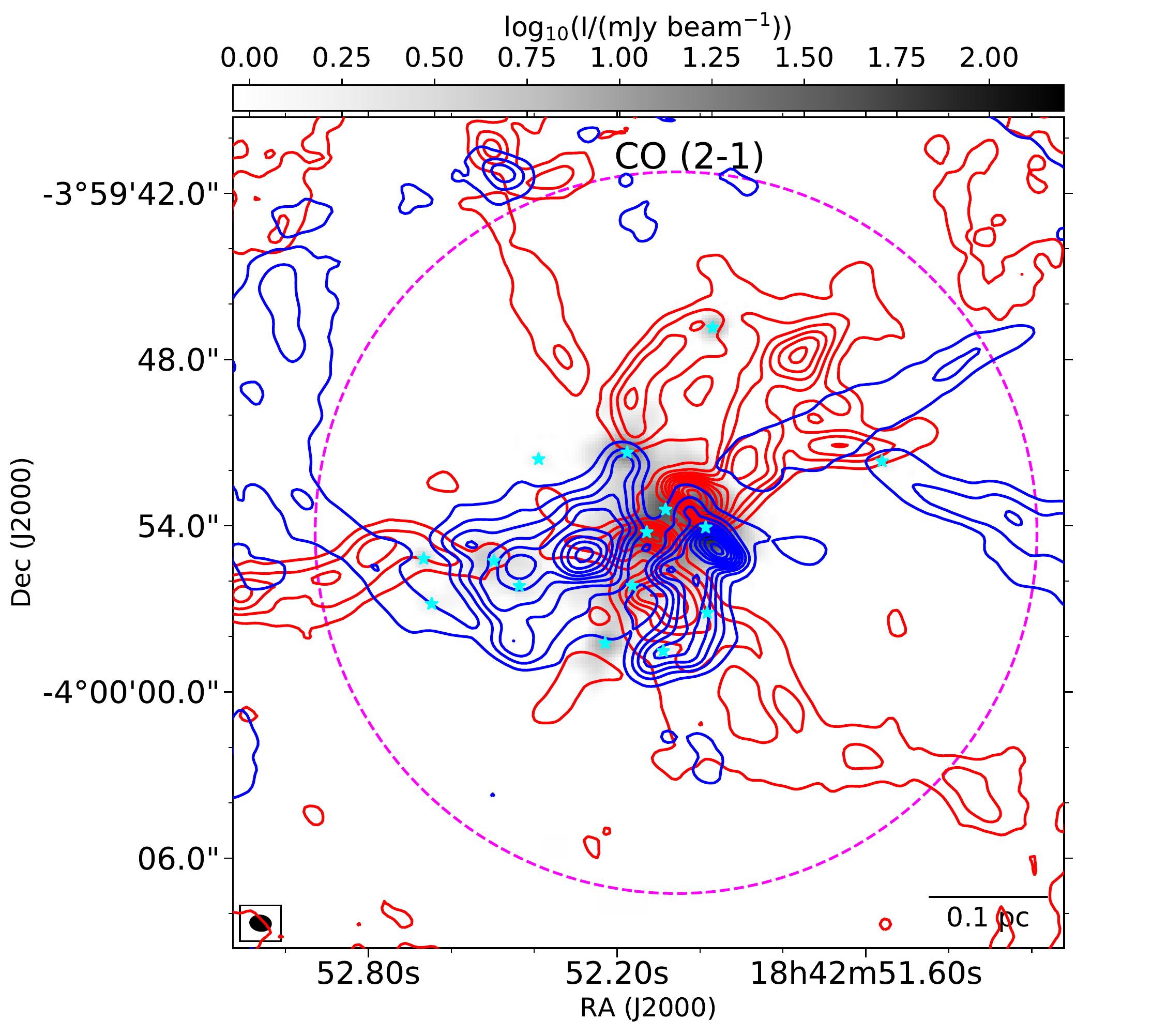}{0.47\textwidth}{(a)}}
\gridline{\fig{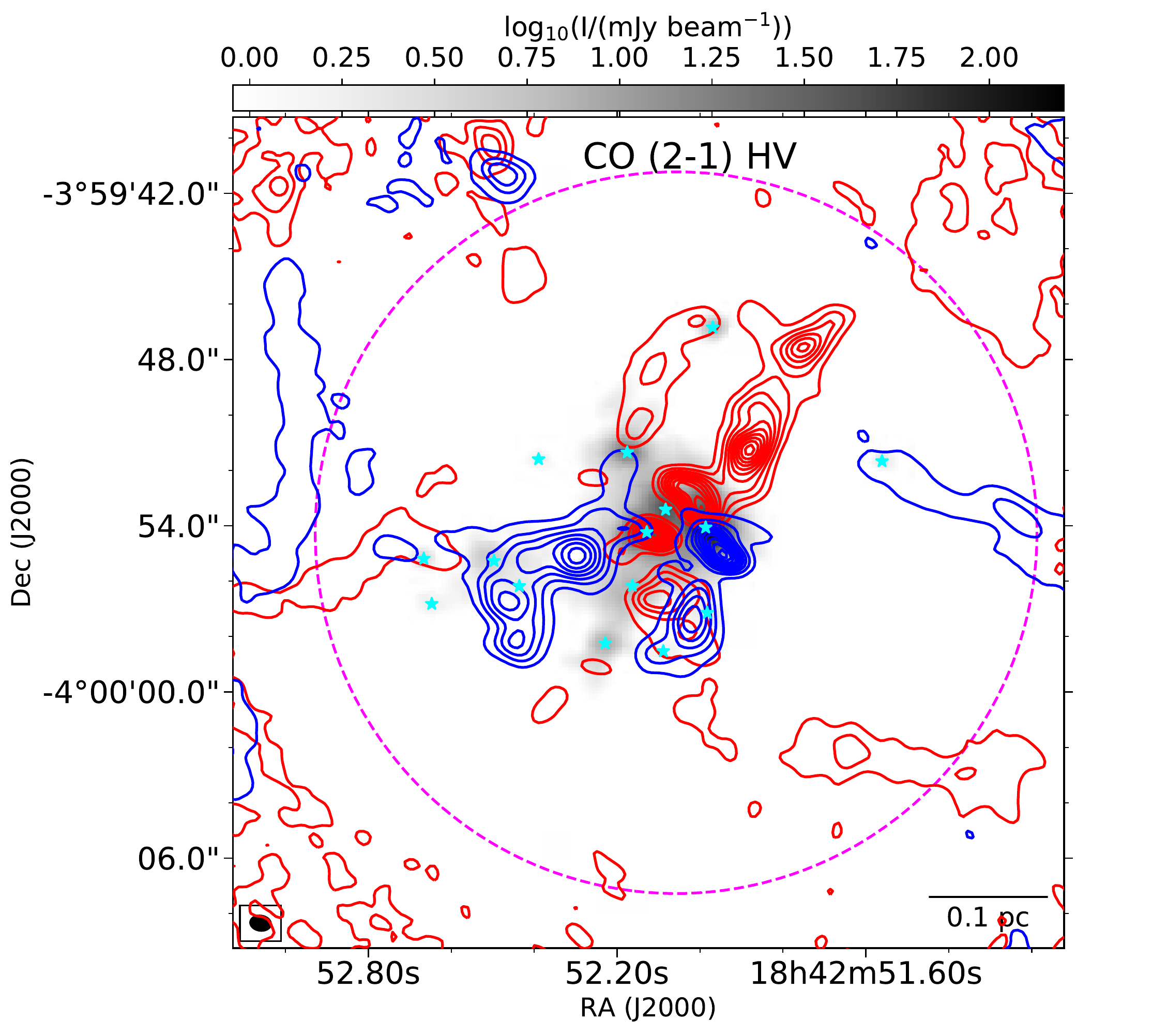}{0.47\textwidth}{(b)}
      }
\caption{Molecular outflows detected in CO (2-1) in MM1. The blueshifted and redshifted outflows are shown in blue and red contours, respectively. Contour levels start from 5\% and continue at 10\% of the peak of the integrated CO emission. The dust continuum is shown in gray scales. Star symbols denote the position of the condensations identified by dendrogram. Magenta contours correspond to the the FWHM field of view of the ALMA observations. (a) CO emissions integrated from 45 to 74 km s$^{-1}$ for the blue robe, and 84 to 115 km s$^{-1} $ for the red robe. (b) High-velocity (outflow velocity $> \pm 20$ km s$^{-1}$) CO emissions integrated from 45 to 59 km s$^{-1}$ for the blue robe, and 99 to 115 km s$^{-1} $ for the red robe. \label{fig:figMM1CO}}
\end{figure}

The molecular outflows in MM4 \citep{2011ApJ...735...64W, 2015ApJ...804..141Z} and MM9 \citep{2016ApJ...828..100F, 2016ApJ...821L...3T, 2018ApJ...867...94K} have been widely studied, so we only present the outflows of MM1 in CO (2-1) in this paper (see Figure \ref{fig:figMM1CO}). The outflow emissions in MM1 are detected throughout the velocity coverage of our observations ($\sim$41 km s$^{-1}$ to $\sim$116 km s$^{-1}$). Figure \ref{fig:figMM1CO}(a) shows the integrated emissions of CO in velocity ranges from 45 km s$^{-1}$ to 74 km s$^{-1}$ for the blue-shifted lobe, and 84 km s$^{-1}$ to 115 km s$^{-1}$ for the red-shifted lobe. The data channels with velocities $<$84 km s$^{-1}$ and $>$74 km s$^{-1}$ are excluded in the integration to avoid the contamination of surrounding diffuse gas. We also excluded channels with velocities $<$45 km s$^{-1}$ or $>$115 km s$^{-1}$ in the integration due to domination of instrumental noise in these edge channels. Figure \ref{fig:figMM1CO}(b) shows the integrated emission of the high-velocity (HV) outflow gas with velocities $>$20 km s$^{-1}$ with respect to MM1's ambient velocity. The outflows in MM1 show clustered overlapping structures. The HV outflow components are highly collimated and show some jet-like structures. Due to the complex structure of the outflows and the lack of shock tracers (e.g., SiO) to trace the vicinity outflow gas, we refrain from detailed analysis of the MM1 outflow and leave it to future studies. 

\subsection{Magnetic field strength of MM1-Core1 and MM4-Core4} \label{section:DCF}
The Davis-Chandrasekhar-Fermi \citep[DCF; ][]{1951PhRv...81..890D, 1953ApJ...118..113C} method and its modified form have been widely used to estimate the plane-of-sky magnetic field strength ($B_{\mathrm{pos}}$) by interpreting the observed angular dispersion ($\Delta\theta$) of polarization position angles as a MHD-wave component of the field perturbed by turbulent motions. With the small angle approximation, the $B_{\mathrm{pos}}$ is estimated as \citep{2004ApJ...600..279C}:
\begin{equation}\label{eq:eqcf}
B_{\mathrm{pos}} = Q_c \sqrt{\mu_0 \rho }\frac{\sigma_{\mathrm{turb}}}{\Delta\theta}
\end{equation} 
in SI units or CGS units, where $\mu_0$ is the permeability of vacuum, $\rho = \mu_{\mathrm{H_2}} m_{\mathrm{H}} n_{\mathrm{H_2}}$ is the density of the gas, $\sigma_{\mathrm{turb}}$ is the line-of-sight non-thermal velocity dispersion, and $Q_c$ is a correction factor. 

Further approaches have been made on modifying the DCF method toward more accurately quantifying the angular dispersion in the DCF formula through the angular dispersion function (ADF) analysis \citep{2008ApJ...679..537F, 2009ApJ...696..567H, 2009ApJ...706.1504H, 2016ApJ...820...38H} or the unsharp masking analysis \citep{2017ApJ...846..122P}. Specifically, \citet{2016ApJ...820...38H} derived the ADF for polarimetric images obtained from an interferometer, taking into account variations in the large-scale magnetic field, the effects of signal integration along the line of sight and within the beam, and the large-scale filtering effect of interferometers. The ADF is given by \citep{2016ApJ...820...38H}:
\begin{align}
1 - \langle \cos \lbrack \Delta \Phi (l)\rbrack \rangle =  \sum_{j=1}^{\infty}a_{2j} l^{2j} + \left[ \frac{N}{1+N \langle B_0^2 \rangle/ \langle B_{\mathrm{t}}^2 \rangle } \right] \nonumber \\
\times \Bigg\{ \frac{1}{N_1} \left[ 1 - e^{-l^2/2(\delta^2+2W_1^2)}\right] \nonumber \\+ \frac{1}{N_2} \left[ 1 - e^{-l^2/2(\delta^2+2W_2^2)}\right] \nonumber \\- \frac{2}{N_{12}} \left[ 1 - e^{-l^2/2(\delta^2+W_1^2+W_2^2)}\right] \Bigg\},
\end{align}
where $\Delta \Phi (l)$ is the angular difference of two line segments separated by a distance $l$, $\delta$ is the turbulent correlation length, the summation is the Taylor expansion of the ordered component of the ADF, $B_{\mathrm{t}}$ is the turbulent component of the magnetic field, $B_0$ is the ordered magnetic field, $W_1$ and $W_2$ are the standard deviation of the two Gaussian profiles of the synthesized beam and the large-scale filtering effect (i.e., the FWHM divided by $\sqrt{8 \ln{2}}$), and $N$ is the number of turbulent cells probed by the telescope beam given by: 
\begin{equation}
N_1 = \frac{(\delta^2 + 2W_1^2)\Delta\arcmin}{\sqrt{2\pi}\delta^3},
\end{equation}
\begin{equation}
N_2 = \frac{(\delta^2 + 2W_2^2)\Delta\arcmin}{\sqrt{2\pi}\delta^3},
\end{equation}
\begin{equation}
N_{12} = \frac{(\delta^2 + W_1^2 + W_2^2)\Delta\arcmin}{\sqrt{2\pi}\delta^3},
\end{equation}
\begin{equation}
N = \left( \frac{1}{N_1} + \frac{1}{N_2} - \frac{2}{N_{12}} \right)^{-1},
\end{equation}
where $\Delta\arcmin$ is the effective thickness of the concerned region through which the signals are integrated along the line of sight. The turbulent component of the ADF is
\begin{equation*}
\begin{aligned}
b^2(l) = \left[ \frac{N}{1+N \langle B_0^2 \rangle/ \langle B_{\mathrm{t}}^2 \rangle } \right]
\times \Big[ \frac{1}{N_1}  e^{-l^2/2(\delta^2+2W_1^2)} \\+ \frac{1}{N_2} e^{-l^2/2(\delta^2+2W_2^2)} - \frac{2}{N_{12}}  e^{-l^2/2(\delta^2+W_1^2+W_2^2)} \Big],
\end{aligned}
\end{equation*}
The ordered magnetic field strength can be derived by \citep{2009ApJ...706.1504H}: 
\begin{equation}
B_0 \simeq \sqrt{\mu_0 \rho} \sigma_{\mathrm{turb}} \left[ \frac{\langle B_{\mathrm{t}}^2 \rangle}{\langle B_0^2 \rangle} \right]^{-1/2}
\end{equation}
in SI units or CGS units. 

\begin{figure}[!tbp]
\gridline{\fig{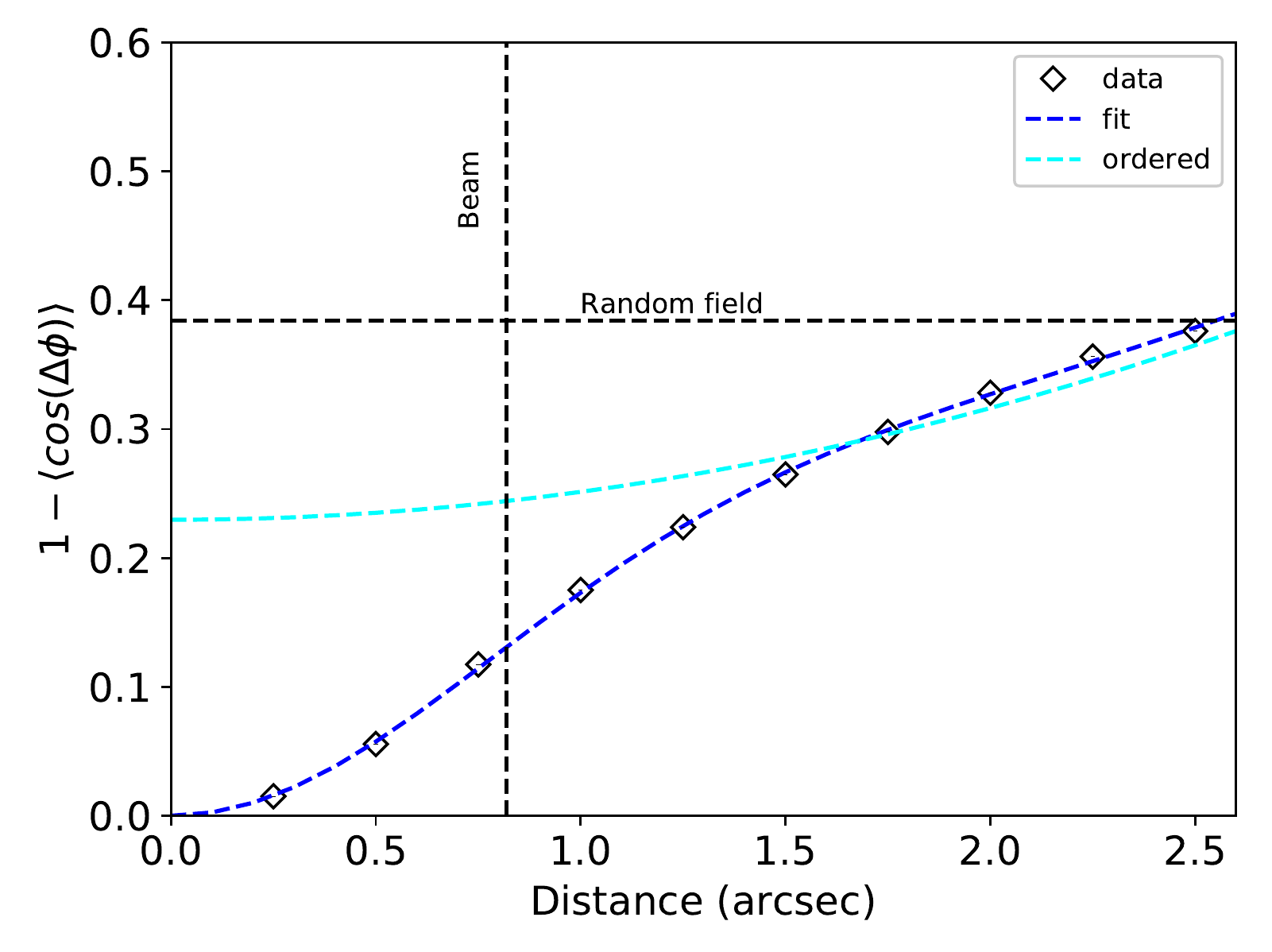}{0.5\textwidth}{(a)}}
\gridline{\fig{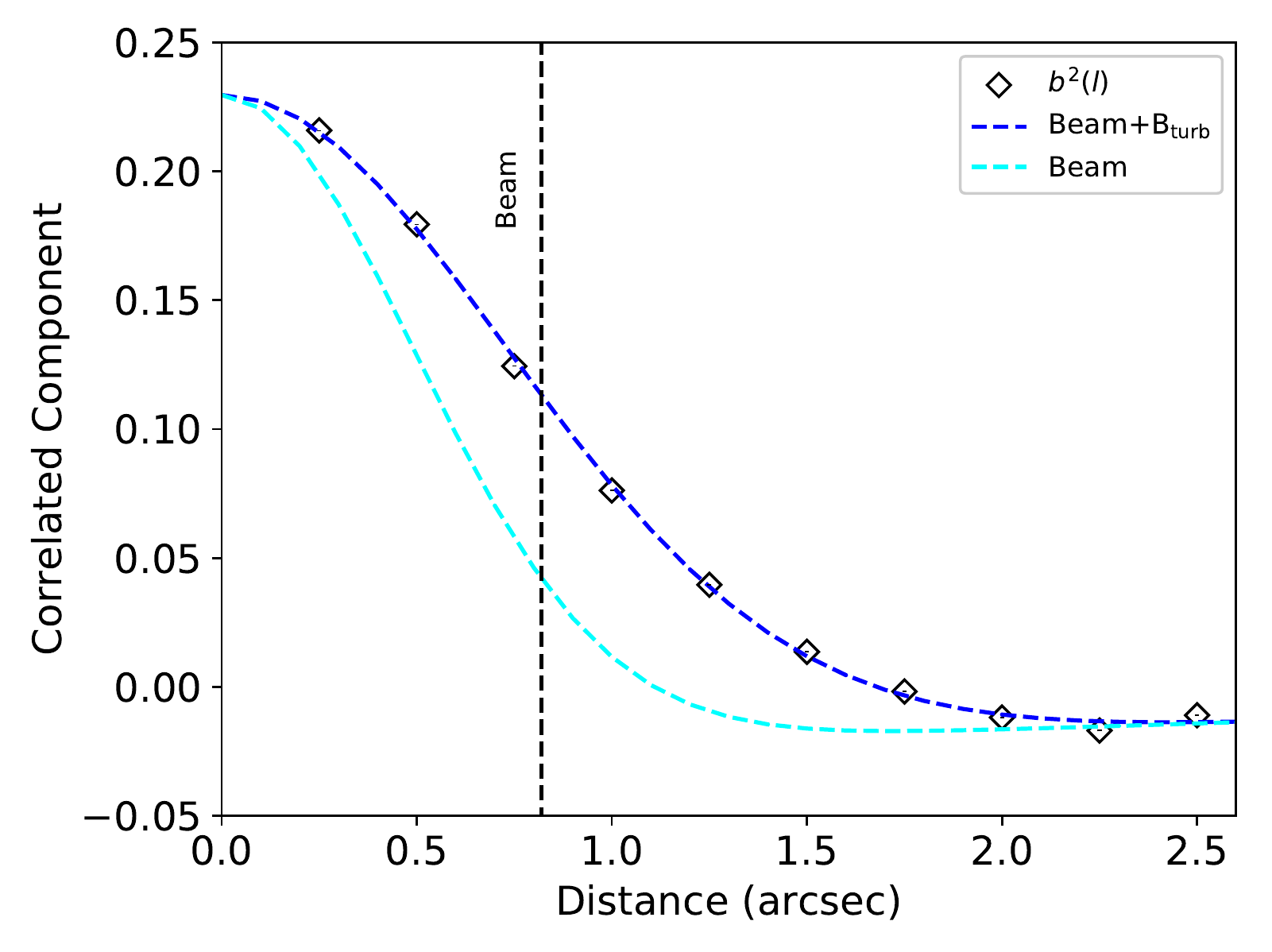}{0.5\textwidth}{(b)}
      }
\caption{(a) Angular dispersion function for MM1-Core1. The angle dispersion segments are shown in diamond symbols with error bars. The error bars indicate the statistical uncertainties of the angular dispersion function propagated from the observational uncertainty. The blue dashed line shows the fitted ADF. The cyan dashed line shows the large-scale component ($1 - \langle \cos \lbrack \Delta \Phi (l)\rbrack \rangle - b^2(l)$) of the best fit. The horizonal dashed line indicates the value corresponding to a random field. (b) Correlated component ($b^2(l)$) of the ADF. The correlated component of the best fit is shown in the blue dashed line. The correlated component solely due to the beam is shown in the cyan dashed line. \label{fig:figADFMM1}}
\end{figure}

\begin{figure}[!tbp]
\gridline{\fig{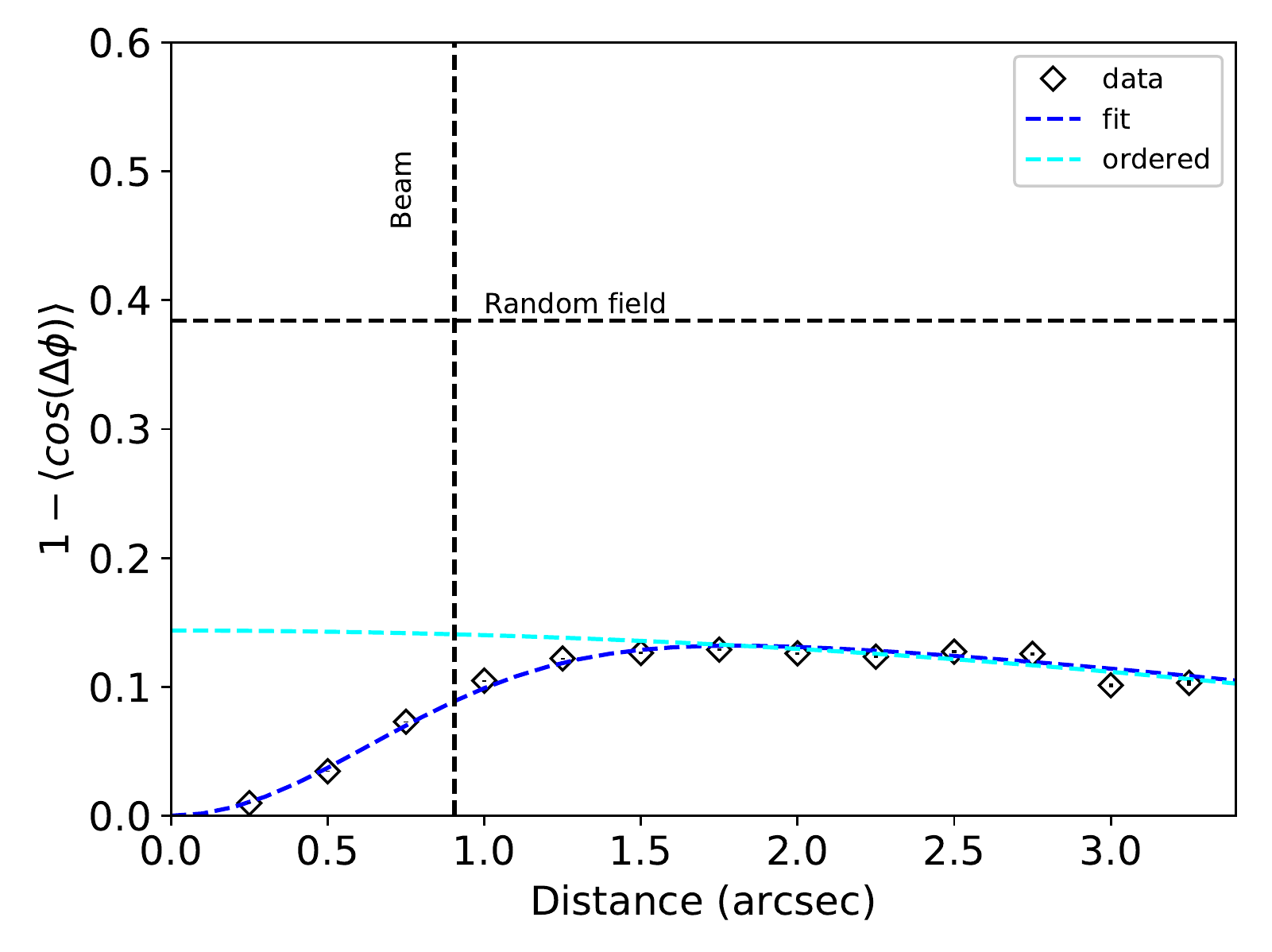}{0.5\textwidth}{(a)}}
\gridline{\fig{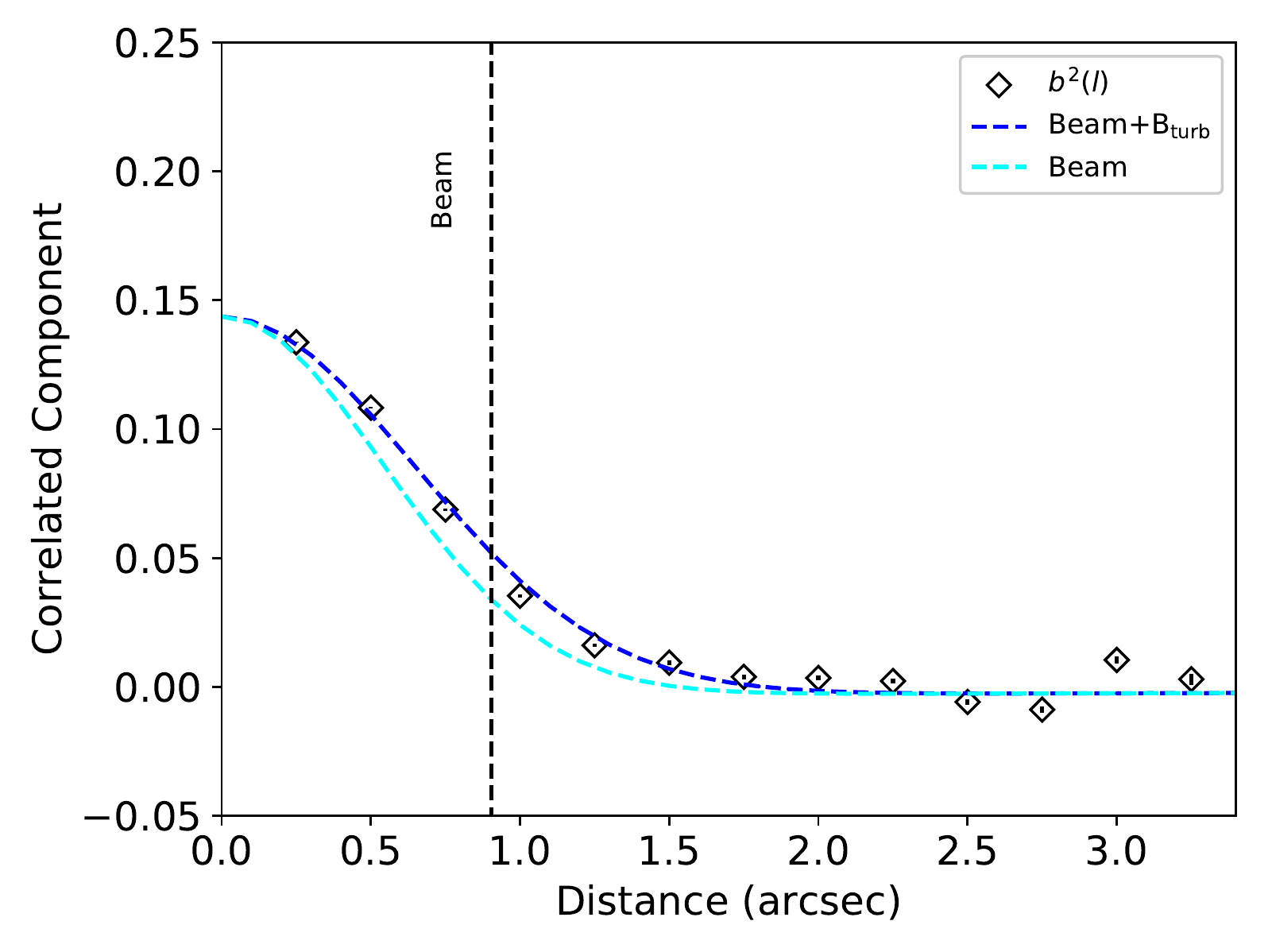}{0.5\textwidth}{(b)}
      }
\caption{Same as Figure \ref{fig:figADFMM1} but for MM4-Core4. \label{fig:figADFMM4}}
\end{figure}

The $\Delta\arcmin$ is estimated as the width at half of the maximum of the normalized auto-correlation function of the integrated normalized polarized flux \citep{2009ApJ...706.1504H}. The $\Delta\arcmin$ of MM1-Core1 and MM4-Core4 are estimated to be 2$\arcsec$.5 and 2$\arcsec$.2, respectively, which are consistent with the FWHMs derived from dendrogram. 

The information about the nonthermal (turbulent) velocity dispersion ($\sigma_{\mathrm{turb}}$) of the gas, which is required to calculate the $B_{\mathrm{pos}}$, is determined from previous NH$_3$ observations. The line-of-sight velocity dispersions of NH$_3$ ($\sigma_{\mathrm{NH_3}}$) for MM1-Core1 and MM4-Core4 are 1.4 km s$^{-1}$ \citep[][]{2008ApJ...672L..33W} and 0.47 km s$^{-1}$ \citep[][]{2012ApJ...745L..30W}, respectively. The $\sigma_{\mathrm{turb}}$ is calculated with the relation
\begin{equation}
\sigma^2_{\mathrm{turb}} = \sigma^2_{\mathrm{NH_3}} - \frac{k_\mathrm{B} T}{m_\mathrm{NH_3}},
\end{equation}
where $m_\mathrm{NH_3}$ is the mass of the NH$_3$ molecule. The nonthermal velocity dispersions are estimated to be 1.4 km s$^{-1}$ and 0.46 km s$^{-1}$ for MM1-Core1 and MM4-Core4, respectively. The corresponding 3D turbulent velocity dispersion $\sigma_{\mathrm{turb,3D}} = \sqrt{3}\sigma_{\mathrm{turb}}$ are 2.42 km s$^{-1}$ and 0.80 km s$^{-1}$ for MM1-Core1 and MM4-Core4, respectively.

Figures \ref{fig:figADFMM1} and \ref{fig:figADFMM4} show the angular dispersion functions and the turbulent component of the ADF derived from the polarization segments with $PI/\sigma_{PI} > 3$ in MM1-Core1 and MM4-Core4 (see Figure \ref{fig:figG28B}), respectively. To avoid large uncertainties due to sparse sampling at large spatial scales \citep[see Appendix of ][]{2019ApJ...877...43L}, we fit the ADFs of MM1-Core1 and MM4-Core4 over $l < 2.6\arcsec$ and $l < 3.4\arcsec$, respectively, where the angular dispersion function is relatively smooth and shows well-correlated relations with $l$. The best fit is obtained via $\chi^2$ minimization. The $B_{\mathrm{pos}}$ for MM1-Core1 and MM4-Core4 are calculated to be 1.6 mG and 0.32 mG, respectively. The parameters derived from the ADF method are listed in Table \ref{tab:ADF}. We note that since the polarization observations in MM4-Core4 only cover part of this core, an assumption of constant turbulent-to-ordered magnetic field ratio in the core needs to be adopted. According to \citet{2016ApJ...820...38H}, the $B_{\mathrm{pos}}$ estimated from the ADF method is accurate within a factor of $\sim$3. 

\begin{deluxetable*}{cccccccc}[t!]
\tablecaption{Physical parameters relevant to the ADF analysis \label{tab:ADF}}
\tablecolumns{8}
\tablewidth{0pt}
\tablehead{
\colhead{Source} &
\colhead{$n_{\mathrm{H_2}}$}  &
\colhead{$\sigma_{\mathrm{turb}}$}  &
\colhead{$(\langle  B_{\mathrm{t}}^2 \rangle $/$ \langle B_0^2 \rangle)^{\frac{1}{2}}$} &
\colhead{$B_{\mathrm{pos}}$} &
\colhead{$\delta$} &
\colhead{$N$} &
\colhead{$Q_c'$\tablenotemark{a}}
\\
\colhead{} &
\colhead{($10^6$ cm$^{-3}$)} &
\colhead{(km s$^{-1}$)} &
\colhead{} & 
\colhead{(mG)} &
\colhead{($\arcsec$)} & 
\colhead{} &
\colhead{} 
}
\startdata
MM1-Core1 & 3.2 &1.4  & 1.2& 1.6& 0.53& 4.1 &0.50\\ 
MM4-Core4 & 1.1 &0.46  & 1.2& 0.32& 0.38& 6.9 &0.38\\ 
\enddata
\tablenotetext{a}{Equivalent correction factor $Q_c' = 1/\sqrt{N}$ for the signal-integration effect.}
\end{deluxetable*}

Since the line-of-sight component of the magnetic field is unknown, the 3D mean ordered magnetic field strength $B$ is calculated with the statistical relation $B = 4 B_\mathrm{pos}/\pi$ \citep{2004ApJ...600..279C}. Then the 3D Alfv\'{e}n velocity is estimated as 
\begin{equation}
V_{\mathrm{A,3D}} = \frac{B}{\sqrt{\mu_0 \rho}}
\end{equation}
in SI units or CGS units. The $V_{\mathrm{A,3D}}$ for MM1-Core1 and MM4-Core4 are 1.5 km s$^{-1}$ and 0.51 km s$^{-1}$, respectively. The Alfv\'{e}n Mach number $Mach_{\mathrm{A}} = \sigma_{\mathrm{turb,3D}} / V_{\mathrm{A,3D}}$ of MM1-Core1 and MM4-Core4 are estimated to be 1.61 and 1.57, respectively, which means that the two cores are in super-Alfv\'{e}nic states (weakly magnetized)

\subsection{Comparison of orientations of magnetic field, intensity gradient, and local gravity} \label{section:pil}
Star-forming regions are subject to various forces such as gravity, gas pressure, magnetic force, and other possible forces. \citet{2012ApJ...747...79K} has used ideal magnetohydrodynamics (MHD) force equations to describe the interaction of the forces and developed a technique (the polarization-intensity gradient-local gravity method, hereafter the PIL method) to measure the local magnetic field strength. In this method, the gradient of intensity is assumed to trace the resulting direction of particle motion in the MHD force equation.  Following the approach of the PIL method, we calculate the angular differences among the magnetic field, the intensity gradient (IG), and the local gravity (LG), and study the relative importance of the magnetic field and local gravity at different positions.

The surrounding mass distribution needs to be taken into account when calculating the local gravitational force at a given position in a map \citep{2012ApJ...747...79K}. Adopting constant temperatures for MM4 and MM9, and adopting the previously mentioned temperature profile (see section \ref{section:cont}) for MM1, we derived the gas mass ($m_{i}$) for each pixel of the map with $I > 3\sigma_{I}$ for the three clumps using Equation \ref{eq:M}. The local gravity force at position $\boldsymbol{r_j}$ is then given by
\begin{equation}
\boldsymbol{g_{j}(r)} = G \sum\limits_{i=1}^{n} \frac{m_{i}m_{j}}{\vert \boldsymbol{r_j} - \boldsymbol{r_i} \vert ^2} \boldsymbol{e_{ij}},
\end{equation}
where $\boldsymbol{e_{ij}}$ is the unit vector indicating the direction between position $\boldsymbol{r_i}$ and $\boldsymbol{r_j}$. The orientations of the local gravity force ($\theta_{\mathrm{LG}}$) of the three clumps are presented in Figure \ref{fig:figG28LGLIC}. The local gravity orientation maps show similar radial patterns toward local emission peaks, which is quite reasonable because the emission centers contain more mass and have larger gravitational potential. 


\begin{figure}[!tbp]
\gridline{\fig{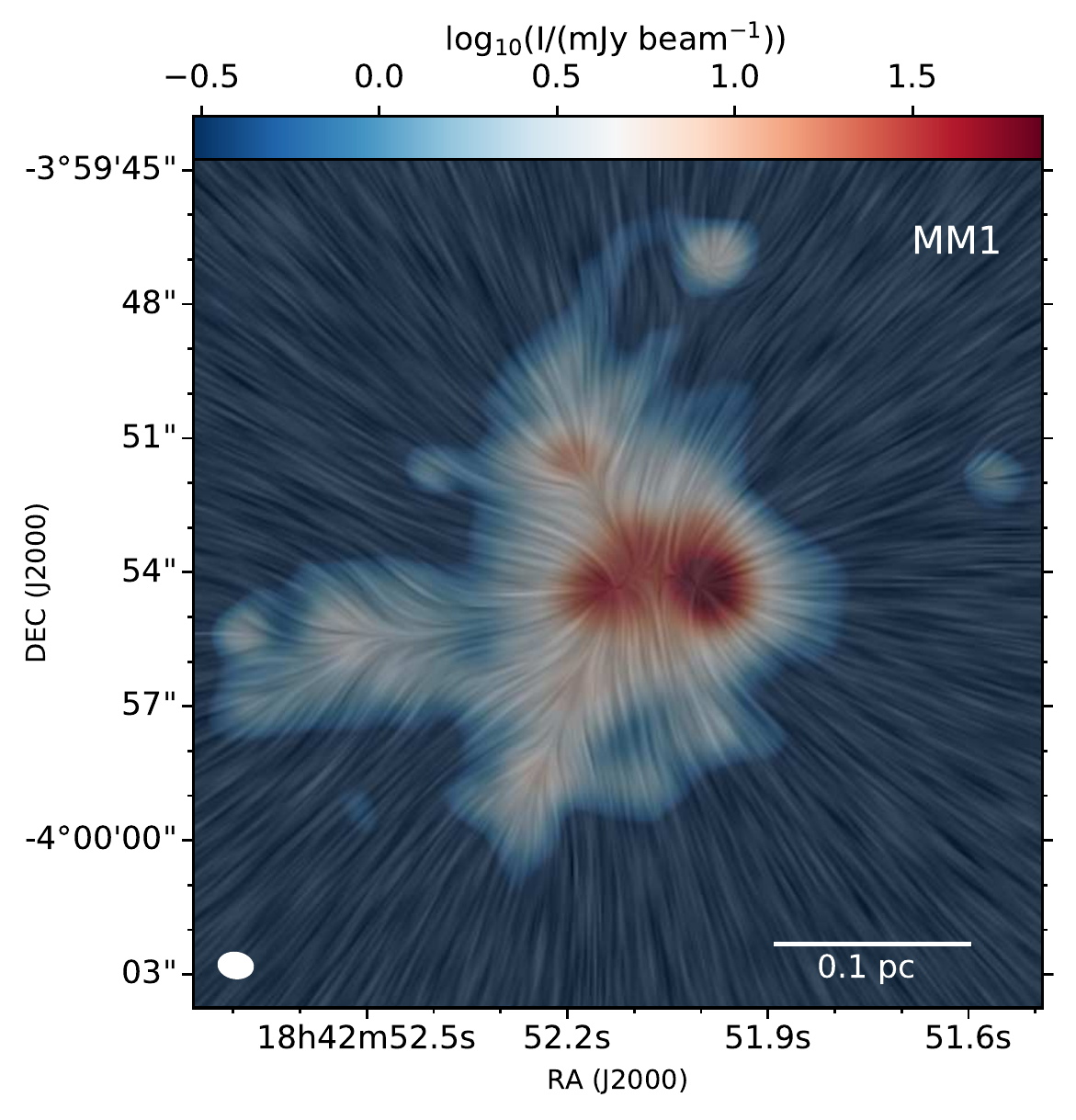}{0.47\textwidth}{}}
\gridline{\fig{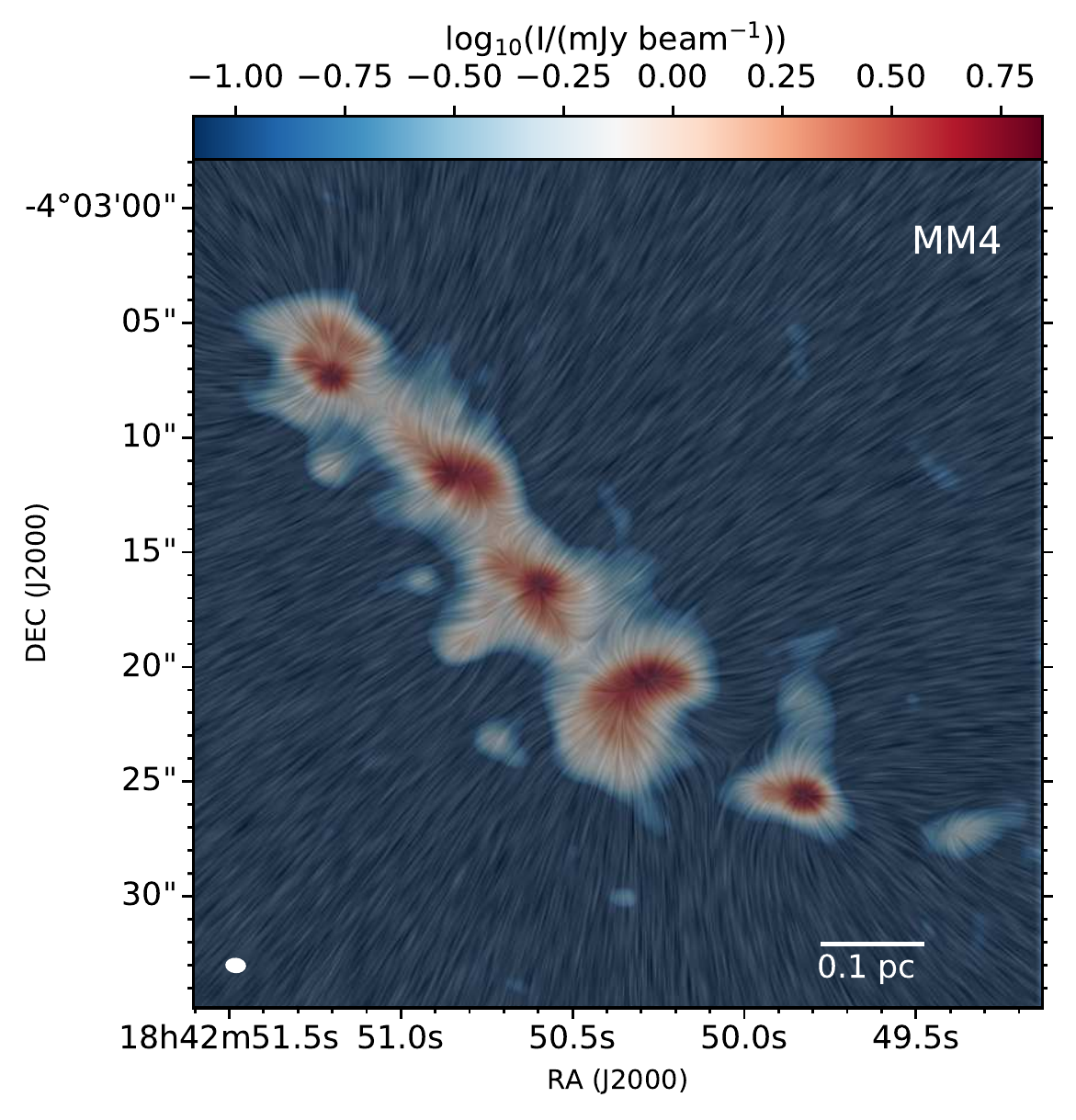}{0.47\textwidth}{}}
\caption{Local gravity orientation maps. The Stokes $I$ of the ALMA 1.3 mm continuum is shown in color scales. The overlaid patterns are produced with the line integral convolution method \citep[LIC, ][]{1993_LIC_C} and indicate the orientation of the local gravity ($\theta_{\mathrm{LG}}$). \label{fig:figG28LGLIC}}
\end{figure}

\begin{figure}[!tbp]
\ContinuedFloat
\includegraphics[width=.47\textwidth]{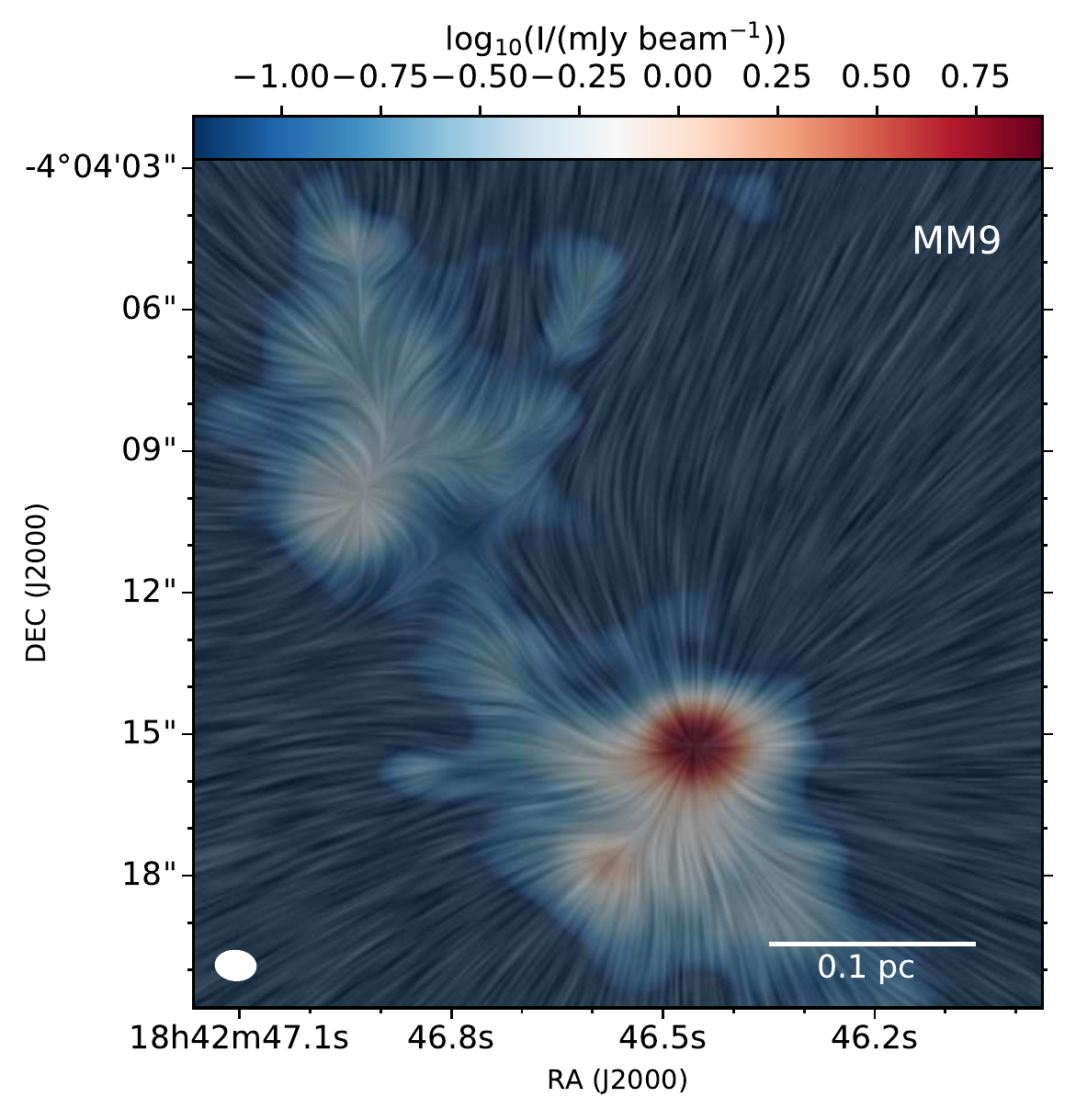}
\caption{(Continued)}
\end{figure}

\subsubsection{Intensity gradient versus local gravity}
The angle $\psi$ measures the angular difference between the position angles of the intensity gradient ($\theta_{\mathrm{IG}}$) and that of the local gravity ($\theta_{\mathrm{LG}}$). Figure \ref{fig:figG28psi} shows the $\psi$ maps for the three clumps. The average values of $\psi$ in MM1, MM4, and MM9 are 30$\degr$, 22$\degr$, and 28$\degr$, respectively. The small average value of $\psi$ indicates that the intensity gradient tend to be aligned with the local gravity and that the gravity plays an important role in regulating the gas motion. On the other hand, there seems to be large $\psi$ values near some condensations where accretion or rotation are likely ongoing. 


\begin{figure}[!tbp]
\gridline{\fig{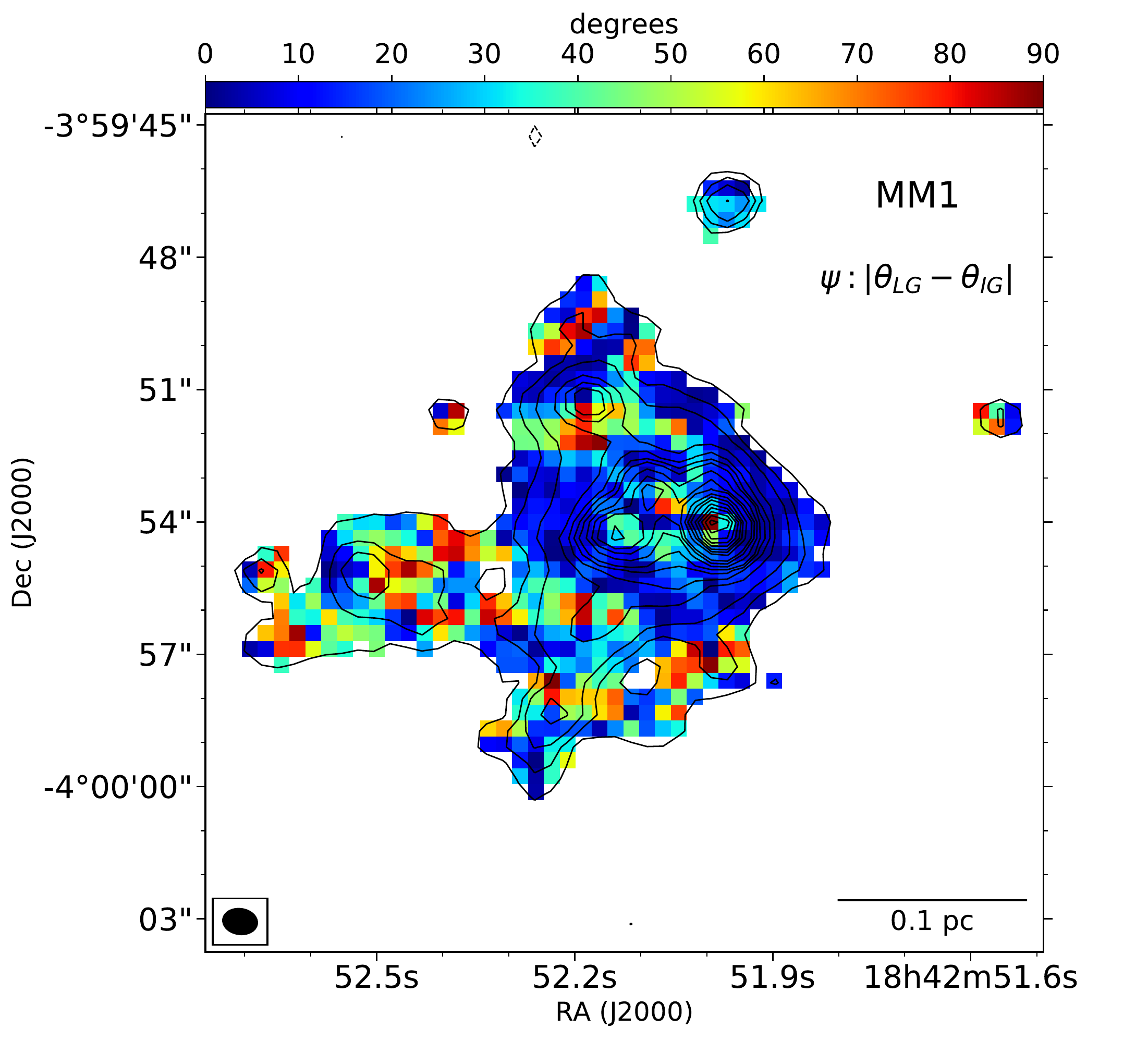}{0.47\textwidth}{}}
\gridline{\fig{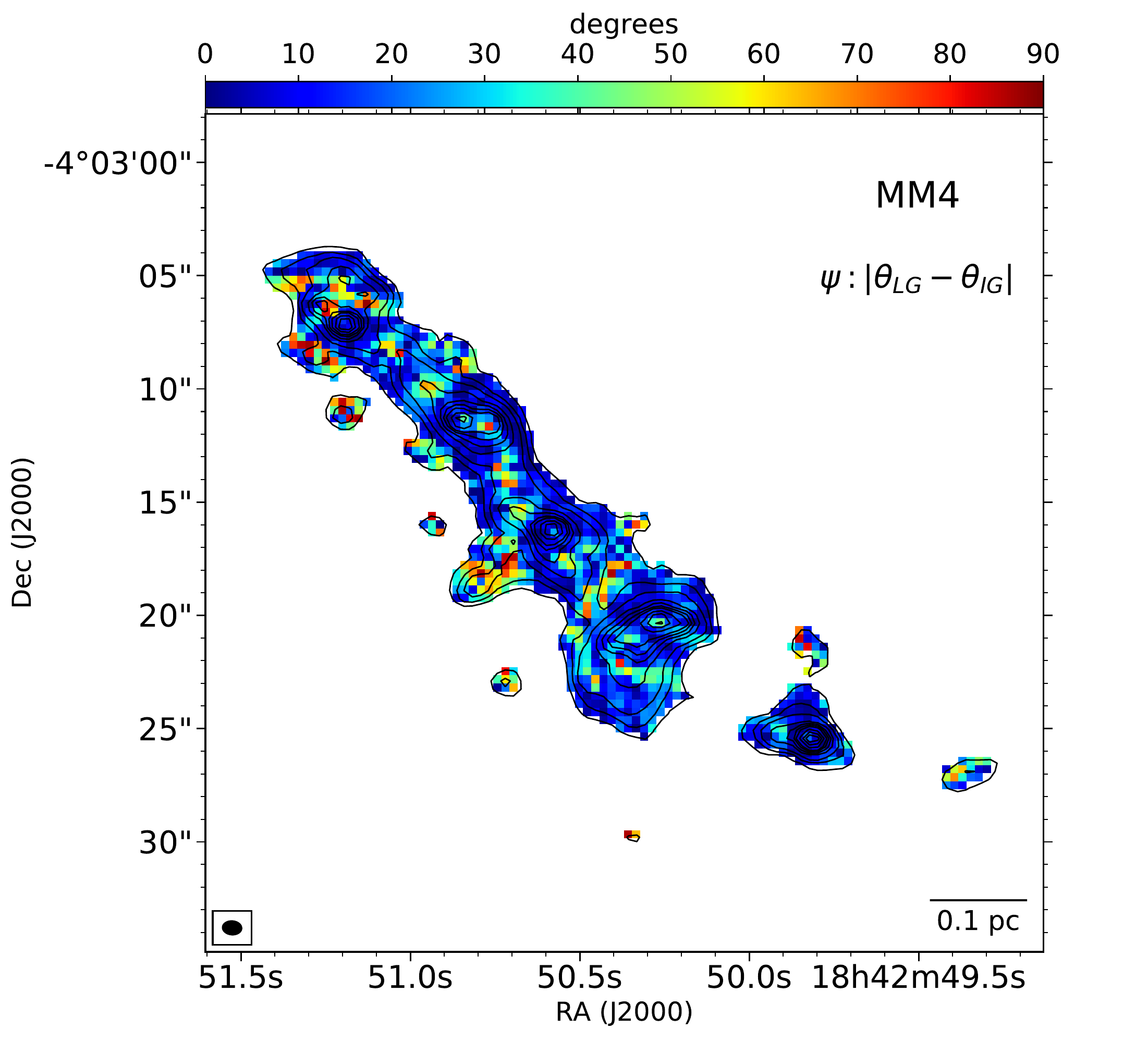}{0.47\textwidth}{}}
\caption{Angle difference between orientations of the intensity gradient and the local gravity ($\psi = \vert \theta_{\mathrm{IG}} - \theta_{\mathrm{LG}} \vert$). Contour levels are the same as those in Figure \ref{fig:figG28B}. \label{fig:figG28psi}}
\end{figure}

\begin{figure}[!tbp]
\ContinuedFloat
\includegraphics[width=.47\textwidth]{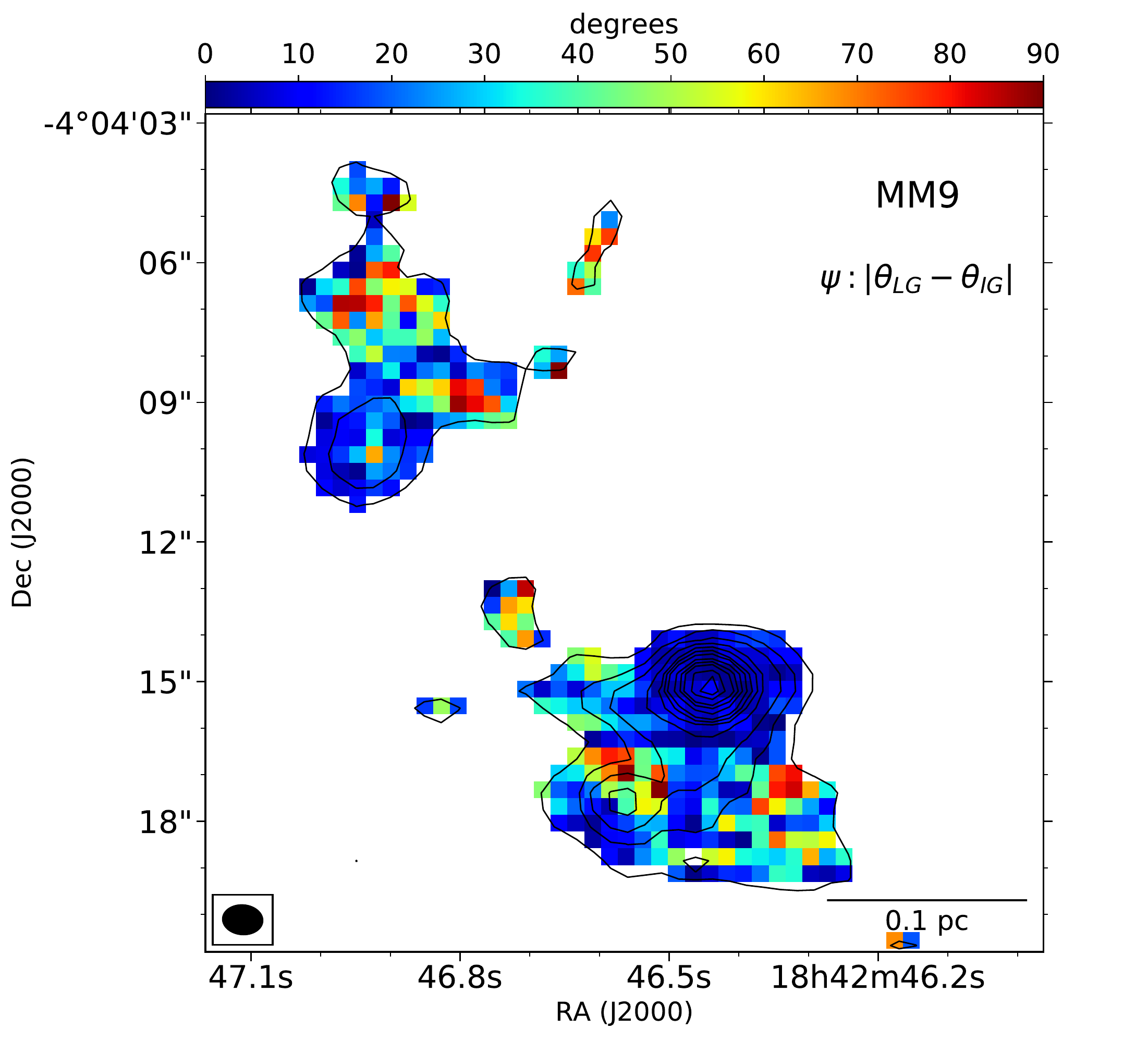}
\caption{(Continued)}
\end{figure}


\subsubsection{Magnetic field versus intensity gradient}
\begin{figure}[!tbp]
\gridline{\fig{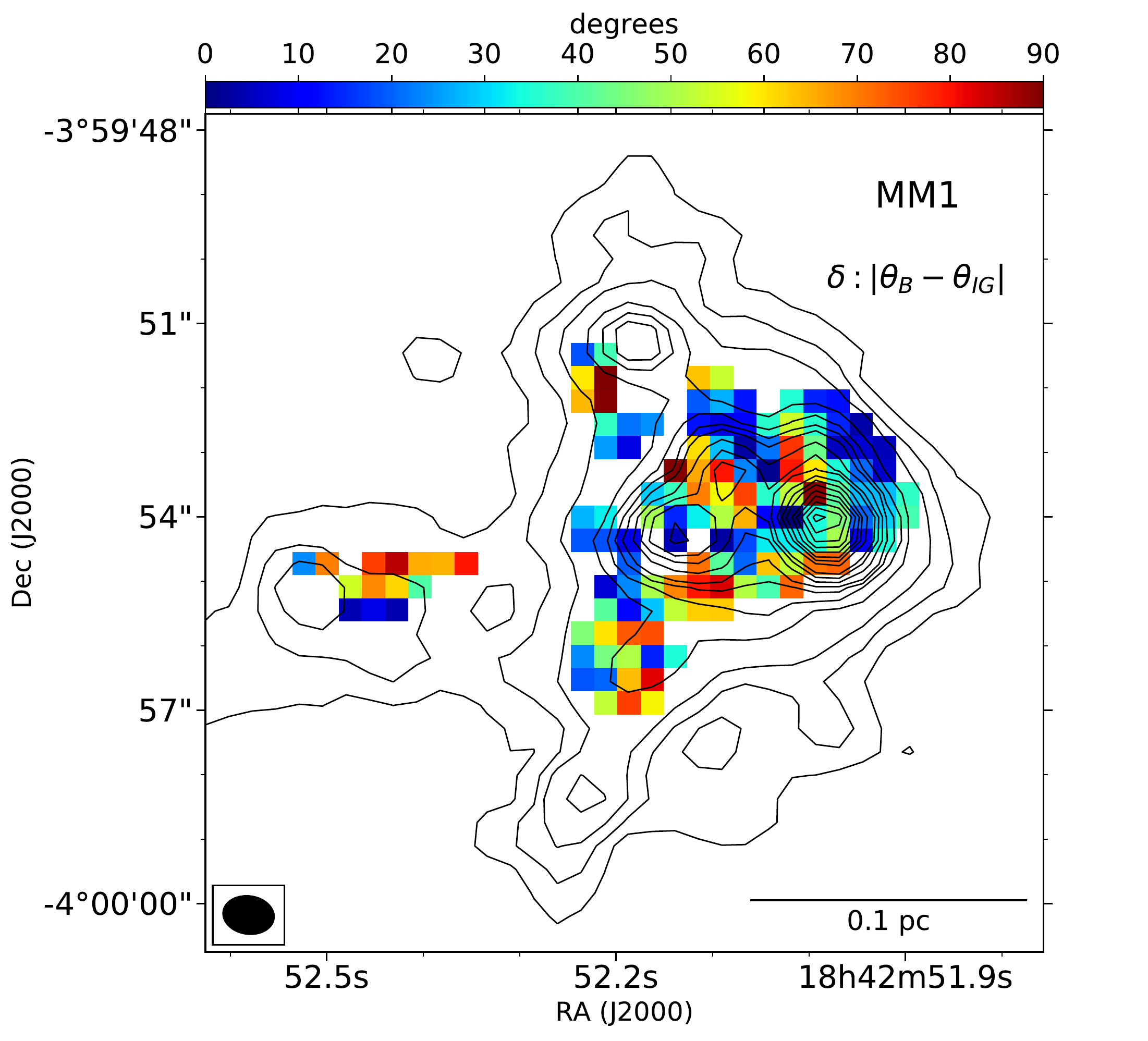}{0.47\textwidth}{}}
\gridline{\fig{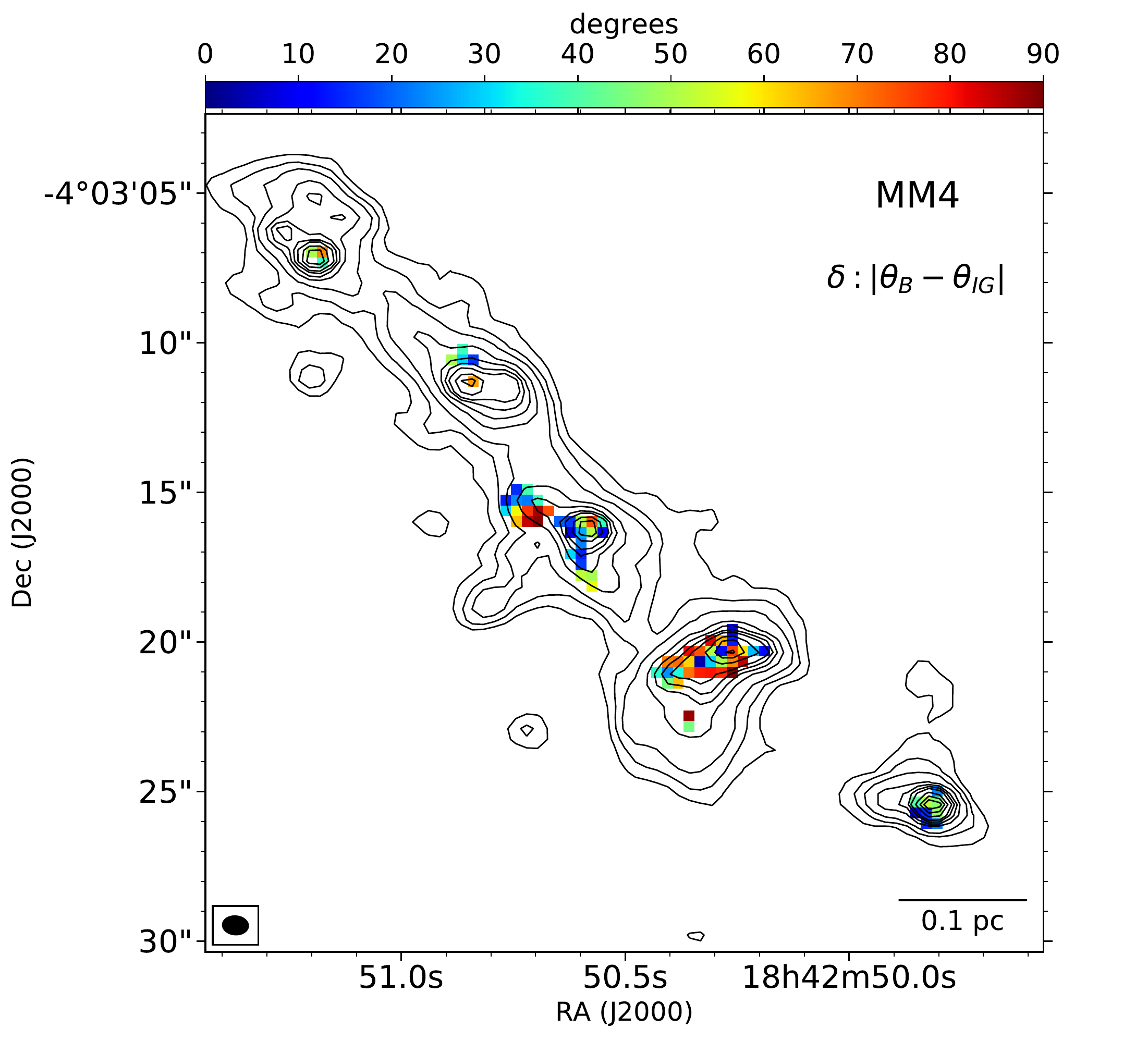}{0.47\textwidth}{}
      }
\caption{Angle difference between orientations of the magnetic field and the intensity gradient ($\delta = \vert \theta_{\mathrm{B}} - \theta_{\mathrm{IG}} \vert$). Values of $\delta$ are computed at where $PI/\sigma_{PI} > 3$. Contour levels are the same as those in Figure \ref{fig:figG28B}. \label{fig:figG28delta}}
\end{figure}

The angle $\delta$ measures the angular difference between the position angles of the intensity gradient ($\theta_{\mathrm{IG}}$) and that of the magnetic field ($\theta_{\mathrm{B}}$). Figure \ref{fig:figG28delta} shows the $\delta$ maps for MM1 and MM4. The average values of $\delta$ in MM1 and MM4 are 40$\degr$ and 46$\degr$, respectively. The intermediate average values of $\delta$ suggests that the magnetic field plays a moderate role in resisting the gas collapse.

\subsubsection{Magnetic field versus local gravity}
\begin{figure}[!tbp]
\gridline{\fig{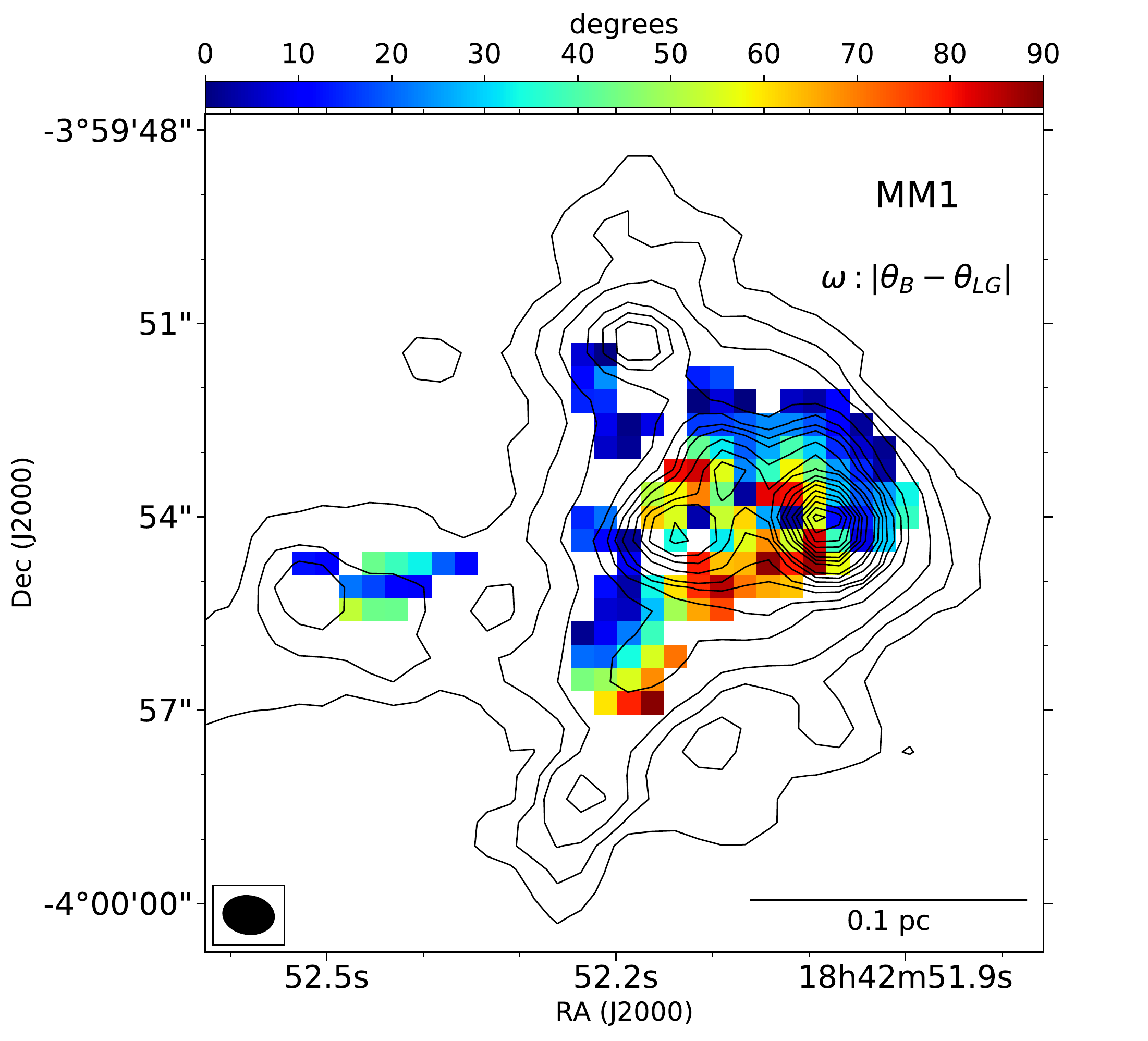}{0.47\textwidth}{}}
\gridline{\fig{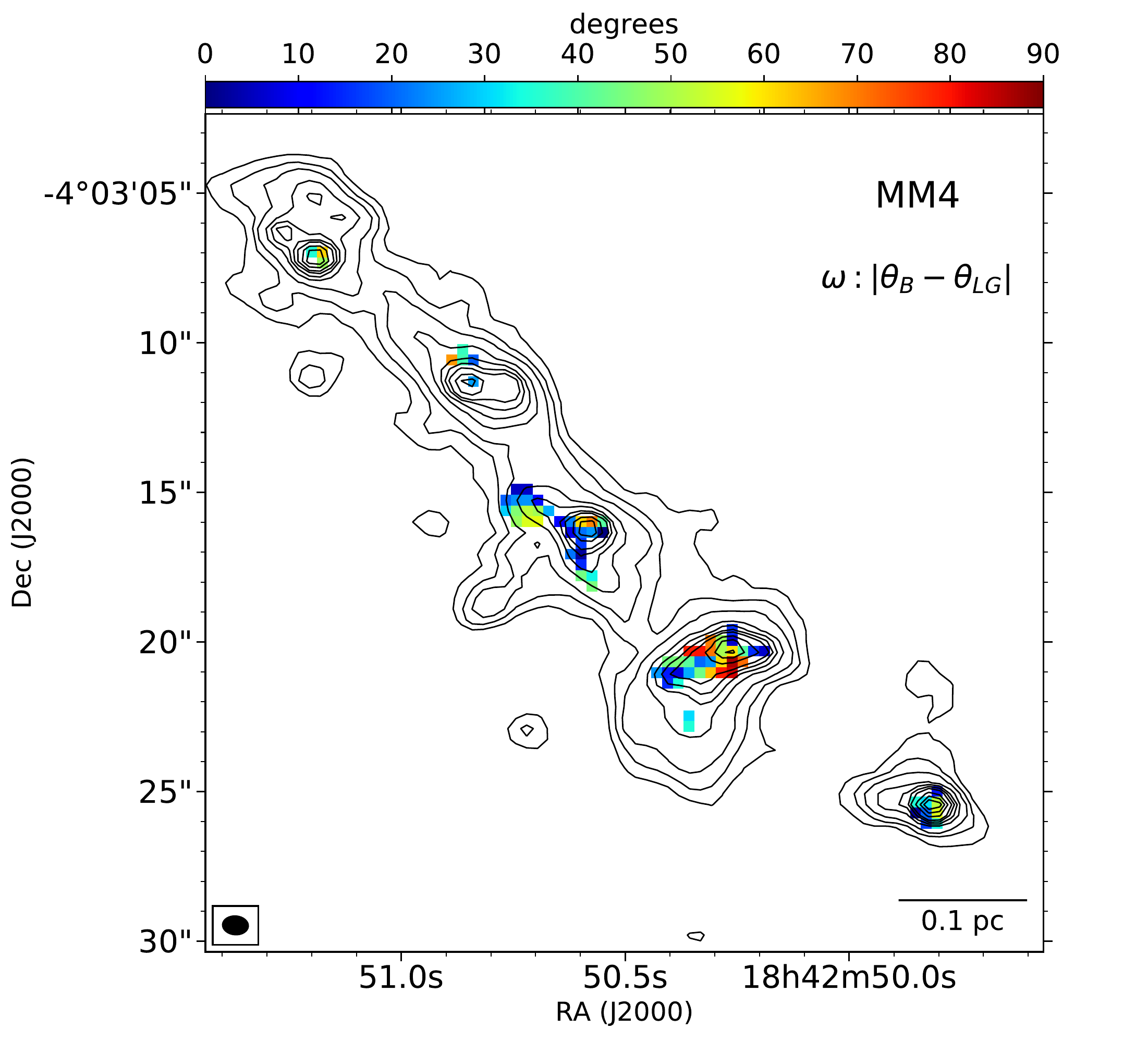}{0.47\textwidth}{}
      }
\caption{Angle difference between orientations of the magnetic field and the local gravity ($\omega = \vert \theta_{\mathrm{B}} - \theta_{\mathrm{LG}} \vert$). Values of $\omega$ are computed at where $PI/\sigma_{PI} > 3$. Contour levels are the same as those in Figure \ref{fig:figG28B}. \label{fig:figG28omega}}
\end{figure}

The angle $\omega$ measures the angular difference between the position angles of the local gravity ($\theta_{\mathrm{LG}}$) and that of the magnetic field ($\theta_{\mathrm{B}}$). Figure \ref{fig:figG28omega} shows the $\omega$ maps for MM1 and MM4. The average values of $\omega$ in MM1 and MM4 are 34$\degr$ and 36$\degr$, respectively. In both MM1 and MM4, the magnetic field and the local gravity are apparently poorly aligned toward the brightest emission peaks, which might be a result of field tangling of complex small scale magnetic field along the line of sight or the perturbation from rotation and accretion. In MM1, the well-aligned magnetic field and local gravity in the northeastern region suggests that the magnetic field morphology is shaped by gravity in this region, while the magnetic field and the local gravity in the southwestern region are poorly aligned, suggesting that the magnetic field has kept its own dynamics. 

\begin{figure}[!tbp]
\gridline{\fig{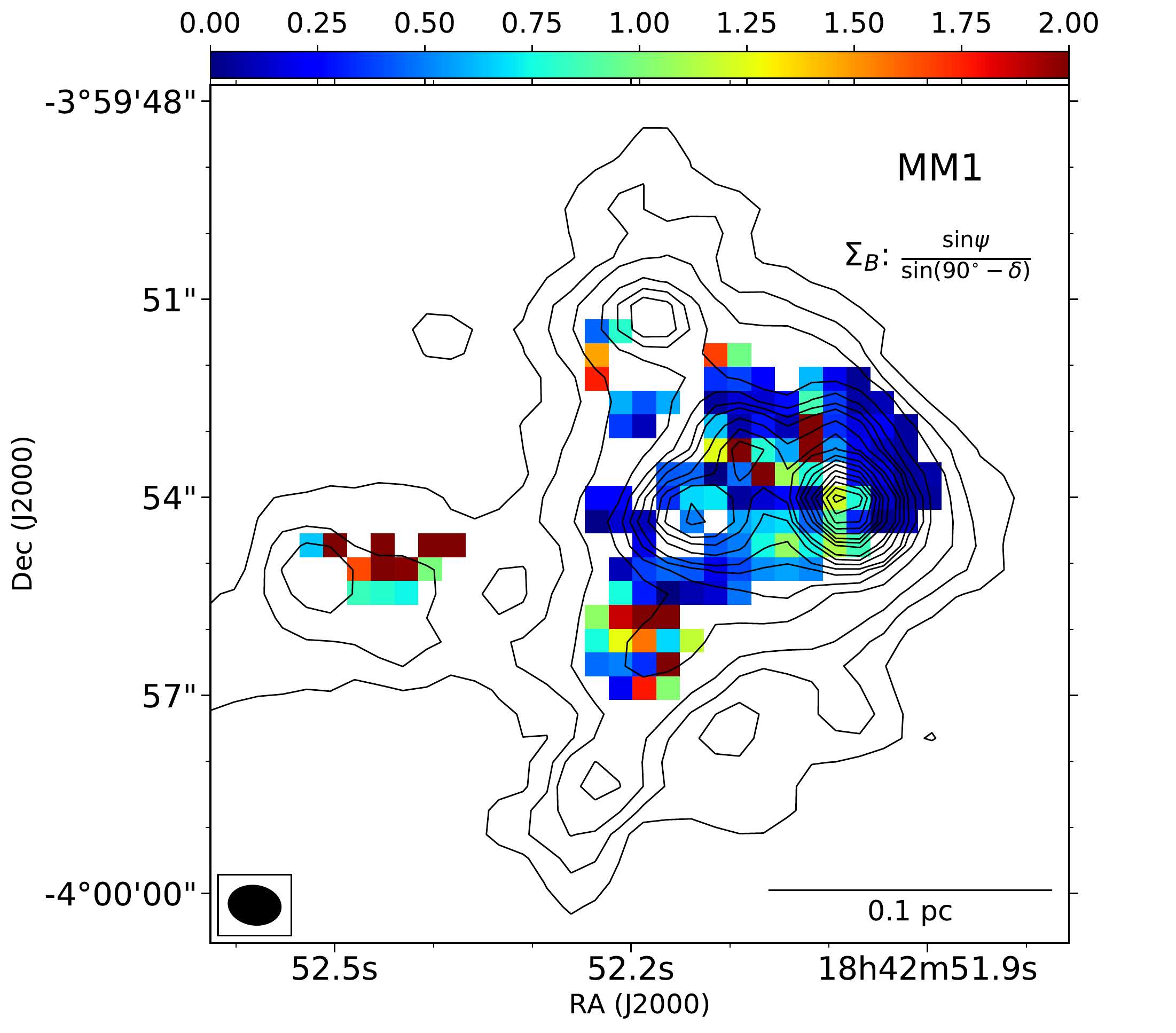}{0.47\textwidth}{(a)}}
\gridline{\fig{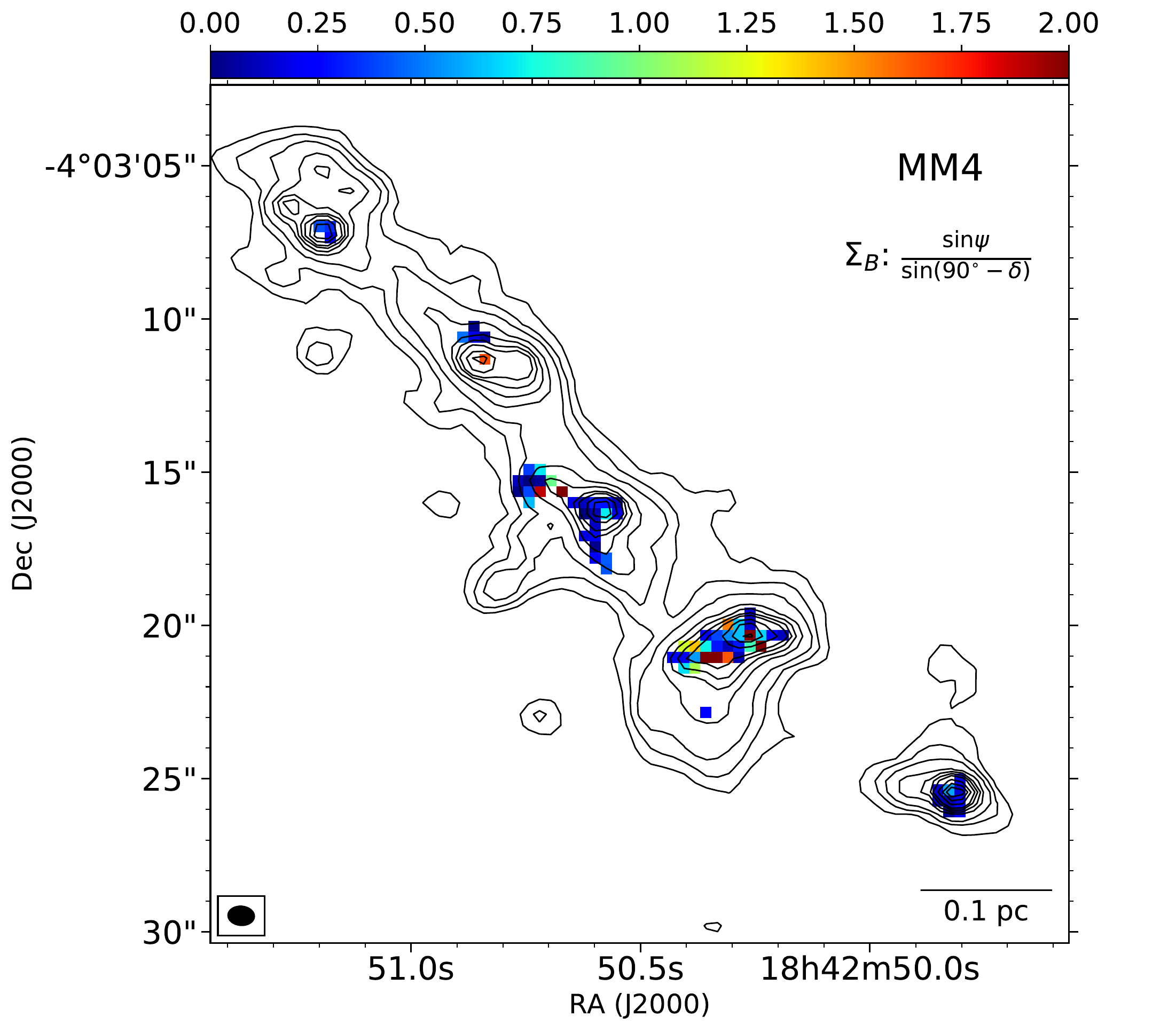}{0.47\textwidth}{(b)}
      }
\caption{Maps of $\Sigma_{\mathrm{B}}$. Values of $\Sigma_{\mathrm{B}}$ are computed at where $PI/\sigma_{PI} > 3$. Contour levels are the same as those in Figure \ref{fig:figG28B}. \label{fig:figG28sigma}}
\end{figure}

If the hydrostatic gas pressure is neglectable, the local ratio between magnetic field force $F_{\mathrm{B}}$ and the gravity force $F_{\mathrm{G}}$ can be measured by the magnetic field-to-gravity force ratio \citep{2012ApJ...747...79K}
\begin{equation}
\Sigma_{\mathrm{B}} = \frac{\sin{\psi}}{\sin{(90\degr - \delta)}} = \frac{F_{\mathrm{B}}}{\vert F_{\mathrm{G}} \vert}.
\end{equation}
Figure \ref{fig:figG28sigma} shows the $\Sigma_{\mathrm{B}}$ maps for MM1 and MM4. The median values of $\Sigma_{\mathrm{B}}$ in MM1 and MM4 are 0.47 and 0.27 respectively. The small values of $\Sigma_{\mathrm{B}}$ indicate that overall the magnetic field cannot solely balance the gravitational force.


\section{Discussion} \label{section:discussion}
\subsection{Uncertainty of the DCF method for super-Alfv\'{e}nic cases}
The accuracy of the DCF method has been under investigation in the literature. The reliability of the DCF method can be tested with with numerical simulations \citep{2001ApJ...546..980O, 2001ApJ...559.1005P, 2001ApJ...561..800H, 2008ApJ...679..537F}. Specifically, \citet{2001ApJ...561..800H} and \citet{2008ApJ...679..537F} investigated the effect of large angular dispersion and the effect of energy equipartition between the turbulent magnetic energy and the turbulent kinetic energy. Their results showed that the DCF method could overestimate the magnetic field strength for super-Alfv\'{e}nic models.

The equipartition between the turbulent magnetic energy and the turbulent kinetic energy is a basic assumption of the DCF method. However, since the perturbation on the magnetic field cannot follow the strong kinetic motions for super-Alfv\'{e}nic cases, the energy equipartition would be broken and the magnetic field strength would be overestimated \citep{2001ApJ...561..800H, 2008ApJ...679..537F}.

Another basic assumption of the DCF method is that the dispersion of polarization position angles corresponds to the ratio between the turbulent and ordered components of the magnetic field. For weak fields with Alfv\'{e}n Mach number $>$1, the magnetic field lines can be easily distorted and the angular dispersion can approach that expected for a random field. As a consequence, the measured angular dispersion cannot reflect the true extent of the perturbed magnetic field and the magnetic field strength would be overestimated \citep{2001ApJ...561..800H, 2008ApJ...679..537F}.

The magnetic field strength of MM1-Core1 and MM4-Core4 is derived from the ADF method (see Section \ref{section:DCF}). The ADF method improves the original DCF method by taking into account the signal-integration effect and the large-scale field. However, the ADF method does not correct for the aforementioned effects of energy equipartition and large angular dispersion. Since the two cores are in super-Alfv\'{e}nic states (see Section \ref{section:DCF}), it is possible that the ADF method has overestimated the magnetic field strength in the two cores.

\subsection{Dynamical state}
The star formation process are governed by the complex interplay of self-gravity and competing processes (e.g., turbulence, thermal pressure, magnetic fields, stellar feedback, and rotation). For decades, the virial theorem has been used to study whether star-forming regions are stable against gravitational collapse. We briefly review the concept of the virial theorem and point out possible mistakes in some previous virial studies in Appendix \ref{virial}. Here we calculate the relevant parameters in MM1-Core1 and MM4-Core4, and discuss the implication on massive star formation.

\subsubsection{Dynamic states in MM1-Core1 and MM4-Core4} \label{sec:virial_g28}
We computed the virial masses and virial parameters for the two massive dense cores MM1-Core1 and MM4-Core4. The thermal velocity dispersion (1D sound speed, $\sigma_{\mathrm{th}}$) is derived by 
\begin{equation}
\sigma_{\mathrm{th}} = \sqrt{\frac{k_\mathrm{B} T}{\mu_\mathrm{p} m_\mathrm{H}}},
\end{equation}
where $\mu_\mathrm{p} = 2.33$ is the conventional mean molecular weight per free particle. The 3D sound speed is estimated as $\sigma_{\mathrm{th,3D}} = \sqrt{3}\sigma_{\mathrm{th}}$. We adopted radial density profiles $\rho \propto r^{-a}$  \citep[$a = 1.6$ and $a = 2.1$ for MM1-Core1 and MM4-Core4, respectively,][]{2009ApJ...696..268Z} for the two cores. The kinetic virial mass is derived by
\begin{equation}
M_{\mathrm{k}} = \frac{3(5-2a)\sigma_{tot}^2R}{(3-a)G},
\end{equation}
where $\sigma_{\mathrm{tot}} = \sqrt{\sigma_{\mathrm{th}}^2 + \sigma_{\mathrm{turb}}^2}$ is the total 1D gas velocity dispersion, $R =$ FWHM$_{\mathrm{mean}}$ is the radius of the core, and $G$ is the gravitational constant. Similarly, the thermal virial mass $M_{\mathrm{th}}$ and the turbulent virial mass $M_{\mathrm{turb}}$ can be derived by replacing $\sigma_{tot}$ with $\sigma_{th}$ or $\sigma_{turb}$, respectively. The ordered magnetic virial mass is given by
\begin{equation}\label{eq:MB}
M_{\mathrm{B}} = \frac{\pi R^2 B}{\sqrt{\frac{3(3-a)}{2(5-2a)}\mu_0\pi G}},
\end{equation}
in SI units or CGS units. The total virial mass accounting for both the ordered magnetic field and the kinetic motions is given by
\begin{equation}\label{eq:MkB}
M_{\mathrm{k+B}} = \sqrt{M^2_{\mathrm{B}} + (\frac{M_{\mathrm{k}}}{2})^2} + \frac{M_{\mathrm{k}}}{2}.
\end{equation}
The corresponding total virial parameter is
\begin{equation}\label{eq:alphakB}
\alpha_{k+B} = \frac{M_{\mathrm{k+B}}}{M}.
\end{equation}
As indicated in Table \ref{tab:ADF}, the turbulent magnetic energy is comparable to the ordered magnetic energy. If we take into account the turbulent magnetic energy in the virial analysis, the modified magnetic virial mass $M_{\mathrm{B}}^{\mathrm{mod}}$, total virial mass $M_{\mathrm{k+B}}^{\mathrm{mod}}$, and total virial parameter $\alpha_{\mathrm{k+B}}^{\mathrm{mod}}$ could be derived from Equations \ref{eq:MB}, \ref{eq:MkB}, and \ref{eq:alphakB} by replacing $B$ with $B(1+B^2_{\mathrm{t}}/B^2_{0})^{\frac{1}{2}}$.

\begin{deluxetable*}{cccccccccccccc}[t!]
\tablecaption{Physical parameters relevant to the virial analysis \label{tab:virial}}
\tablecolumns{14}
\tablewidth{0pt}
\tablehead{
\colhead{Source} &
\colhead{$\sigma_{\mathrm{th,3D}}$} &
\colhead{$\sigma_{\mathrm{turb,3D}}$} & 
\colhead{$V_{\mathrm{A,3D}}$ } & 
\colhead{$M$} &  
\colhead{$M_{\mathrm{th}}$} &
\colhead{$M_{\mathrm{turb}}$} &
\colhead{$M_{\mathrm{B}}$} &  
\colhead{$M_{\mathrm{B}}^{\mathrm{mod}}$} &  
\colhead{$M_{\mathrm{k+B}}$} &  
\colhead{$M_{\mathrm{k+B}}^{\mathrm{mod}}$} &  
\colhead{$\alpha_{\mathrm{k+B}}$}  &
\colhead{$\alpha_{\mathrm{k+B}}^{\mathrm{mod}}$}  \\
\colhead{} & \colhead{(km s$^{-1}$)} &
\colhead{(km s$^{-1}$)} & \colhead{(km s$^{-1}$)} & \colhead{$M_{\odot}$} & \colhead{$M_{\odot}$} & \colhead{($M_{\odot}$)} & \colhead{($M_{\odot}$)}& \colhead{($M_{\odot}$)}& \colhead{($M_{\odot}$)}& \colhead{($M_{\odot}$)}& \colhead{}& \colhead{}
}
\startdata
MM1-Core1&  0.33&  2.42& 2.2& 212.4 & 1.9& 102.8 & 61.1 & 95.5 & 132.8 & 161.2 & 0.63 & 0.76\\
MM4-Core4&  0.24&  0.80& 0.75& 42.6& 0.58& 6.6 & 7.3 & 11.4 & 11.8 & 15.6 & 0.28 & 0.37\\ 
\enddata
\end{deluxetable*}

Table \ref{tab:virial} lists the calculated parameters of the two cores. The nonthermal kinetic energy, thermal kinetic energy, and ordered magnetic energy can be compared by the square of the 3D turbulent velocity dispersion ($\sigma_{\mathrm{turb,3D}}$), 3D thermal velocity dispersion ($\sigma_{\mathrm{turb,3D}}$), and 3D Alfv\'{e}n velocity ($V_{\mathrm{A,3D}}$). Both cores have $(\sigma_{\mathrm{turb,3D}})^2 \gg (\sigma_{\mathrm{th,3D}})^2$ (super-sonic) and $(V_{\mathrm{A,3D}})^2 \gg (\sigma_{\mathrm{th,3D}})^2$, which means the thermal energy only plays a neglectable role in the energy balance. The two cores also have $(\sigma_{\mathrm{turb,3D}})^2 > (V_{\mathrm{A,3D}})^2$, suggesting that the turbulent energy exceeds the ordered magnetic energy. Another way to assess the role of the three forces is to compare the maximum mass that can be supported by each force. The mass ratios show similar trends to the energy ratios in the two cores, except that the ordered magnetic virial mass is larger than the turbulent virial mass in MM4-Core4, which is in contrast to the energy ratio in this core ($E_{\mathrm{turb}} > E_{\mathrm{B}}$). 

The modified total virial parameters $\alpha_{k+B}^{\mathrm{mod}}$ in MM1-Core1 and MM4-Core4 are found to be less than 1 (i.e., sub-virial). As shown in many previous studies, the main uncertainty comes from the mass estimation. If the mass is overestimated by a factor of 2 (i.e., the magnetic field strength overestimated by a factor of $\sqrt{2}$, if we ignore other sources of uncertainties), the $\alpha_{k+B}^{\mathrm{mod}}$ of MM1-Core1 and MM4-Core4 would be 1.3 and 0.58, respectively. Since the $\alpha_{k+B}^{\mathrm{mod}}$ in MM4-Core4 is much less than 1, it is very unlikely that MM4-Core4 is in virial equilibrium even with consideration of the large uncertainty of the analysis. However, we cannot rule out the possibility that MM1-Core1 is in virial equilibrium. More accurate estimations of the dust opacity might be required to reduce the uncertainty of the estimated virial parameter of MM1-Core1. Due to the limited angular resolution of our polarization observations and the lack of observations of appropriate turbulence tracers, we refrain from discussing the virial parameters at condensation scale. 

\subsubsection{Non-equilibrium massive star formation}

Previous observations show that a significant amount of massive star forming clumps and cores are in sub-virial states \citep{2011A&A...530A.118P, 2013ApJ...779..185K, 2013ApJ...779...96T, 2015ApJ...804...37L, 2015ApJ...804..141Z, 2016ApJ...833..209O, 2017ApJ...841...97S, 2018ApJ...855....9L}. However, there are no direct measurements of the magnetic field strength in their analyses, while the magnetic field could provide significant support to regulate the high-mass star formation \citep{2014A&A...567A.116F, 2015ApJ...799...74P}. 

Our observations and calculations offer direct evidence of the role of magnetic fields in the dynamical state of two massive dense cores. As discussed in Section \ref{sec:virial_g28}, MM4-Core4 is in a sub-virial state and MM1-Core1 is likely in a sub-virial state. The sub-virial state indicates that massive dense cores could undergo dynamical collapse in non-equilibrium in the early stage of star formation, which is in agreement with the competitive accretion model. Considering that MM1-Core1 is in a later evolution stage than MM4-Core4 \citep{2009ApJ...696..268Z}, the higher virial parameter of MM1-Core1 might suggest enhanced supports from the gas motion and the magnetic field as cores evolve.

\subsection{Comparing outflow axis with magnetic field orientation in condensation}
The importance of magnetic fields in star formation can be studied by comparing the orientations of molecular outflows and ambient magnetic fields. Results from MHD simulations \citep{2017ApJ...834..201L} suggest that strong ambient magnetic fields tend to align the axis of protostellar outflows because of magnetic braking \citep{2003ApJ...599..363A}, while the outflow-field orientation in the weak-field case is more random. 

Observationally, a compilation of all of the outflow-versus-magnetic-field angles derivable to date in low-mass protostellar cores shows that overall the magnetic fields in low-mass dense cores are randomly aligned with outflows \citep{2019FrASS...6....3H}. \citet{2018A&A...616A.139G} found a bimodal distribution of the angles between the envelope-scale magnetic field orientation and outflow axis in low-mass class 0 objects, suggesting that the magnetic field could play an important role in regulating the direction of outflows in the early stage of low-mass star formation. 

In the case of massive star formation region, an SMA survey toward high-mass clumps found no strong correlation between the outflow axis and the magnetic field orientation in evolved dense cores with a total of 21 outflow samples \citep{2014ApJ...792..116Z}. Since molecular outflows are believed to be driven by MHD winds originated in or near the circumstellar disk \citep{2000prpl.conf..789S, 2006MNRAS.365.1131P}, the observed misalignment between the magnetic field orientation and the outflow axis indicates a less dynamically important role of magnetic fields from the core scale down to the disk scale. 

With ALMA CO (2-1) and SiO (5-4) observations, \citet{2019ApJ...874..104K} has determined the plane-of-sky position angle of the outflows ($\theta_{\mathrm{outflow}}$) in IRDC G28.34 (except those in MM1). Our polarization observations have resolved the magnetic field structure down to scales of $\sim$0.02 pc, which allows us to compare the outflow axis with the magnetic field orientation at condensation scales. Since MM1 is in a later stage of star formation and the position angle of the outflows in MM1 is not determined due to the complexity of the CO emission, we limit the comparison in MM4 and MM9. 

\begin{figure*}[!tbp]
\gridline{\fig{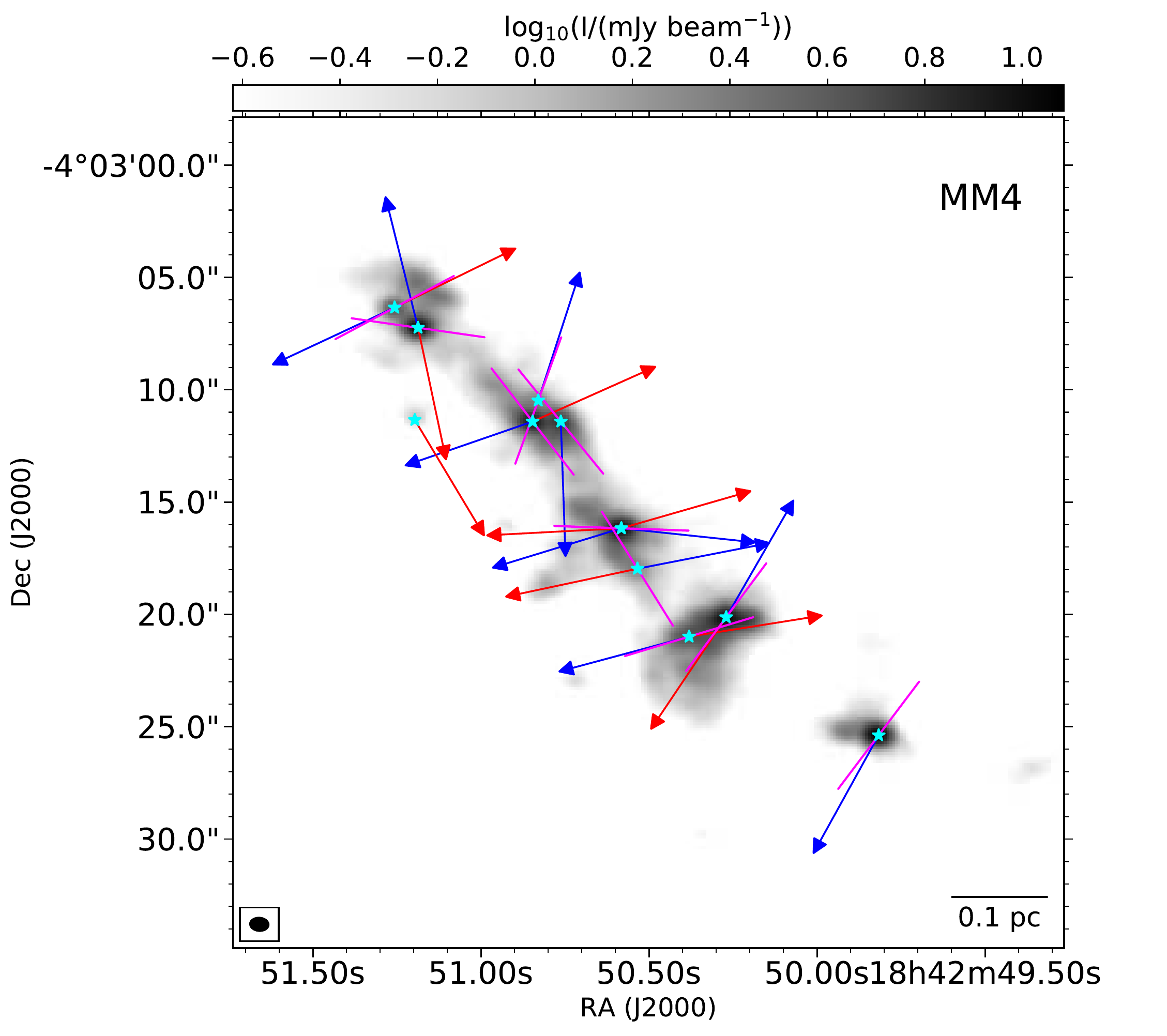}{0.47\textwidth}{}
\fig{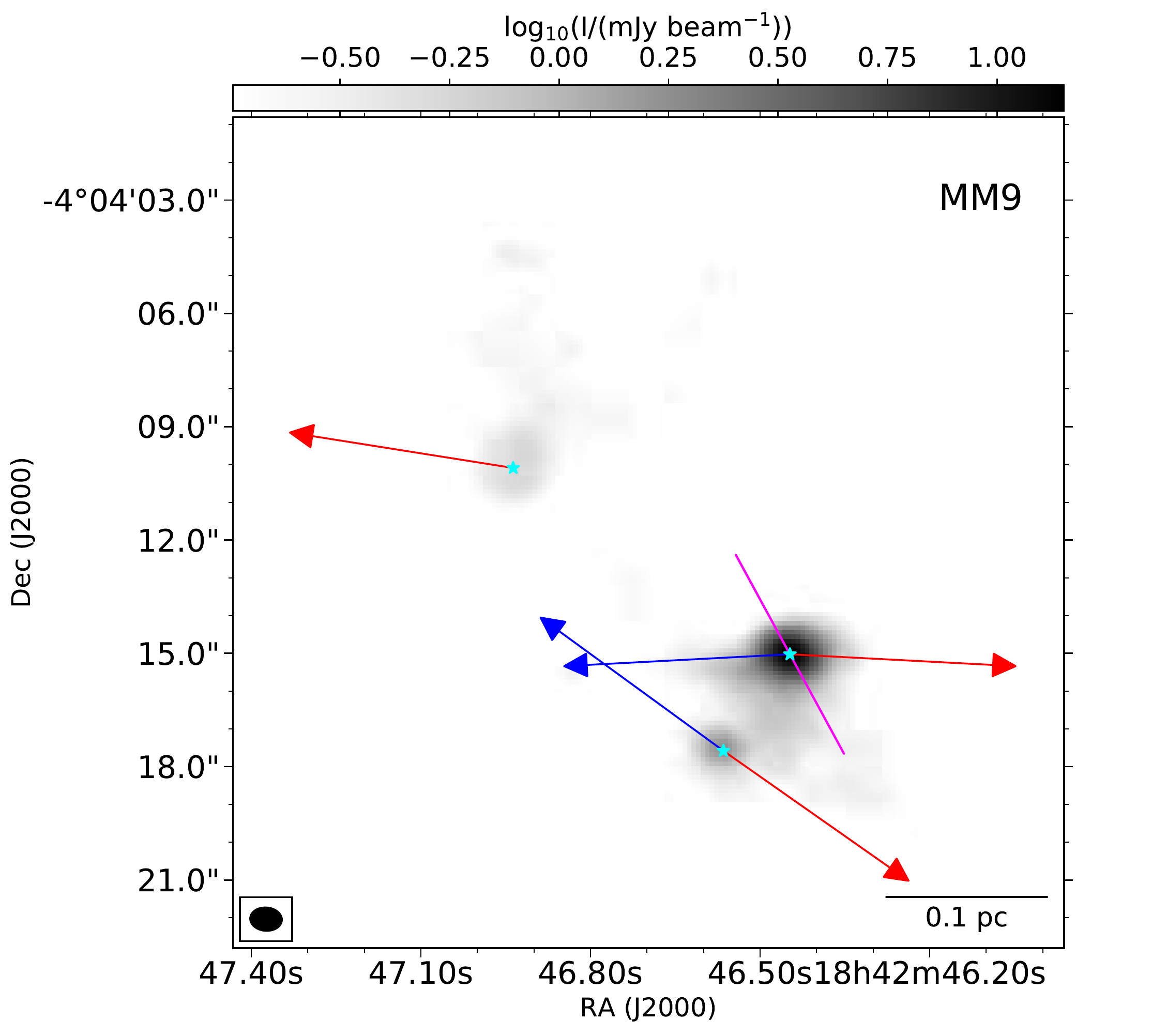}{0.47\textwidth}{}
      }
\caption{Summary of orientations of magnetic fields and outflows in MM4 and MM9. The Stokes $I$ of the 1.3 mm continuum is shown in grey scale. The star symbols denote the position of dense condensations \citep{2015ApJ...804..141Z, 2019ApJ...873...31K} that are associated with outflows. The large magenta bars indicate the average magnetic orientation for each condensation. The blue- and red-shifted outflow lobes are shown as blue and red dashed arrows, respectively. Bars and arrows are of arbitrary unit length.  \label{fig:figG28out_B}}
\end{figure*}

\begin{figure}[htbp]
\centering
\includegraphics[scale=.5]{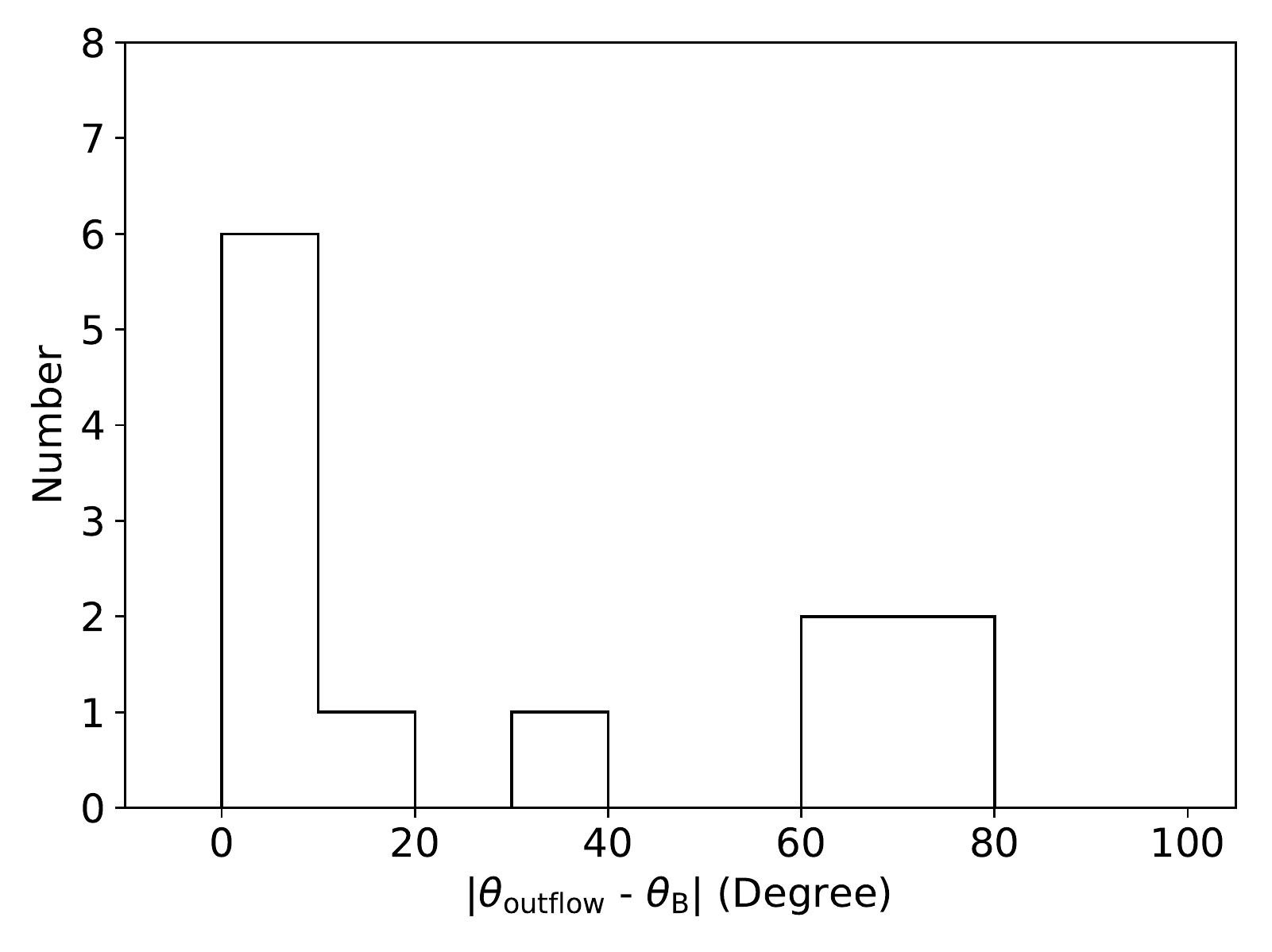}
\caption{Angular difference between the outflow axis and the $\sim$0.02 pc scale magnetic field orientation in the condensations where outflows originate for the 12 outflow samples in MM4 and MM9. \label{fig:figG28angout_B}}
\end{figure}

The magnetic field orientation in each condensation  ($\theta_{\mathrm{B}}$) in MM4 and MM9 is measured within one ALMA synthesized beam. There are 15 outflows in MM4 and MM9, among which 12 of them are associated with polarization detections in the condensation (see Figure \ref{fig:figG28out_B}). Figure \ref{fig:figG28angout_B} shows the angular difference between the outflow axis and the magnetic field orientation in the condensations where outflows originate for the 12 outflow samples in MM4 and MM9. The angular difference between the outflow axis and the field orientation appears to be in an approximate bimodal distribution, where the outflow axis tends to be either parallel or perpendicular to the orientation of magnetic fields. In half of the objects, the magnetic field is aligned within 10$\degr$ of the outflow axis, suggesting that magnetic fields could be dynamically important from the condensation scale to the disk scale in the early stage of high-mass star formation. However, the observed magnetic field and outflow position angles are projected on the plane of sky. Observations toward larger samples of magnetic fields in IRDCs are essential to help us to rule out the possibility that the observed distribution is due to the projection effect.

\subsection{Fragmentation and clustered star formation in MM1}
The two competing massive star formation models have different predictions on the fragmentaion of molecular clouds. The competitive accretion model proposes that the initial fragments in molecular clouds should be in thermal Jeans mass. The fragments and protostars therein near the gravitational center can thus accrete at a higher rate and form massive stars. Alternatively, the turbulent core accretion model suggests massive stars are formed via the monolithic collapse of a massive core that is supported by turbulent pressure rather than thermal pressure.

\begin{figure}[htbp]
\centering
\includegraphics[scale=.63]{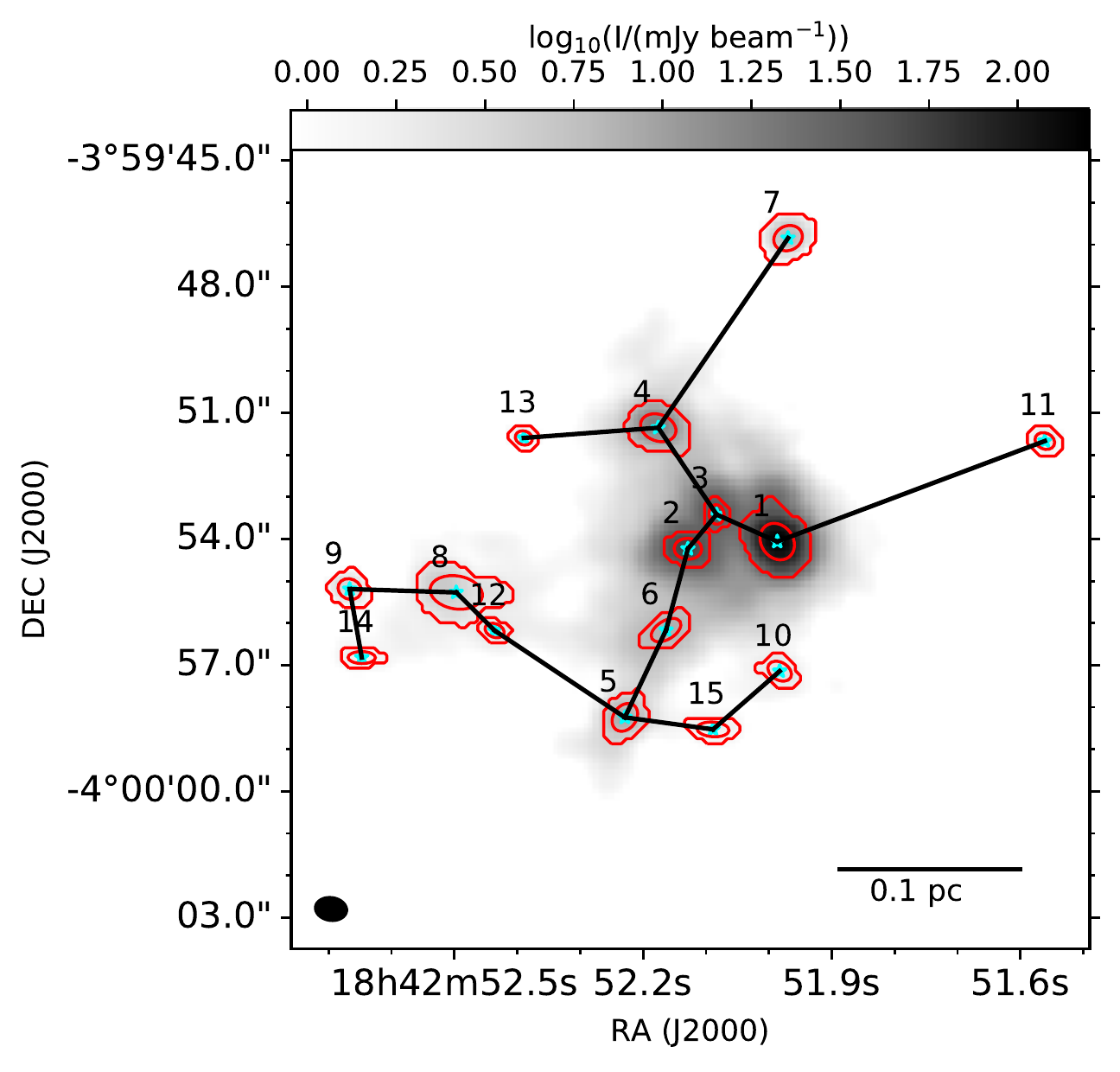}
\caption{Results of the MST (shown in black segments) in MM1. Red contours and ellipses are the same as those in Figure \ref{fig:figMM1dendro}. Star symbols mark the position of the condensations identified by dendrogram.  \label{fig:figG28MM1separ}}
\end{figure}

Our observations may help to distinguish between the two models. To quantify the separation of condensations in MM1, we applied the minimum spanning tree (hereafter MST) technique on these condensations using the python package MiSTree \citep{2019JOSS....4.1721N}. The MST method connects the condensations with straight lines and minimizes the sum of the line length. Figure \ref{fig:figG28MM1separ} show the MST for MM1. The separation of condensations in MM1 ranges from 0.025 pc to 0.16 pc. 

Since the maximum recoverable scale of our ALMA observations is $\sim$7$\arcsec$ for MM1, we focus on interpreting the fragmentation in the core MM1-Core1 (including Condensations 1, 2, 3, 4, 5, and 6) that has a size of $\sim$6$\arcsec$. The separation of condensations in MM1-Core1 ranges from 0.025 pc to 0.058 pc and the average separation ($L_{av,MST}$) of these condensations is calculated to be 0.044 pc. As the measured separations are projected on the plane-of-sky, we multiplied the observed separation by a factor of $\pi /2$ to correct for the statistical projection effect \citep{2019ApJ...886..102S} and the corrected average separation ($L_{av,MST,cor}$) is estimated to be 0.069 pc. The mass of the condensations in MM1-Core1 ranges from 3.7 to 35.5 M$_{\odot}$ (see Table \ref{tab:dendro}) and the average mass ($M_{av}$) of these condensations is calculated to be 14.3 M$_{\odot}$. 

With a temperature of 30 K and a density of 3.2 $\times$ 10$^6$ cm$^{-3}$, the thermal Jeans length
\begin{equation}\label{equation:jeanslength}
\lambda_{\mathrm{J,th}} = \sigma_{\mathrm{th,3D}}\sqrt{\frac{\pi}{G\mu_{\mathrm{H_2}} m_{\mathrm{H}} n_{\mathrm{H_2}}}}
\end{equation}
and the thermal Jeans mass
\begin{equation}\label{equation:jeansmass}
M_{\mathrm{J,th}} = \frac{4}{3}\pi \mu_{\mathrm{H_2}} m_{\mathrm{H}} n_{\mathrm{H_2}} (\frac{\lambda_{\mathrm{J,th}}}{2})^3
\end{equation}
of MM1-Core1 are estimated to be 0.019 pc and 0.76 M$_{\odot}$, respectively. The $M_{av}$ is about 19 times larger than the $M_{\mathrm{J,th}}$ and the $L_{av,MST,cor}$ is about 3 times larger than the $\lambda_{\mathrm{J,th}}$. Considering that the resolution of our observations ($\sim$0.02 pc) is insufficient to resolve the foreshortened thermal Jeans length (2$/\pi$ times $\lambda_{\mathrm{J,th}}$) of 0.012 pc, we can not rule out the possibility that the detected condensations would further fragment into smaller structures with separations $\sim$ $\lambda_{\mathrm{J,th}}$. However, even if each condensation in MM1-Core1 further fragments into several smaller structures, the average mass of the fragments is unlikely to be close to the $M_{\mathrm{J,th}}$ since $M_{av}$ is much larger than $M_{\mathrm{J,th}}$. 

Replacing the thermal velocity dispersion with the turbulent velocity dispersion in Equations \ref{equation:jeanslength} and \ref{equation:jeansmass}, the turbulent Jeans length and the turbulent Jeans mass in MM1-Core1 are estimated to be 0.14 pc and 309 M$_{\odot}$, respectively, which are much larger than the $L_{av,MST,cor}$ and the $M_{av}$. 

Thus, we conclude that the fragmentation in MM1-Core1 cannot be solely explained by the thermal Jeans fragmentation nor the turbulent Jeans fragmentation, which are inconsistent with either the competitive accretion model or the turbulent core accretion model. These discrepancies indicate that the physical condition in MM1-Core1 might have deviated from the initial condition of the core that controls the fragmentation process. On the other hand, we find that the mass of the condensations near the core center is more massive than that of the condensations in the outer region (see Table \ref{tab:dendro}). Thus, we expect MM1 to form a cluster of stars with several massive stars near the center and a population of low-mass stars in the outer region. This picture seems to agree with the competitive accretion model.
%

\section{Summary} \label{section:summary}

With ALMA 1.3 mm dust polarization observations, we have presented a study of the magnetic field of three massive clumps in the IRDC G28.34. The main conclusions are as follows:
\begin{enumerate}

\item Polarized dust emission are detected in all the three clumps. The magnetic field morphology varies in different regions. Trends of decreasing polarization percentage with increasing intensity are detected in MM1 and MM4. The $P$-$I$ relation in MM1 is shallower than that in MM4, which might be explained by the improved grain alignment efficiency due to enhanced internal radiation in more evolved regions. 
\item From the dust polarization maps, we measured the plane-of-sky magnetic field strength in two massive dense cores, MM1-Core1 ($M = 212.4 M_{\odot}$) and MM4-Core4 ($M = 43.0 M_{\odot}$), with the auto-correlation function method. The $B_{\mathrm{pos}}$ is found to be $\sim$1.6 mG and $\sim$0.32 mG in MM1-Core1 and MM4-Core4, respectively. 
\item We studied the dynamical state in MM1-Core1 and MM4-Core4 by calculating the virial parameter that includes both the turbulence and the magnetic field. The results of the virial analysis suggest that MM4-Core4 is in a sub-virial state and MM1-Core1 is likely in a sub-virial state, which signifies a dynamical massive star formation in non-equilibrium.
\item We compared the orientations of the magnetic field, local gravity, and intensity gradient with the polarization-intensity gradient-local gravity method. We found that the intensity gradient closely follows the local gravity in the three clumps. The magnetic field is found to be randomly aligned with the intensity gradient, but better aligned with the local gravity in MM1 and MM4. The average magnetic field to gravity force ratio are found to be smaller than 1 in MM1 and MM4, which suggests that the magnetic force cannot solely prevent the gravitational collapse.
\item With a total of 12 outflow samples in MM4 and MM9, the angle between orientations of the magnetic field at condensation scales and the outflow axis shows an approximate bimodal distribution with half of the outflows aligned within 10$\degr$ of the magnetic field, which suggests that magnetic fields could play an important role from the condensation scale to the disk scale in the early stage of massive star formation.
\item We identified the dense condensations in MM1 with the dendrogram method and characterized the separation of these condensations using the minimum spanning tree technique. The mass of the condensations and the average minimum separation between condensations in MM1-Core1 are found to be larger than the values predicted by thermal Jeans fragmentation and smaller than the values predicted by turbulent Jeans fragmentation. Thus we propose that the physical condition in MM1-Core1 might has deviated from the initial condition that controls the fragmentation. 

\end{enumerate}

\acknowledgments 
We thank Dr. Diego Falceta-Gon\c{c}alves, Dr. J. Michail, and Dr. Fabio Santos for sharing their codes for plotting the LIC maps and Dr. Shuo Kong for sharing the catalog of the outflows in G28.34. We are indebt to the anonymous referee whose constructive comments improved the presentation and clarity of the paper. K.Q. and J.L. are supported by National Key R\&D Program of China No. 2017YFA0402600. K.Q. and J.L. acknowledge the support from National Natural Science Foundation of China (NSFC) through grants U1731237, 11590781, and 11629302. J.L. acknowledges the support from the program of China Scholarship Council (No. 201806190134) and from the Smithsonian Astrophysical Observatory pre-doctoral fellowship. JMG is supported by the Spanish grant AYA2017-84390-C2-R (AEI/FEDER, UE). ZYL is supported in part by NASA 80NSSC18K1095 and NSF AST-1716259 and 1815784. This paper makes use of the following ALMA data: ADS/JAO.ALMA\#2016.1.00248.S and ADS/JAO.ALMA\#2017.1.00793.S. ALMA is a partnership of ESO (representing its member states), NSF (USA) and NINS (Japan), together with NRC (Canada), MOST and ASIAA (Taiwan), and KASI (Republic of Korea), in cooperation with the Republic of Chile. The Joint ALMA Observatory is operated by ESO, AUI/NRAO and NAOJ. The National Radio Astronomy Observatory is a facility of the National Science Foundation operated under cooperative agreement by Associated Universities, Inc. This research made use of astrodendro, a Python package to compute dendrograms of Astronomical data (http://www.dendrograms.org/), APLpy, an open-source plotting package for Python \citep{2012ascl.soft08017R}, Astropy, a community-developed core Python package for Astronomy \citep{2013A&A...558A..33A}, and Matplotlib, a Python 2D plotting library for Python \citep{2007CSE.....9...90H}.
\facility{Atacama Large Millimeter/Submillimeter Array (ALMA).}
\software{APLpy \citep{2012ascl.soft08017R}, Astropy \citep{2013A&A...558A..33A}, Matplotlib \citep{2007CSE.....9...90H}.}


\appendix
\section{Virial theorem}\label{virial}
Ignoring the surface kinetic energy, the virial theorem is written as: 
\begin{equation}
\frac{1}{2}\frac{d^2 I}{dt^2} = 2E_{\mathrm{k}} + E_{\mathrm{B}} + E_{\mathrm{G}},
\end{equation}
where $I$ is the moment of inertia, $E_{\mathrm{k}}$ is the kinetic energy, $E_{\mathrm{B}}$ is the magnetic energy, and $E_{\mathrm{G}}$ is the gravitational energy. For a sphere with a radial density profile $\rho \propto r^{-a}$ (for a uniform density, $a = 0$), the gravitational energy is given by
\begin{equation}
E_{\mathrm{G}} = - \frac{(3-a)}{(5-2a)} \frac{GM^2}{R},
\end{equation}
where $G$ is the gravitational constant, $M$ is the mass of the considered structure, and $R$ is the radius of the structure. The magnetic energy is given by
\begin{equation}
E_{\mathrm{B}} = \frac{B^2 V}{2\mu_0},
\end{equation}
where $V = 4 \pi R^3/3$ is the volume of the structure. The kinetic energy is the sum of the thermal energy ($E_{\mathrm{th}}$) and the turbulent energy ($E_{\mathrm{turb}}$): 
\begin{equation}
E_{\mathrm{k}} = E_{\mathrm{th}} + E_{\mathrm{turb}} = \frac{3}{2} M\sigma_{tol}^2,
\end{equation}
where $\sigma_{\mathrm{tot}}$ is the total 1D gas velocity dispersion. The thermal energy and the turbulent energy can be derived by 
\begin{equation}
E_{\mathrm{th}} = \frac{3}{2} M\sigma_{\mathrm{th}}^2,
\end{equation}
and 
\begin{equation}
E_{\mathrm{turb}} = \frac{3}{2} M\sigma_{\mathrm{turb}}^2,
\end{equation}
respectively, where $\sigma_{\mathrm{th}}$ is the 1D thermal velocity dispersion and $\sigma_{\mathrm{turb}}$ is the 1D turbulent velocity dispersion. 

For a non-magnetized ($E_{\mathrm{B}} = 0$) sphere, the structure is stable when $2E_{\mathrm{k}} + E_{\mathrm{G}} < 0$, which requires $M < M_{\mathrm{k}}$, where
\begin{equation}
M_{\mathrm{k}} = \frac{3(5-2a)\sigma_{tot}^2R}{(3-a)G}
\end{equation}
is the kinetic virial mass. Similarly, we can define the thermal virial mass $M_{\mathrm{th}}$ and the turbulent virial mass $M_{\mathrm{turb}}$ by replacing $\sigma_{tot}$ with the corresponding velocity dispersion to account for the thermal component and the turbulent component of the virial mass separately. \citet{1992ApJ...395..140B} introduced a kinetic virial parameter 
\begin{equation}
\alpha_{k} = \frac{2E_{\mathrm{k}}}{\vert E_{\mathrm{G}} \vert} = \frac{3(5-2a)\sigma_{tot}^2R}{(3-a)GM} = \frac{M_{\mathrm{k}}}{M}
\end{equation}
to represent the ratio of the kinetic energy and the gravitational energy. We note that $M < M_{\mathrm{k}}$ (kinetically super-virial) is equivalent to $\alpha_{k} > 1$. For comparison with $M_{\mathrm{k}}$, a Bonnor-Ebert sphere \citep{1956MNRAS.116..351B, 1957ZA.....42..263E} in isothermal hydrostatic equilibrium is stable when $M < M_{\mathrm{BE}}$, where $M_{\mathrm{BE}}$ is the Bonnor-Ebert critical mass given by 
\begin{equation}
M_{\mathrm{BE}} = 2.43 \frac{\sigma_{tot}^2R}{G}.
\end{equation}
For a uniform sphere (a=0), $M < M_{\mathrm{BE}}$ is equivalent to $\alpha_{k} > 2.06$. 


If the kinetic energy is neglectable ($E_{\mathrm{k}} = 0$), a stable sphere has $E_{\mathrm{B}} + E_{\mathrm{G}} < 0$, which implies $M < M_{\mathrm{B}}$, where
\begin{equation}
M_{\mathrm{B}} = \frac{\pi R^2 B}{\sqrt{\frac{3(3-a)}{2(5-2a)}\mu_0\pi G}} 
\end{equation}
is the magnetic virial mass in SI units or CGS units. We noticed some previous studies \citep[e.g.,][]{2008ApJ...684..395H, 2011A&A...530A.118P, 2016ApJ...833..209O, 2017ApJ...841...97S, 2019FrASS...6....3H} have written the magnetic virial mass of a uniform (a=0) sphere as
\begin{equation}
M'_{\mathrm{B}} = \frac{5RV_{\mathrm{A,3D}}^2}{6G} = \frac{5RB^2}{6\mu_0 \rho G} = \frac{10\pi R^4B^2}{9\mu_0 GM},
\end{equation}
where $V_{\mathrm{A,3D}} = \frac{B}{\sqrt{\mu_0 \rho}}$ is the 3D Alfv\'{e}n velocity in SI units or CGS units. The $M'_{\mathrm{B}}$ is derived by writing the magnetic energy as 
\begin{equation}
E'_{\mathrm{B}} = \frac{B^2 M}{2\mu_0 \rho}
\end{equation}
and solving the critical mass for $E'_{\mathrm{B}} + E_{\mathrm{G}} = \frac{B^2 M}{2\mu_0 \rho} - \frac{3}{5} \frac{GM^2}{R} < 0$. However, we point out that since $M$, $\rho$, and $R$ are not independent ($M = 4 \pi R^3 \rho/3$), the magnetic virial mass should be written as a function of $R$ or $\rho$, but not both $R$ and $\rho$. Thus $M'_{\mathrm{B}}$ does not accurately estimate the magnetic virial mass. For a magnetically supercritical ($M > M_{\mathrm{B}}$) structure, $M'_{\mathrm{B}} = \frac{10\pi R^4B^2}{9\mu_0 GM} < \frac{10\pi R^4B^2}{9\mu_0 GM_{\mathrm{B}}} = M_{\mathrm{B}}$ would underestimate the magnetic virial mass, and vice versa. For comparison with $M_{\mathrm{B}}$, the magnetic critical mass of an isothermal disk \citep{1978PASJ...30..671N} has been derived to be 
\begin{equation}
M_{B,disk} = \frac{\pi R^2 B}{\sqrt{\mu_0 \pi G}}.
\end{equation}
\citet{2004ApJ...600..279C} introduced a parameter $\lambda$ to state the mass-to-flux ratio in units of its critical value: 
\begin{equation}
\lambda = \frac{(M/\Phi)_{observed}}{(M/\Phi)_{critical}} = \frac{M}{M_{B,disk}},
\end{equation}
where $(M/\Phi)_{observed}$ is the observed mass-to-magnetic-flux ratio: 
\begin{equation}
(\frac{M}{\Phi})_{observed} = \frac{\mu_{\mathrm{H_2}} m_{\mathrm{H}}N(H_2)}{B},
\end{equation}
and $(M/\Phi)_{critical}$ is the critical mass-to-magnetic-flux ratio:
\begin{equation}
(\frac{M}{\Phi})_{critical} = \frac{1}{\sqrt{\mu_0 \pi G}}.
\end{equation}
For magnetically subcritical structure with mass less than the magnetic critical mass, the magnetic field alone can prevent the structure from collapse.

Taking into account both the magnetic energy and the kinetic energy, a structure is stable when $2E_{\mathrm{k}} + E_{\mathrm{B}} + E_{\mathrm{G}} < 0$. The critical virial mass is calculated to be 
\begin{equation}
M_{\mathrm{k+B}} = \sqrt{M^2_{\mathrm{B}} + (\frac{M_{\mathrm{k}}}{2})^2} + \frac{M_{\mathrm{k}}}{2}.
\end{equation}
It should be noted that $M_{\mathrm{k}} + M_{\mathrm{B}}$ is systematically larger than $M_{\mathrm{k+B}}$ \citep{1992ApJ...395..140B} and the largest ratio between $M_{\mathrm{k}} + M_{\mathrm{B}}$ and $M_{\mathrm{k+B}}$ is 1.25 when $M_{\mathrm{B}} = 0.67 M_{\mathrm{k}}$, Therefore some previous studies using $M_{\mathrm{B}} + M_{\mathrm{k}}$ to represent the total virial mass may have slightly overestimated the support from the magnetic field and gas motion. Following the definition of $\alpha_{k}$, we can define a total virial parameter 
\begin{equation}
\alpha_{k+B} = \frac{M_{\mathrm{k+B}}}{M}.
\end{equation}
%

\end{document}